\documentclass[%
 reprint,
superscriptaddress,
nofootinbib,
 amsmath,amssymb,aps,prd
floatfix,
]{revtex4-1}

\usepackage{graphicx}
\usepackage{dcolumn}
\usepackage{bm}
\usepackage{hyperref}
\usepackage{color}
\usepackage{comment}
\usepackage{bbm}
\usepackage{setspace}

\newcommand{\aeq}{a_\text{eq}}
\newcommand{\MPl}{M_\text{Pl}}
\newcommand{\Heq}{H_\text{eq}}
\newcommand{\dd}{{\rm d}}

\newcommand{\pare}[1]{\left(#1\right)}

\definecolor{deepblue}{rgb}{0.2,0.2,0.8}

\definecolor{deepred}{rgb}{0.8,0.2,0.2}

\definecolor{mulberry}{rgb}{0.5,0,0.5}

\definecolor{teal}{rgb}{0,0.5,0.5}

\definecolor{cerulean}{rgb}{0, 0.75, 1}

\definecolor{orange}{rgb}{1,0.5,0}

\definecolor{green}{rgb}{0,0.8,0.2}

\newcommand{\vect}[1]{\boldsymbol{\mathbf{#1}}}

\begin{document}


\title{
The Large-Misalignment Mechanism for the Formation of Compact Axion Structures:
Signatures from the QCD Axion to Fuzzy Dark Matter}

\author{Asimina Arvanitaki}
 \email{aarvanitaki@perimeterinstitute.ca}
 \affiliation{Perimeter Institute for Theoretical Physics, Waterloo, Ontario N2L 2Y5, Canada}
\author{Savas Dimopoulos}
 \email{savas@stanford.edu}
 \affiliation{Stanford Institute for Theoretical Physics, Stanford University, Stanford, California 94305, USA}
\author{Marios Galanis}
 \email{mgalanis@stanford.edu}
 \affiliation{Stanford Institute for Theoretical Physics, Stanford University, Stanford, California 94305, USA}
 \author{Luis Lehner}
 \email{llehner@perimeterinstitute.ca}
 \affiliation{Perimeter Institute for Theoretical Physics, Waterloo, Ontario N2L 2Y5, Canada}
\author{Jedidiah O. Thompson}
 \email{jedidiah@stanford.edu}
 \affiliation{Stanford Institute for Theoretical Physics, Stanford University, Stanford, California 94305, USA}
\author{Ken Van Tilburg}
 \email{kvt@kitp.ucsb.edu}
 \affiliation{Center for Cosmology and Particle Physics, Department of Physics, New York University, New York, New York 10003, USA}
\affiliation{School of Natural Sciences, Institute for Advanced Study, Princeton, New Jersey 08540, USA}
\affiliation{Kavli Institute for Theoretical Physics, University of California, Santa Barbara, California 93106, USA}

\date{\today}

\begin{abstract}
Axions are some of the best motivated particles beyond the Standard Model. We show how the attractive self-interactions of dark matter (DM) axions over a broad range of masses, from  $10^{-22}$~eV to $10^7$~GeV, can lead to nongravitational growth of density fluctuations and the formation of bound objects. This structure formation enhancement is driven by parametric resonance when the initial field misalignment is large, and it affects axion density perturbations on length scales of order the Hubble horizon when the axion field starts oscillating, deep inside the radiation-dominated era. This effect can turn an otherwise nearly scale-invariant spectrum of adiabatic perturbations into one that has a spike at the aforementioned scales, producing objects ranging from dense DM halos to scalar-field configurations such as solitons and oscillons. We call this class of cosmological scenarios for axion DM production ``the large-misalignment mechanism."

We explore observational consequences of this mechanism for axions with masses up to 10~eV. For axions heavier than $10^{-5}$~eV, the compact axion halos are numerous enough to significantly impact Earth-bound direct detection experiments, yielding intermittent but coherent signals with repetition rates exceeding one per decade and crossing times less than a day. These episodic increases in the axion density and kinematic coherence suggest new approaches for axion DM searches, including for the QCD axion. Dense structures made up of axions from $10^{-22}$~eV to $10^{-5}$~eV are detectable through gravitational lensing searches, and their gravitational interactions can also perturb baryonic structures and alter star formation. At very high misalignment amplitudes, the axion field can undergo self-interaction-induced implosions long before matter-radiation equality, producing potentially-detectable low-frequency stochastic gravitational waves.
\end{abstract}

\maketitle

\begin{spacing}{0.95}
\tableofcontents
\end{spacing}

\section{Introduction} 
\label{sec:introduction}

\begin{figure*}[ht]
\includegraphics[width=0.97\textwidth]{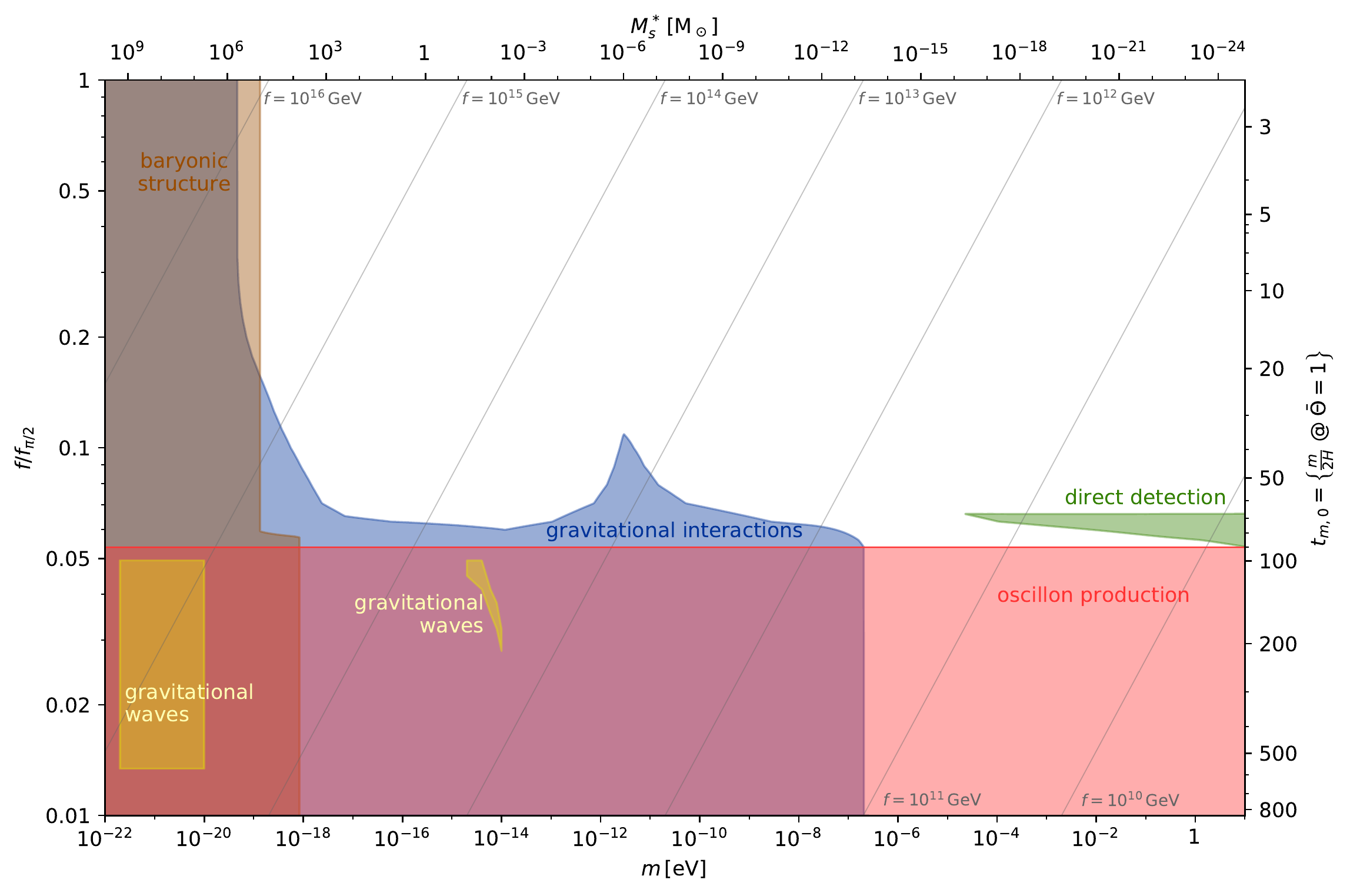}
\caption{Summary of signatures for axions with mass $m$, decay constant $f$, cosine potential of Eq.~\ref{eq:cosinepot}, and an initial axion misalignment chosen such that the axion accounts for all DM. The left axis shows $f$ normalized relative to $f_{\pi/2}$, the value for which the initial axion misalignment is $|\phi_0|/f=\pi/2$; for $f/f_{\pi/2}$ decreases, the misalignment has to be closer to $\pi$. Diagonal gray lines represent contours of constant $f$. The top axis displays the typical halo scale mass $M_s^*$ whose density is maximally enhanced by the effects of the attractive axion self-interactions (see Eq.~\ref{eq:Msstar}). The right axis shows the time $t_{m,0}$ in Compton units for which the amplitude of the axion field oscillation is $\bar \Theta\equiv \bar \phi/f  =1$. For axions lighter than $10^{-5}\,\mathrm{eV}$, the enhanced-density halos can be detectable through their gravitational (lensing) interactions (blue).  Axions heavier than $10^{-5}\,\mathrm{eV}$ can produce ``femto-halos'' lighter than $10^{-15}\,\mathrm{M}_\odot$ that have important consequences for direct detection experiments (green). Axions lighter than $10^{-18}\,\mathrm{eV}$ can affect baryonic structures and accelerate star formation in the early Universe (brown). At low $f$, self-interaction-induced collapse into oscillons happens prior to matter-radiation equality (red), a process that produces gravitational waves, which may be detectable in the yellow region. Signature contours are extracted from Figs.~\ref{fig:GRinteractions},  \ref{fig:directdetection}, \ref{fig:starformation}, and~\ref{fig:GWest} of Sec.~\ref{sec:signatures}, and translated to $f/f_{\pi/2}$ via the numerical results for $\mathcal{B}$ as a function of $m$ in Fig.~\ref{fig:boost}. 
}
\label{fig:summary}
\end{figure*}

The overwhelming majority of the energy density in the Universe appears to interact only gravitationally, in all available observational and experimental data so far. A quarter of this energy density is in the form of dark matter (DM), a matter component that does not emit or interact strongly with light. Two of the main pieces of evidence for DM are the fluctuations in the cosmic microwave background (CMB) and the formation of gravitational structures over a large range of length scales, from the size of the largest superclusters of galaxies down to the smallest observable dwarf galaxies. These two bodies of evidence are in mutual quantitative agreement with one another. 

Among the best motivated particle physics candidates for DM are \emph{axions}, CP-odd scalar fields. The most famous one is the QCD axion \cite{Peccei:1977hh, axion1, axion2}, responsible for addressing the strong CP problem as it explains the smallness of the neutron's electric dipole moment. Axions are also ubiquitous in extensions of the Standard Model such as string theory, where they arise as the byproducts of complex topology~\cite{Arvanitaki:2009fg}. 

Axions have a natural production mechanism of near-pressureless energy density, through what is known as the misalignment production mechanism~\cite{Preskill:1982cy, Abbott:1982af, Dine:1982ah}. The dynamics of the axion field  $\phi$ are described by four-dimensional partial differential field equations which depend on the potential of the axion. 
Inflation irons out all spatial wrinkles, converting the axion into a spatially homogeneous but time-dependent field. Near the minimum of its potential (here at $\phi = 0$), the potential of the axion is well approximated by a quadratic function of $\phi$, which then  behaves cosmologically as a damped harmonic oscillator:
\begin{align}\label{eq:freeaxion}
\ddot{\phi}+3 H \dot{\phi} + m^2 \phi=0,
\end{align}
where $H$ is Hubble parameter and $m$ the axion mass. Initially, the axion field value is frozen due to Hubble friction; the axion only starts oscillating once $3 H \lesssim m$. The energy density associated with this oscillation redshifts exactly like cold DM: $\rho_\phi \propto a^{-3}$. However, there is no reason to expect that the axion will start close to the minimum.
If the axion misalignment is large, the quadratic approximation to its potential is no longer adequate and higher order terms must be included. The axion potential generically contains quartic terms which convert its equation to that of a nonlinear damped \emph{anharmonic} oscillator:
\begin{align}\label{eq:nonfreeaxion}
\ddot{\phi}+3 H \dot{\phi} + m^2 \phi -\lambda \phi^3 + \dots =0
\end{align}
The all-important negative last term describes an attractive self-interaction.
When $|\phi^2| \gtrsim m^2/\lambda$, nonlinearities at all orders in the axion field become relevant, and can cause a \emph{delay in the onset of oscillations}: $H_\mathrm{osc} \ll m$. In this scenario, the lower Hubble friction and the attractive quartic self-interaction conspire to usher in a qualitatively new phenomenon: a parametric resonance amplification of semi-relativistic axion fluctuations around the spatially constant $\phi$ background.
In this work, we show that these attractive self-interactions can cause DM structure to grow at scales that are comparable with the axion Compton wavelength when the field starts oscillating. This leads to {\em both denser and more numerous small halos}
than in $\Lambda$CDM. We stress that such behavior is  only possible when the field amplitude of the axion is large enough for the attractive non-linearity to be significant, so we term this the ``large-misalignment'' mechanism for axion DM.

For definiteness, we will mainly focus on a simple periodic potential that is well motivated for several axion models, namely the cosine potential:
\begin{align} \label{eq:cosinepot}
V = m^2 f^2 \left[1-\cos\left(\frac{\phi}{f}\right)\right],
\end{align}
where $f$ is the axion decay constant. Nonperturbative effects generically generate periodic axion potentials; the form of Eq.~\ref{eq:cosinepot} arises from the one-instanton contribution, which is typically dominant in weakly coupled theories. Periodic potentials will in general have attractive (negative) self-interactions because these tame the rapid growth of the quadratic potential and foretell the presence of an upper bound. As we will discuss, the above potential is also nearly that of the QCD axion at temperatures above the QCD phase transition, albeit with a time-dependent mass. We stress that \emph{the observable consequences of this work emerge solely from this attractive self-interaction}, and do not qualitatively depend on the detailed form of the potential. In fact, some of our signatures will be more naturally realized with nonperiodic potentials.  The quartic interaction for the cosine is given by $V \supset \lambda \phi^4 / 4$ with $\lambda = - m^2 / 6 f^2$. 

If the axion's initial misalignment amplitude $\phi_0$ is in the ``large-misalignment'' range $ |\phi_0|/f > \pi/2 $, we show that there will be enhanced structure around a comoving wavelength:  
\begin{align}
\lambda_* \equiv \frac{2\pi}{\sqrt{2 m \aeq^2 \Heq}}\approx 0.69\,\mathrm{Mpc} ~ \sqrt{\frac{10^{-22}\,\mathrm{eV}}{m}},
\end{align}
generating numerous halos with scale mass of $M_s \sim M_s^*$:
\begin{align}\label{eq:Msstar}
\hspace{-0.5em} M_s^*  \equiv \frac{4 \pi \rho_\mathrm{DM}^0}{3}\left(\frac{\lambda_*}{2}\right)^3 \approx 5 \times 10^9\,\mathrm{M_\odot}\left[\frac{10^{-22}\,\mathrm{eV}}{m}\right]^{3/2}.
\end{align}
The halo scale density $\rho_s$ is an increasing function of $|\phi_0|/f$, and can be much larger than the scale density $\rho_s^\mathrm{CDM}$ of CDM halos of the same mass by a parametric factor:
\begin{align}
\mathcal{B} \equiv \frac{\rho_s}{\rho_s^\mathrm{CDM}} \sim \exp\left\lbrace \xi \frac{m}{H_\mathrm{osc}} \right\rbrace.\label{eq:Bgeneral}
\end{align}
The parametric form of this ``density boost factor'' $\mathcal{B}$ is valid for generalized axion potentials as well; $\xi$ is an $\mathcal{O}(1)$ model-dependent constant. The corresponding scale radius is $r_s=87~ \text{pc}\left(\frac{M_s}{5\times 10^9 M_\odot}\right)^{1/3}\left(\frac{10^5}{\mathcal{B}}\right)^{1/3}$.

We present our analysis of the development and dynamics of these enhanced structures in Sec.~\ref{sec:evolution}. To fix ideas, we mainly focus on a cosine potential and study the evolution and signatures of axion DM structures when $|\phi_0|/f>\pi/2$ as a function of the axion mass and decay constant.\footnote{Requiring that the present-day axion density accounts for all the DM automatically fixes the initial value $\phi_0$ of the axion field as a function of $m$ and $f$.}
First, we provide a fully relativistic treatment of the growth of density fluctuations in linear perturbation theory. Starting from a standard spectrum of primordial density perturbations, we show that growth in density contrast can be understood as the result of a parametric resonance instability at the level of the equations of motion, which are valid in the early universe up to axion masses of $\mathcal{O}(10^7)~\mathrm{GeV}$ (Sec.~\ref{sec:linear}). We also present a perturbative Newtonian approximation, where the boost in structure growth can be attributed to a negative pressure resulting from the nonlinearities in the potential of Eq.~\ref{eq:cosinepot}.  In Sec.~\ref{sec:nonlinear}, we describe the nonlinear evolution of the axion density fluctuations. For moderate enhancements in the density contrast with respect to large scales, compact halos will form after matter-radiation equality (Sec.~\ref{sec:gravcollapse}). Depending on their density, these compact halos may be {\em solitons}---gravitationally bound scalar field configurations of minimum energy (App.~\ref{sec:boundstates})---and can even have  a gravothermal cusp (Sec.~\ref{sec:gravcooling}). At yet larger density contrasts, we demonstrate in Sec.~\ref{sec:quarticcollapse} that our mechanism can produce {\em oscillons}---metastable configurations solely supported by axion self-interactions (App.~\ref{sec:boundstates})---during radiation domination. Further, we show that these dense structures are expected to survive tidal stripping in the Milky Way (Sec.~\ref{sec:tidalstripping}). 

\begin{figure*}[ht]
\includegraphics[width=0.97\textwidth]{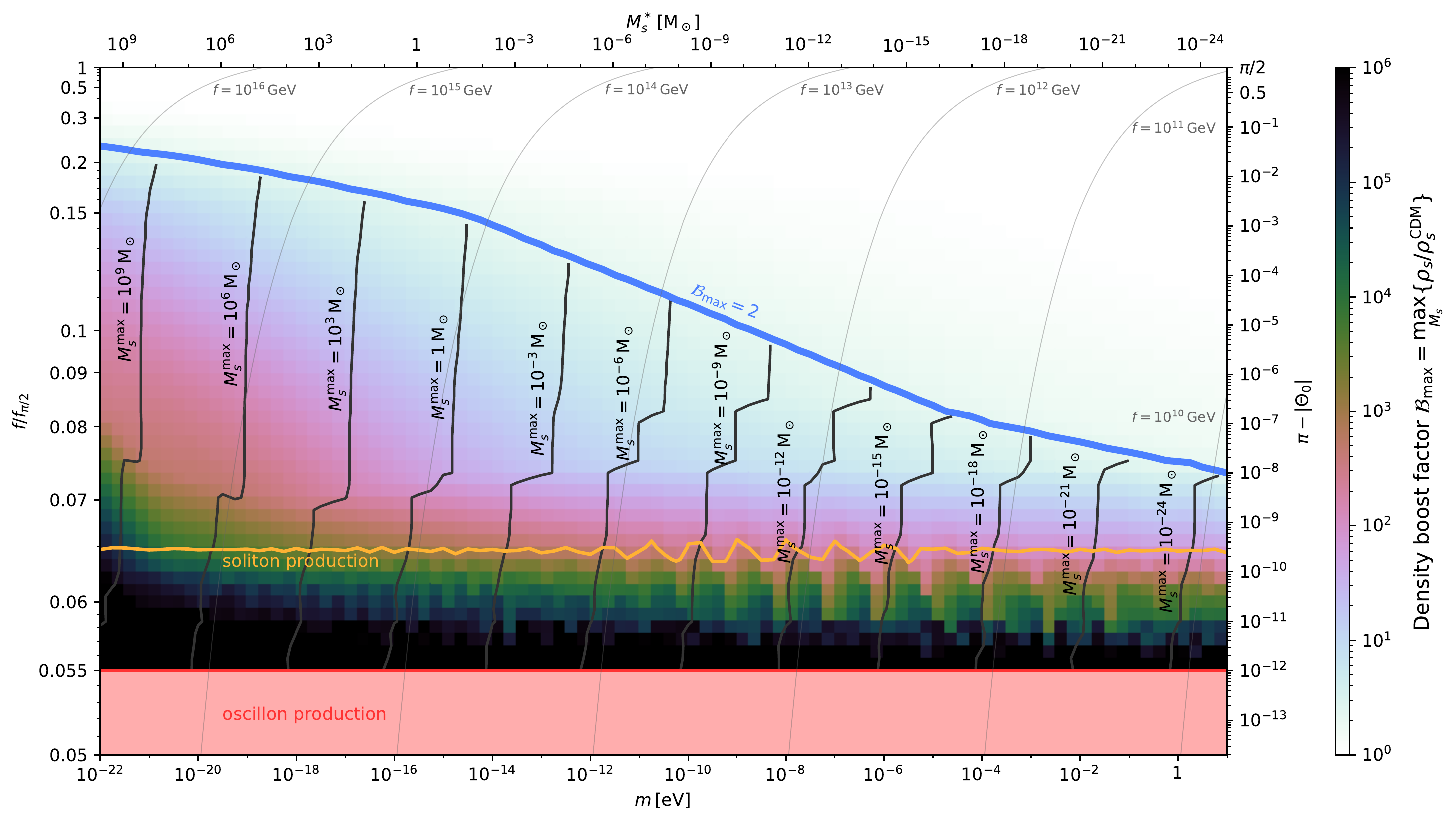}
\caption{Summary of properties of compact structures resulting from the linear and nonlinear evolution of axion density perturbations in Sec.~\ref{sec:evolution}. The maximum density boost factor $\mathcal{B}_\mathrm{max}$ is shown as a color map (legend on right) as a function of axion mass $m$ and misalignment angle $\Theta_0$ (right axis), or equivalently $f/f_{\pi/2}$ (left axis). For parameter space where $\mathcal{B}_\mathrm{max}>2$ (below the thick blue contour), dark gray contours indicate the halo scale mass $M_s^\mathrm{max}$ that exhibits the maximum density boost relative to the CDM prediction, parametrically tracking the reference scale mass $M_s^*$ of Eq.~\ref{eq:Msstar} (top axis). Below the orange contour ($f/f_{\pi/2} \lesssim 0.065$), solitons are produced; in the red region ($f/f_{\pi/2} < 0.055$), early collapse into oscillons also occurs. We assumed the axion cosine potential of Eq.~\ref{eq:cosinepot} and a scale-invariant curvature power of $\mathcal{P}_{\Phi} \approx 2.1 \times 10^{-9}$.}
\label{fig:boost}
\end{figure*}

Armed with the understanding of the behavior of these more numerous and higher-density halos, we focus in Sec.~\ref{sec:signatures} on
several observable consequences that follow in cosmological histories with a boost in structure on small scales (cfr.~Eq.~\ref{eq:Msstar}). These are summarized in Fig.~\ref{fig:summary} in the parameter space of $m$ and $f$ as extracted from  from Figs.~\ref{fig:GRinteractions},  \ref{fig:directdetection}, \ref{fig:starformation}, and~\ref{fig:GWest} of Sec.~\ref{sec:signatures}, translated via the results of Fig.~\ref{fig:boost}.\footnote{For clarity, the oscillatory behavior in Fig.~\ref{fig:boost} is suppressed by Gaussian smoothing over neighboring $m$ bins, and we used Eq.~\ref{eq:Msstar} for the $M_s$--$m$ correspondence, \emph{not} the $M_s^\mathrm{max}$ results of Fig.~\ref{fig:boost}.} Compact axion halos and other potentially long-lived axion structures have irreducible gravitational couplings, so one may look for their local gravitational perturbations on stellar structures or their gravitational lensing (Sec.~\ref{sec:GRinteractions}). Extremely small minihalos---``femto-halos'', their mass being $\lesssim 10^{-15}~\mathrm{M_\odot}$---can dramatically alter the signatures and sensitivity of direct detection efforts to search for nonminimal couplings of the axion (Sec.~\ref{sec:directdetection}). Early-forming minihalos can also influence the formation of the first stars and leave other imprints on baryonic structure (Sec.~\ref{sec:baryons}).  The implosion and subsequent explosion of oscillons can lead to a low-frequency stochastic gravitational wave background (Sec.~\ref{sec:GW}).

We next focus on the QCD axion in Sec.~\ref{sec:QCD}, which is one of the best-motivated particles beyond the Standard Model. 
This axion, which has a temperature-dependent potential, will collapse into 
halos of mass $M_s \sim 10^{-18}\,\mathrm{M_\odot}$ for axion decay constants $f_a \lesssim 2\times 10^{10}\,\mathrm{GeV}$, with important consequences for direct detection searches  of high-mass, cosmic QCD axions, potentially improving prospects for their discovery in the laboratory. We stress that these femto-halos are produced from a standard spectrum of small primordial perturbations. In contrast,  ultra dense QCD axion miniclusters~\cite{Hogan:1988mp, Tkachev:1986tr, Kolb:1993zz, Kolb:1993hw, Kolb:1994fi, Tkachev:2014dpa} rely on large density fluctuations caused by a late post-inflationary Peccei-Quinn (PQ) phase transition. Their internal density is so high that they encounter Earth too infrequently to positively impact direct dark matter searches.

For the cosine potential of Eq.~\ref{eq:cosinepot}, significant enhancement in structure growth via our mechanism requires the axion field to start very close to $|\phi_0|/f \approx \pi$, with self-interaction-induced collapse requiring apparent tunings of 1 part in $10^{12}$.  This apparent tuning is not, however, necessarily an actual tuning.  We discuss this in Sec.~\ref{sec:flat}, and in this section we also discuss other forms of axion potentials, such as those in some axion monodromy models \cite{Silverstein:2008sg, McAllister2014, Kaloper:2016fbr, Olle:2019kbo}. In this latter case, the structure growth can be even more extreme and lead to long-lived oscillon configurations, all without any tuning whatsoever (apparent or actual).  We offer concluding remarks and discussion in Sec.~\ref{sec:discussion}. 

The appendices of this paper deal with further 
details that are relevant for a complete understanding of our proposed mechanism.  In App.~\ref{sec:boundstates} we review the spectrum of bound, metastable scalar field configurations (\emph{solitons} and \emph{oscillons}) because in much of our parameter space they will be formed inside the DM overdensities we predict.  In App.~\ref{sec:numerics} we discuss the implementation and results of various numerical simulations we utilized to help understand the nonlinear behavior of the axion field in regimes particularly relevant to this work.  App.~\ref{sec:isocurvature} discusses possible constraints coming from the production of isocurvature fluctuations in the CMB, although these constraints are only present in some models.  Finally, we summarize in App.~\ref{sec:gwprospects}, the projected sensitivities and detection prospects for ultra-low-frequency gravitational waves, which can be produced particularly by very light ($m \lesssim 10^{-14}\,\mathrm{GeV}$) large-misalignment axions.

We note that some of the components of this paper have been previously touched upon in the literature (see e.g.~Refs.~\cite{strobl1994anharmonic, Greene:1998pb, Johnson:2008se, Amin:2011hj, Lozanov:2017hjm, Amin:2018xfe, Lozanov:2019ylm, Amin:2019ums, Olle:2019kbo}).  In particular, the linear perturbation effects under consideration in this work were previously discussed in Refs.~\cite{cedeno2017cosmological, desjacques2018impact, zhang2017cosmological,
zhang2017evolution, schive2017halo}. These works however focused on the regime of $m \sim 10^{-22}\,\mathrm{eV}$ and observables such as the matter power spectrum and the Lyman-$\alpha$ forest. We here extend their analyses and provide a comprehensive treatment of the linear and nonlinear evolution for any axion mass $m$ and decay constant $f$. As we shall see, much larger nonlinearities are permitted (by current data) for larger axion masses (and thus smaller structures).
This leads to qualitative differences in phenomenology and observable consequences. On the other hand, a large body of literature has studied the effective theory and potential observables of ``axion stars'' (i.e.~solitons and oscillons) but has for the most part disregarded their formation mechanism (see e.g.~Refs.~\cite{Olle:2019kbo, Seidel:1991zh, Braaten:2015eeu, Braaten:2016dlp, Chavanis:2011zm, Chavanis:2011zi, Eby:2017azn, Eby:2017teq, Eby:2018ufi, Visinelli:2017ooc, Schiappacasse:2017ham, Mukaida:2016hwd, Salmi:2012ta, Bogolyubsky:1976yu}).  We provide such a mechanism here, and calculate for the first time the enhanced contrast in adiabatic fluctuations for the QCD axion. Ref.~\cite{zurek2007astrophysical} studied a scenario wherein a late-time phase transition in an arbitrary-mass axion potential sources large isocurvature fluctuations and associated small-scale structures; such a structure formation history has a qualitatively different matter power spectrum and no tunable density contrast.

We also note that claimed constraints on ultralight DM due to Lyman-$\alpha$ forests \cite{Irsic:2017yje, Leong:2018opi} or the DM distribution of present-day dwarf galaxies \cite{Bar:2018acw, Safarzadeh:2019sre} do not necessarily apply.  The attractive self-interactions and gravitational thermalization both have significant effects which must be taken into account, and reanalyses are required to understand the true constraints.  We expand upon these effects and discuss more realistic constraints in Sec.~\ref{sec:baryons} (Lyman-$\alpha$) and Sec.~\ref{sec:gravcooling} (dwarf galaxies).

Throughout this paper, we take the dark matter energy density fraction in the Universe to be $\Omega_\mathrm{DM} = 0.23$, the scale factor at matter-radiation equality $a_\mathrm{eq}= 1/3250$, the present-day Hubble constant $H_0 = 67.8\,\mathrm{km\, s^{-1}\, Mpc^{-1}}$, and therefore present-day Universe-average DM density $\rho_\mathrm{DM}^0 = 2.9\times 10^{-8}\,\mathrm{M_\odot \,pc^{-3}}$ and  the Hubble parameter at matter-radiation equality $\Heq = 1.8 \times 10^{-28}\,\mathrm{eV}$. We assume a local DM energy density in the Galaxy of $\rho_\mathrm{DM}^\odot = 0.4\,\mathrm{GeV \, cm^{-3}} = 1.1 \times 10^{-2} \, \mathrm{M_\odot \, pc^{-3}}$. We use the reduced Planck mass $\MPl = 1/\sqrt{8\pi G_N}$, and set the reduced Planck constant and the speed of light to unity $\hbar = c = 1$.


\section{Evolution of density fluctuations}\label{sec:evolution}

In this section, we analyze the growth of adiabatic axion density perturbations in the early Universe and demonstrate how self-interactions can lead to substantial deviations from the CDM prediction. The relevant observable throughout is the gauge-covariant axion energy perturbation $\delta$ (we work in Newtonian gauge, cfr.~Eq.~\ref{eq:RDmetric}). In the CDM framework, after the physical wavelength of a density perturbation with amplitude $\delta$ becomes smaller than the Hubble horizon, $\delta$ grows logarithmically with the scale factor during radiation domination, and linearly with the scale factor during matter domination.  We will find that for a range of comoving scales close to the axion's Compton wavelength at horizon crossing,
there is enhanced growth due to the self-interactions.  Length scales much smaller than this will have their growth suppressed, and density perturbations on much larger scales  will resemble those of CDM. 

Figure~\ref{fig:boost} summarizes the results of both the linear and nonlinear evolution of density perturbations as presented in this section. We show the maximum boost $\mathcal{B}_\mathrm{max} \equiv \max_{M_s}\lbrace \mathcal{B} \rbrace$ in halo scale density relative to the CDM prediction (cfr.~Eq.~\ref{eq:Bgeneral}) as a function of $m$ and $f/f_{\pi/2}$ for the cosine potential of Eq.~\ref{eq:cosinepot}. We also show the corresponding halo scale mass $M_s^\mathrm{max} \equiv \mathrm{argmax}_{M_s} \lbrace \mathcal{B} \rbrace$ for which this maximum density boost factor is achieved, which can be seen to closely track the value $M_s*$ of Eq.~\ref{eq:Msstar} (top horizontal axis). Finally, we also indicate parameter space where production of solitons and oscillons occurs.

In Sec.~\ref{sec:linear}, we discuss the linear regime, where all fractional density perturbations are small: $| \delta | \ll 1$. This is appropriate for all adiabatic perturbations early enough in their history (given a standard primordial curvature power spectrum). In Sec.~\ref{sec:linGR}, we present a full general-relativistic treatment of the density perturbations from the time the axion field starts oscillating and show that the growth of structure is due to a parametric resonance instability well before matter-radiation equality.  We calculate analytically (cfr. Eq.~\ref{eq:Gasymptote} and Eq.~\ref{eq:Gstar}) the $\mathcal{G}\equiv |\delta_k/\delta_k^{\mathrm{CDM}}|^2$ in the power spectrum (the boost $\mathcal{B}$ in density is proportional to $\mathcal{G}^{3/2}$).
Figure~\ref{fig:deltatime} compares the time evolution of adiabatic density perturbations for a large- and small-misalignment axion. The results of our linear analysis for any misalignment are summarized in Fig.~\ref{fig:delta} and~\ref{fig:PR}. In Sec.~\ref{sec:linNewton}, we evolve these parametric-resonance-boosted perturbations past matter-radiation equality (see Fig.~\ref{fig:time}).

When $|\delta|$ becomes $\mathcal{O}(1)$, axion DM structures can form (Sec.~\ref{sec:nonlinear}). The properties of the collapsed structures depend on the amount of growth they receive through axion self-interactions. If the growth is small enough that the perturbations are still linear after matter-radiation equality, their collapse is fueled by gravitational self-interactions. In Sec.~\ref{sec:gravcollapse}, we study the halo spectrum (see Figs.~\ref{fig:haloPS} and~\ref{fig:halospec}) and show that, for moderate structure growth, the collapsing structures can be solitons. Gravitational cooling effects can further change the internal structure of these compact halos and ultimately lead to gravothermal collapse and a central soliton (Sec.~\ref{sec:gravcooling}). In the extreme case where the axion self-interaction induced structure growth is large enough,  structures can grow nonlinear well before matter-radiation equality; their dynamics are dominated by self-interactions, and oscillons are formed (Sec.~\ref{sec:quarticcollapse}). Finally, we show that these compact halos can easily survive tidal stripping within the local galaxy (Sec.~\ref{sec:tidalstripping}).

The range of axion masses for which this section's analysis is relevant is from $10^{-22}\,\mathrm{eV}$ to $10^7\,\mathrm{GeV}$. The lower end is an observational limit from structure formation~(Sec.~\ref{sec:baryons}). The upper limit comes from two requirements: one is that $m \ll f$ which is necessary to ensure that during parametric resonance the axion occupation number is large enough to justify the use of classical wave equations; the second is the condition that the axion is the DM (see discussion around Eq.~\ref{eq:rhopi2}). The requirement that the axion lifetime is longer than the age of the Universe is automatic if the only interactions of the axion are gravity and its self-couplings (Eq.~\ref{eq:cosinepot}), as these are both axion number conserving in the nonrelativistic limit. To have an axion detectable in laboratory experiments we need further interactions that directly couple the axion to photons, electrons, or nuclei. An example is the coupling to the photon given by $ \frac{\alpha}{(2\pi)} \frac{\phi}{f} F \tilde{F}$. In the presence of such a coupling, the longevity of the axion constrains the axion mass to be at most $10\,\mathrm{keV}$ corresponding to $f = 10^{11}\,\mathrm{GeV}$.
Note that axions as heavy as $10^7\,\mathrm{GeV}$ or even $10\,\mathrm{keV}$ are not well described by classical field equations today because the occupation number in a de Broglie wavelength is much smaller than unity. Nevertheless, the classical field description is valid during the crucial era of parametric resonance, when the axion occupation number is large and the initial overdensities are generated. Subsequently, these overdensities grow under the influence of gravity which, by virtue of the equivalence principle, just couples to energy regardless of occupation number or the applicability of the classical approximation.

For simplicity, we will first consider the case of the cosine potential in Eq.~\ref{eq:cosinepot}. We will study entirely analogous phenomena for the temperature-dependent QCD axion potential in Sec.~\ref{sec:QCD}, and present case studies of generalized (but time-independent) axion potentials in Sec.~\ref{sec:flat}. Finally, for those interested in the signatures of compact axion halos, they can directly skip to Sec.~\ref{sec:signatures}, where the observational effects of these halos are described as a function of their scale mass $M_s$ and density $\rho_s$.

\subsection{Linear regime}\label{sec:linear}

In the linear regime (i.e.~$| \delta | \ll 1$), most of the self-interaction-induced growth occurs at very early times, when semi-relativistic modes enter the horizon and the axion potential is poorly approximated by a quadratic.  This means that a full general-relativistic treatment of the perturbations is necessary, which we give in Sec.~\ref{sec:linGR}.  At later times, when nonlinearities in the background axion field are small and the modes of interest are nonrelativistic and well within the horizon, we can patch the general-relativistic solutions onto Newtonian fluid equations, which we describe in Sec.~\ref{sec:linNewton}.

\subsubsection{General relativistic treatment} \label{sec:linGR}

We consider adiabatic perturbations in the axion field and adopt the method of Ref.~\cite{zhang2017evolution}, the only substantive difference being our focus on the potential of Eq.~\ref{eq:cosinepot} and slight changes in notation. The dynamics of interest occur in the radiation-dominated era, where we can study the evolution of the axion field in the background metric
\begin{equation} \label{eq:RDmetric}
\mathrm{d}s^2 = [1 + 2 \Phi(t, \vect{x}) ] \mathrm{d}t^2 - a^2(t) [ 1 - 2 \Phi(t, \vect{x})] \mathrm{d} \vect{x}^2 
\end{equation}
where $a(t) \propto t^{1/2}$ is the scale factor and $\Phi(t, \vect{x})$ are the curvature fluctuations.  We also define the Hubble parameter $H \equiv \dot{a}(t)/a(t) = 1/2t$ where the second equality is true only during radiation domination.  During this era, the energy density in the axion field is a tiny perturbation to the overall energy density in the radiation bath, so we will neglect its backreaction on the metric.
We expand the axion field into modes of comoving wavenumber $\vect{k}$ as:
\begin{equation}
\frac{\phi(t, \vect{x})}{f} = \Theta(t) + \sum\limits_{\vect{k}} \theta_{\vect{k}} (t) e^{-i \vect{k} \cdot \vect{x}}
\end{equation}
where $\Theta$ is the zero mode (spatially-averaged axion field) and $\theta_{\vect{k}}$ are Fourier modes of its perturbations.  


\begin{center} \emph{Zero mode} \end{center}
Before studying the growth of the perturbations, we describe the evolution of the zero-mode.  A field of mass $m$ is frozen by Hubble friction at least until $H \sim m$, which motivates the definition of a dimensionless time $t_m$ given by:
\begin{equation}
t_m \equiv \frac{m}{2 H} \simeq m t
\end{equation}
the latter equality approximately true deep into the radiation-dominated era.
The equation of motion for $\Theta$ in the metric of Eq.~\ref{eq:RDmetric} is given by:
\begin{equation} \label{eq:evolTheta}
\Theta '' + \frac{3}{2 t_m} \Theta ' + \sin ( \Theta ) = 0
\end{equation}
where from hereon primes denote derivatives with respect to $t_m$.  The initial conditions sourced by inflation are a fixed initial misalignment angle $\Theta(t_m = 0) = \Theta_0$ and zero kinetic energy $\Theta'(t_m = 0) = 0$.  We can then see that indeed for $t_m \ll 1$ the field is frozen and for $t_m \gg 1$ the field will roll to and oscillate around the bottom of the potential. 

The energy density contained in the axion field is given by $\rho = m^2 f^2 [ (\Theta ' )^2/2 + 1 - \cos(\Theta) ]$.  For $t_m \gg 1$, an approximate solution to Eq.~\ref{eq:evolTheta} can be found to show that this energy density redshifts as $\rho \propto t_m^{-3/2}$.  We define $\rho_{\pi/2}(t_m)$ as the energy density at late times given an initial misalignment angle $| \Theta_0 | = \pi / 2$.  By the above, we have that
\begin{equation}
\rho_{\pi / 2} = C_{\pi / 2} m^2 f^2 t_m^{-3/2} \label{eq:rhopi2}
\end{equation}
for some constant of proportionality $C_{\pi / 2}$, and a numerical evolution of Eq.~\ref{eq:evolTheta} then gives $C_{\pi / 2} \approx 1.15$.  Requiring that the axion field is the totality of dark matter then implies that an axion with initial misalignment $\pi / 2$ and mass $m$ must have a decay constant $f_{\pi / 2}$ given by:
\begin{equation} \label{eq:fPi2}
\frac{f_{\pi / 2}}{M_{\mathrm{Pl}}} \simeq \frac{3^{1/2}}{2^{5/4} C_{\pi / 2}^{1/2}} \left( \frac{H_{\mathrm{eq}}}{m} \right)^{1/4}.
\end{equation}
At fixed $m$, larger values of $f > f_{\pi/2}$ require the initial misalignment angle to be closer to the bottom of the potential (i.e. $ | \Theta_0 | < \pi / 2$).  Asymptotically for small initial $\Theta_0 \ll 1$ we have $\rho / \rho_{\pi/2} \approx 0.33 \Theta_0^2$, which implies for $f \gg f_{\pi/2}$ a required initial misalignment angle $\Theta_0 \approx {f_{\pi/2}}/{0.33 f}$.

Similarly, $f < f_{\pi / 2}$ requires $| \Theta_0 | > \pi / 2$, our case of interest.  As $|\Theta_0|$ approaches $\pi$, the onset of the field's oscillation is delayed from its typical time of $t_m \sim \mathcal{O}(1)$ to a logarithmically larger value, due to the much smaller gradient near the top of the potential.  The delay results in an enhanced final density $\rho$, and an empirical approximation to the true numeric solution of Eq.~\ref{eq:evolTheta} yields:
\begin{align}
\frac{\rho}{\rho_{\pi/2}} &\simeq 0.2 \left[ t_m^\text{osc} + 4 \ln t_m^{\text{osc}} \right]^2 \label{eq:rhoboost} \\
t_m^\text{osc} &\equiv \ln \left[ \frac{1}{\pi - | \Theta_0 |} \frac{2^{1/4} \pi^{1/2}}{ \Gamma \left( \frac{5}{4} \right) }\right] \label{eq:tC}
\end{align}
where $\Gamma$ is the Euler Gamma function and $t_m^\text{osc}$ corresponds roughly to an effective ``delayed oscillation time". For $10^{-15} \lesssim  \pi - |\Theta_0 | \lesssim 10^{-2}$, this approximation is accurate to within a fractional error of 5\%.

\begin{center} \emph{Finite-wavenumber modes} \end{center}
Now that we understand the evolution of the zero-mode $\Theta$, we turn our attention to the perturbations $\theta_{\vect{k}}$.  We begin by also expanding the curvature perturbations into Fourier modes: $\Phi(t, \vect{x}) = \sum_{\vect{k}} \Phi_{\vect{k}} (t) e^{-i \vect{k} \cdot \vect{x}}$.  
To leading order in perturbative quantities $\theta_{\vect{k}}$ and $\Phi_{\vect{k}}$, modes with different $\vect{k}$ do not interact, and so we may consider each independently.  It is then helpful to introduce another dimensionless time coordinate $t_k$ as well as a dimensionless measure $\tilde{k}$ of the comoving wavenumber $\vect{k}$:
\begin{equation}
t_k \equiv \frac{k/a}{\sqrt{3} H} \qquad \qquad \tilde{k}^2 \equiv \frac{k^2 / a^2}{2 m H} = \frac{3 t_k^2}{4 t_m}
\label{eq:ktilde}
\end{equation}
Note that in a radiation-dominated universe, $\tilde{k}$ is constant and parametrizes how relativistic a perturbation mode is at $t_m \sim 1$, i.e.~roughly when the axion zero mode starts oscillating.

Adiabatic fluctuations in the axion field are sourced by curvature fluctuations $\Phi_{\vect{k}}$, and an exact solution for these may be found in the linear theory \cite{zhang2017evolution}:
\begin{equation} \label{eq:curvpert}
\Phi_{\vect{k}}(t_k) = 3 \Phi_{\vect{k}, 0} \left[ - \frac{\cos(t_k)}{t_k^2} + \frac{\sin(t_k)}{t_k^3} \right]
\end{equation}
where $\Phi_{\vect{k}, 0}$ is the primordial value imprinted by inflation. \textit{Planck} measurements over scales $k < 1\,\mathrm{Mpc}^{-1}$ are consistent with a Gaussian-distributed curvature with dimensionless power spectrum $\mathcal{P}_\Phi(k) = \langle \Phi_{\vect{k}, 0} \Phi_{\vect{k}, 0} \rangle \simeq (2.1 \times 10^{-9}) (k / (0.05~\mathrm{Mpc}^{-1}))^{n_s-1}$ and a slight spectral tilt $n_s-1 \approx -0.03$~\cite{Aghanim:2018eyx}.\footnote{The dimensionless power spectrum of a scalar $s(\vect{r})$ is $\mathcal{P}_s(k) = P_s(k) k^3 / 2\pi^2$, where the power spectrum is $P_s(k) = V^{-1} \langle s(\vect{k})^2 \rangle$ and the Fourier transform is $s(\vect{k}) = \int_V \dd^3r\,s(\vect{r}) e^{-i\vect{k}\cdot \vect{r}}$. $\mathcal{P}_s(k)$ is independent over the averaging volume $V$ as long as $k^3 V \gg 1$.} 
For specificity and to elucidate the scale dependence of our mechanism, we will ignore the spectral tilt and take $|\Phi_{\vect{k}, 0}| \simeq \sqrt{2.1 \times 10^{-9}}$ as a fiducial amplitude.  Note that for $t_k \lesssim 1$ the curvature perturbations are frozen, but for $t_k \gtrsim 1$ they begin oscillating and decay as $\Phi_{\vect{k}} \propto t_k^{-2} \propto a^{-2}$.

\begin{figure*}
\includegraphics[width=0.50\textwidth]{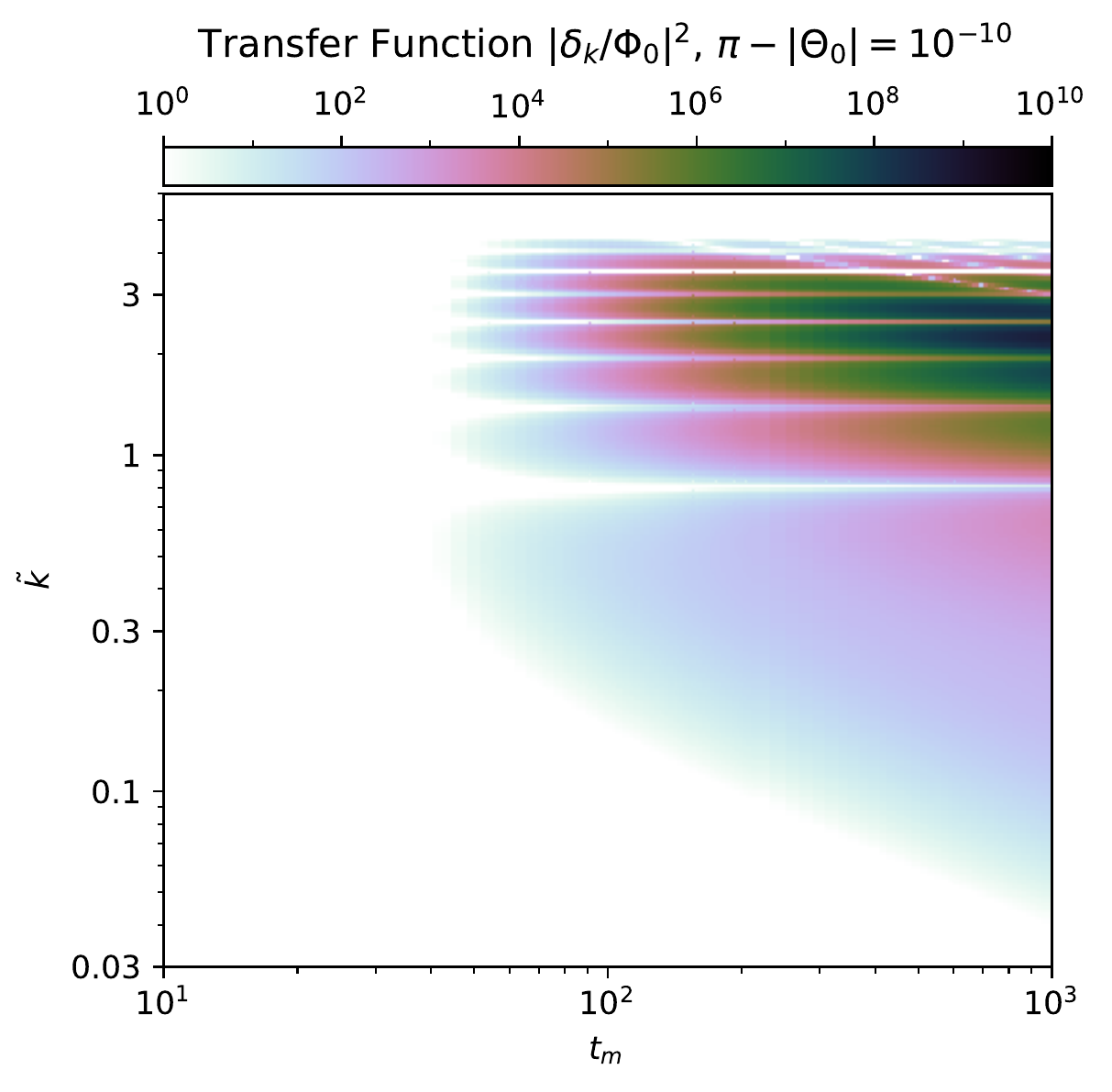}
\hspace{-0.02\textwidth}
\includegraphics[width=0.50\textwidth]{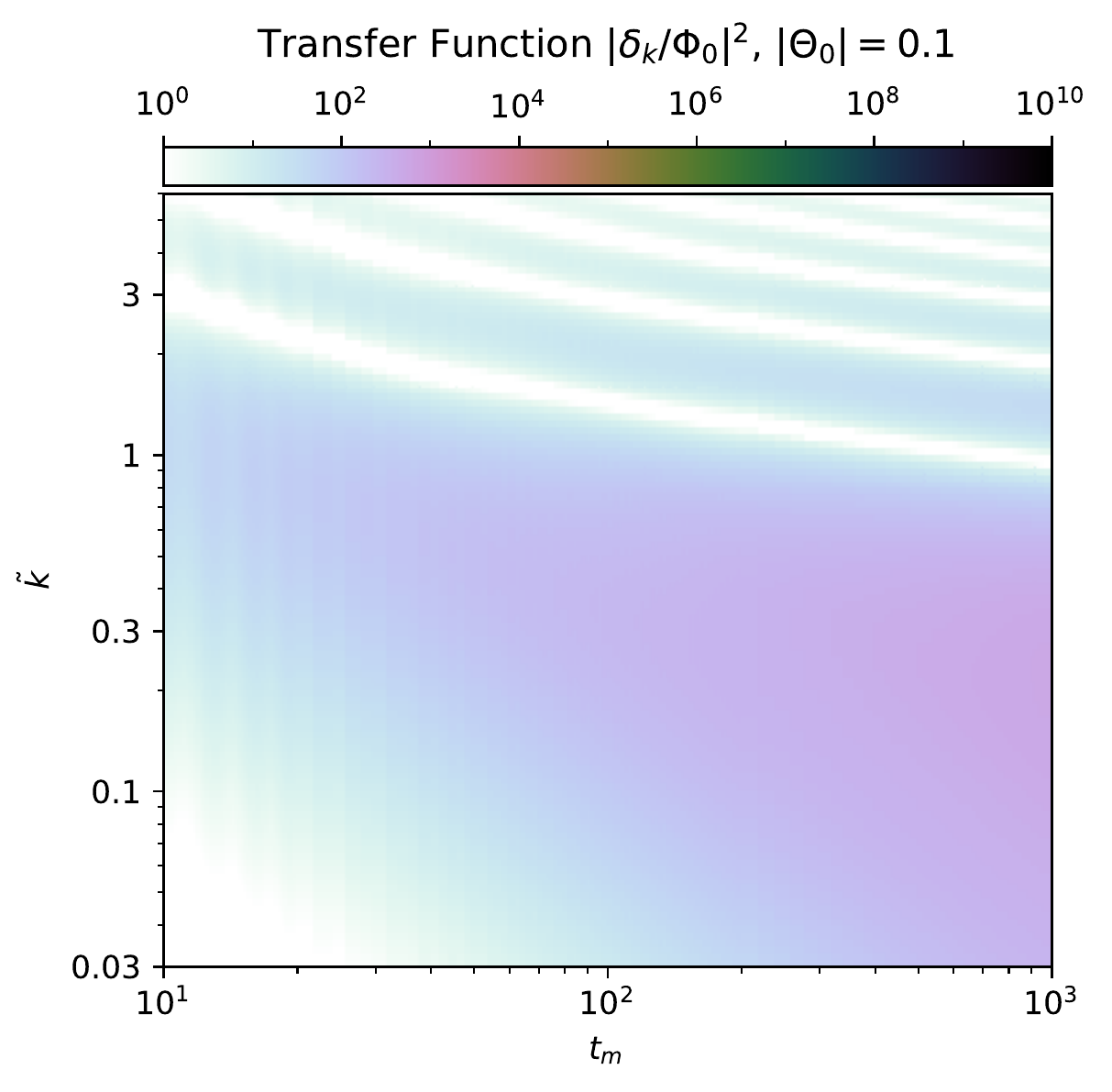}
\caption{Transfer function $|\delta_{\vect{k}}/\Phi_{\vect{k},0}|^2$ of the axion density fluctuation $\delta_{\vect{k}}$ relative to the primordial curvature fluctuation $\Phi_{\vect{k},0}$, as function of rescaled time $t_m = m t$  
and dimensionless wavenumber constant $\tilde{k} = \frac{k/a}{m} \sqrt{t_m}$.  
The left panel has an initial condition of $\pi-|\Theta_0| = 10^{-10}$, while the right panel shows the reference case of a nearly free scalar field with $\Theta_0 = 0.1$. When $\pi-|\Theta_0| = 10^{-10}$, one can see that modes with $\tilde{k}\sim 1$ get enhanced by up to 10 orders of magnitude soon after the axion enters the parametric resonance regime (see main text for details). When $\tilde{k}\ll1$ or $\tilde{k}\gg1$, for both values of the initial axion field, the behavior of the density perturbations is similar; $\delta_k$ is suppressed when $\tilde{k} \gg 1$, while for $\tilde{k} \ll 1$ modes experience logarithmic growth after they enter the horizon in the radiation dominated era.}\label{fig:deltatime}
\end{figure*}

Now we can finally write the relativistic equation of motion for axion perturbations $\theta_{\vect{k}}$ in the background of the zero-mode solution $\Theta$ to Eq.~\ref{eq:evolTheta} and the curvature perturbations of Eq.~\ref{eq:curvpert}:
\begin{equation} \label{eq:evoltheta}
\theta_{\vect{k}}'' + \frac{3}{2 t_m} \theta_{\vect{k}}' + \left[ \cos ( \Theta ) + \frac{\tilde{k}^2}{t_m} \right] \theta_{\vect{k}} = S \left( \tilde{k},t_m \right),
\end{equation}
\begin{equation} \label{eq:thetaforcing}
S \left(\tilde{k} , t_m\right) \equiv 2 \left[ \frac{t_k}{t_m} \frac{\mathrm{d} \Phi_{\vect{k}}}{\mathrm{d} t_k} \Theta' - \Phi_{\vect{k}} \sin(\Theta) \right].
\end{equation}
Here the forcing term $S$ is such that even with initial conditions $\theta_{\vect{k}}'(0) = \theta_{\vect{k}}(0) = 0$, a nonzero $\theta_{\vect{k}}$ will be generated by the curvature fluctuations.  Nonzero initial $\theta_{\vect{k}}(0)$ will be sourced by inflation and manifest as isocurvature fluctuations in the CMB. Their absence in \emph{Planck} measurements of the CMB \cite{Akrami:2018odb} provides a joint constraint on $f$ and the inflationary Hubble scale $H_\mathrm{inf}$, derived later in App.~\ref{sec:isocurvature} and shown in Fig.~\ref{fig:iso}.

\begin{figure}
\includegraphics[trim = 5mm 0 0 0, height=0.480\textwidth]{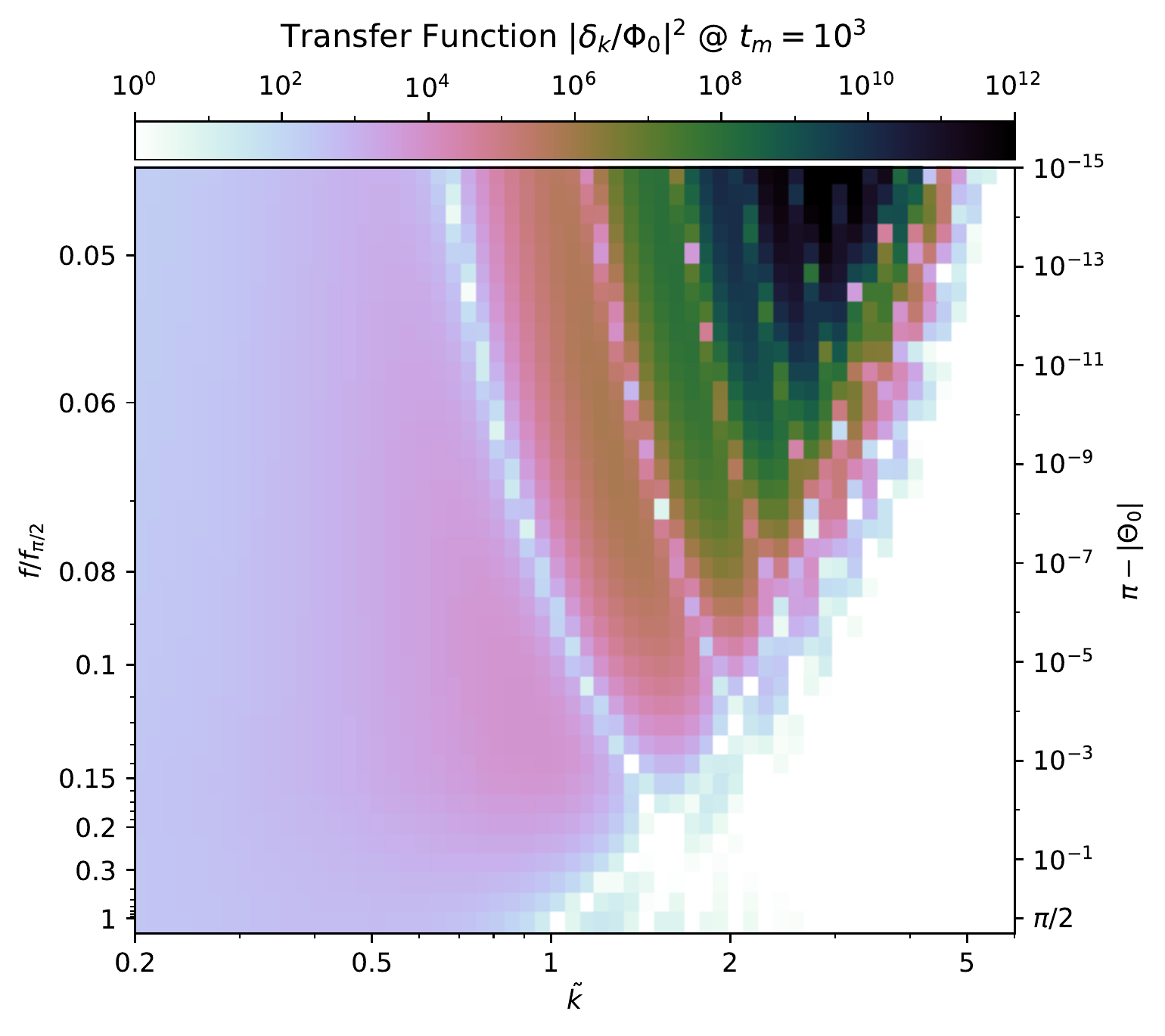}
\caption{Transfer function $|\delta_{\vect{k}}/\Phi_{\vect{k},0}|^2$ of the axion density fluctuation $\delta_{\vect{k}}$ relative to the primordial curvature fluctuation $\Phi_{\vect{k},0}$, at a fixed dimensionless time $t_m = m t = 10^3$, as function of rescaled comoving wavenumber $\tilde{k} = \frac{k/a}{m} \sqrt{t_m}$ and initial misalignment angle $\pi-| \Theta_0 |$ (right axis), or equivalently the axion decay constant $f$ (left axis) relative to the reference value $f_{\pi/2}$ of Eq.~\ref{eq:fPi2}. This plot assumes the axion comprises all of DM and has the cosine potential of Eq.~\ref{eq:cosinepot}, for which large enhancements manifest only for initial misalignments very close to the top of the potential $|\Theta_0| \simeq \pi$. This apparent tuning of initial conditions only serves to delay the onset of oscillation (see Fig.~\ref{fig:PR}); it can be explained by natural dynamics, and is not present for generalized potentials (Sec.~\ref{sec:flat}).}
\label{fig:delta}
\end{figure}

\begin{figure}
\includegraphics[trim = 0 0 0 0, height = 0.480\textwidth]{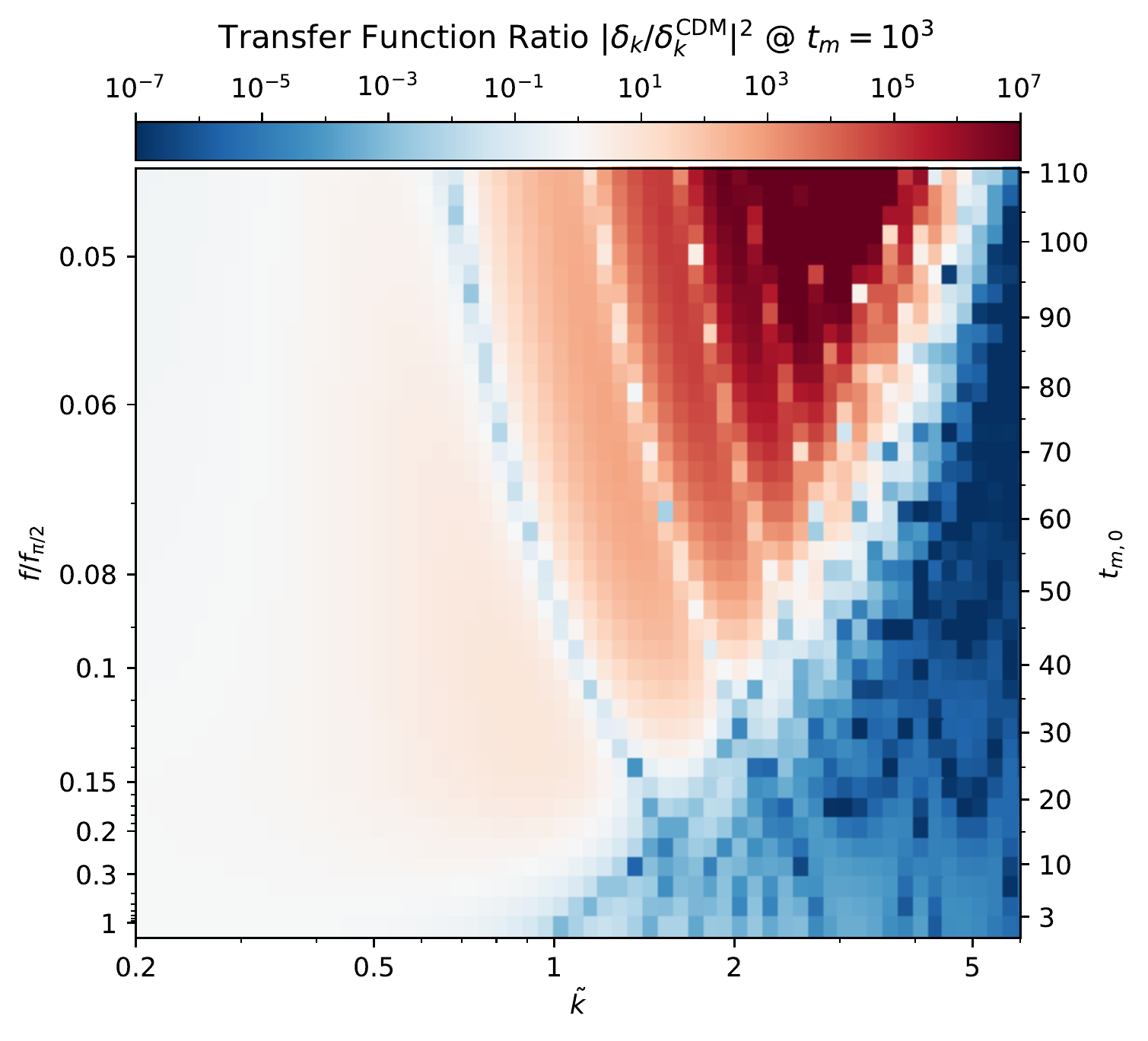}
\includegraphics[trim = 0 0 0 0, height = 0.480\textwidth]{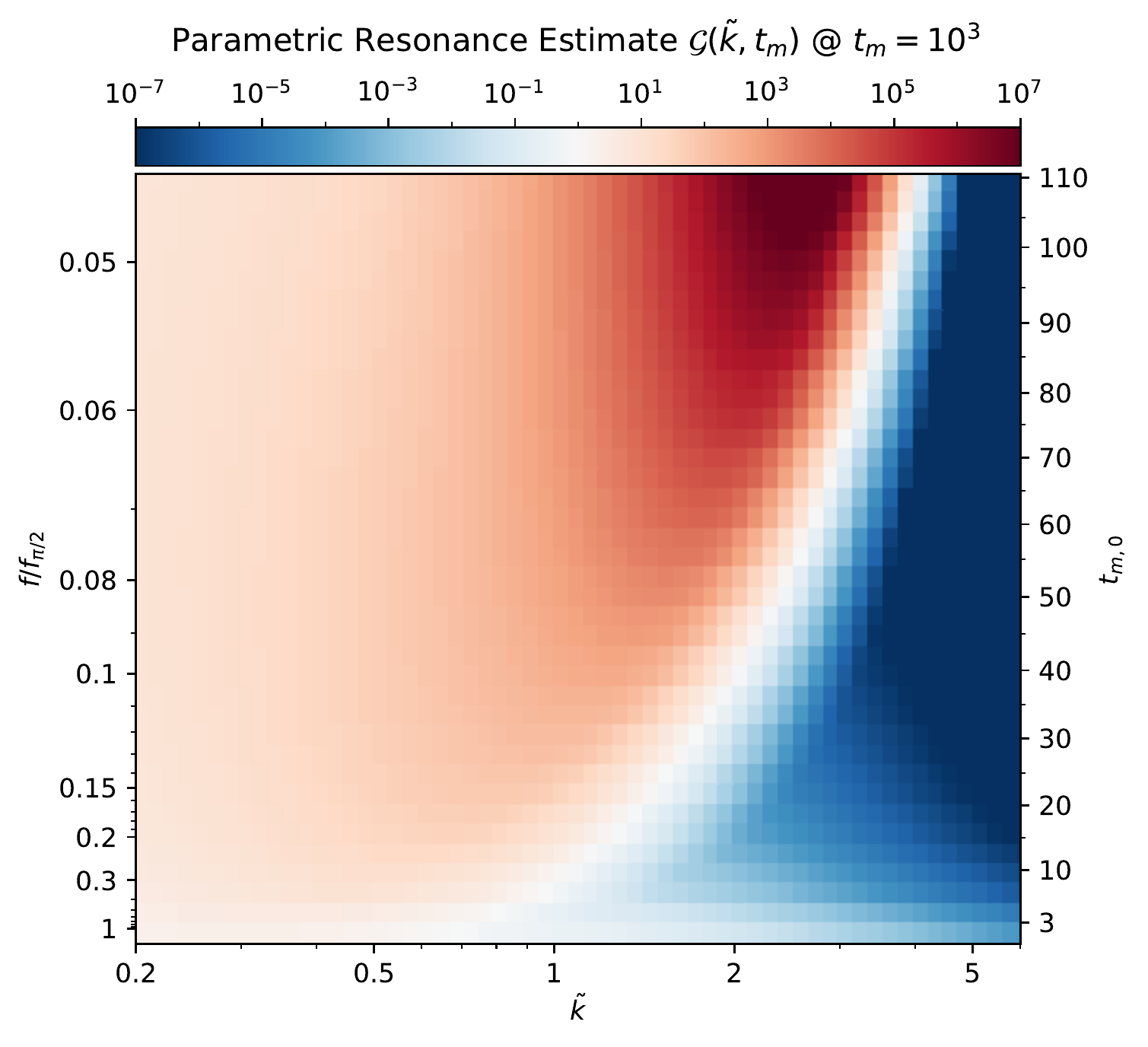}
\caption{\emph{Top panel: } Transfer function ratio of axion perturbations $\delta_{\vect{k}}$ versus CDM perturbations $\delta_{\vect{k}}^\mathrm{CDM}$ as a function of dimensionless wavenumber $\tilde{k}$ and normalized axion decay constant $f/f_{\pi/2}$, at a dimensionless time $t_m = 10^3$ shortly after the modes shown have crossed the horizon, as in Fig.~\ref{fig:delta}. On the right vertical axis, we indicate $t_{m,0}$, defined as the dimensionless time at which the axion amplitude equals unity; $t_{m,0}$ has a one-to-one map with $f/f_{\pi/2}$ and $|\Theta_0|$ discussed around Eqs.~\ref{eq:rhopi2}, \ref{eq:fPi2}, \ref{eq:rhoboost}, \ref{eq:tC}, and \ref{eq:tm0}.
\emph{Bottom panel: } Boost function $\mathcal{G}(\tilde{k},t_m)$ that captures all perturbative parametric resonance growth until $t_m = 10^3$, and parametrizes the curvature forcing suppression for high-$\tilde{k}$ modes.
The analytic function $\mathcal{G}$ is seen to be a reasonably good parametric estimate of the enhancement (and suppression) of the relative matter power spectra $|\delta_{\vect{k}} / \delta_{\vect{k}}^\mathrm{CDM}|^2$ calculated numerically.
}\label{fig:PR}
\end{figure}

\begin{center} \emph{Axion density perturbation results} \end{center}
The gauge-covariant axion energy perturbation at wavenumber $\vect{k}$ is the fractional energy density perturbation minus the velocity potential for the axion species~\cite{zhang2017evolution}, which can be written as: 
\begin{equation} \label{eq:deltadef}
\delta_{\vect{k}} \equiv \frac{ \Theta'  \theta_{\vect{k}}' + \sin(\Theta) \theta_{\vect{k}} - ( \Theta' )^2 \Phi_{\vect{k}}}{\frac{1}{2} (\Theta ' )^2 + \left( 1 - \cos(\Theta) \right)}.
\end{equation}
At late times, when $| \Phi_{\vect{k}} / \Phi_{\vect{k}, 0} | \ll 1$, $| \Theta | \ll 1$, and $t_m \gg 1$,  $\delta_{\vect{k}}$ tends to a Newtonian fractional energy density fluctuation $\delta \rho_{\vect{k}}/\rho$:
\begin{equation} \label{eq:deltaapprox}
\delta_{\vect{k}} \simeq \frac{\Theta ' \theta_{\vect{k}}' + \Theta \theta_{\vect{k}}}{\frac{1}{2} (\Theta ')^2 + \frac{1}{2} \Theta^2}.
\end{equation}
Note that nearly all of the forcing effects from $S$ occur early, as $\Phi_{\vect{k}}$ redshifts as $t_k^{-2} \propto t_m^{-1}$.

At this point, we can numerically solve the full set of equations to obtain $\delta_{\vect{k}}(t_m)$ for any value of $\vect{k}$ and initial misalignment angle $\Theta_0$.  In Fig.~\ref{fig:deltatime}, we show the evolution of $\delta_{\vect{k}}$ (by means of the transfer function $|\delta_{\vect{k}}/\Phi_{\vect{k},0}|^2$) as a function of time $t_m$ at different rescaled wavenumbers $\tilde{k}$, for a large-amplitude axion with $|\Theta_0| = \pi - 10^{-10}$ (left panel) and an axion with a small misalignment amplitude $|\Theta_0| = 0.1$. In Fig.~\ref{fig:delta}, we fix the time at $t_m = 10^3$, to highlight the dependence of the transfer function on both the wavenumber $\tilde{k}$ and the misalignment angle $\Theta_0$, which has a one-to-one map with $f/f_{\pi/2}$ from the discussion around Eq.~\ref{eq:fPi2}. We can classify the qualitative behavior into three wavenumber regimes:

$\mathbf{\tilde{k} \ll 1}$\textbf{:} In this regime, the curvature perturbation $\Phi_{\vect{k}}$ enters the horizon at a time $t_m \sim 1/\tilde{k}^2$, long after the axion has started oscillating (regardless of initial amplitude) at $t_m \sim 1$. The zero-mode $\Theta$ has already been damped down to the harmonic regime $|\Theta | \ll 1$.  In this regime, an axion behaves as a noninteracting, pressureless fluid, whose density perturbations thus grow like those of CDM---logarithmically with time during radiation domination.

$\mathbf{\tilde{k} \gg 1}$\textbf{:} Curvature perturbations with high enough wavenumbers enter the horizon long before the axion stars oscillating. By the time Hubble friction is reduced to a point where both $\Theta$ and $\theta_{\vect{k}}$ can start oscillating ($t_m \gtrsim 1$), the curvature perturbation $\Phi_{\vect{k}}$ and thus the forcing term $S$ have been damped away significantly by the radiation bath, such that $\delta_{\vect{k}}$ is suppressed. In addition, $\delta_{\vect{k}}$ oscillates in time (as opposed to the logarithmic growth for $\tilde{k} \ll 1$), since the behavior of the modes is dominated by a large positive kinetic energy pressure, further suppressing the structure relative to the CDM prediction.

$\mathbf{\tilde{k} \sim 1}$\textbf{:} The qualitative behavior of very high-$\tilde{k}$ and low-$\tilde{k}$ modes is not strongly dependent on the misalignment amplitude. At large misalignment angles $|\Theta_0| \simeq \pi$, an intermediate regime with new phenomenology appears. Unlike the free scalar case, where the $\tilde{k}\sim 1$ case is a smooth interpolation between the high- and low-$\tilde{k}$ regimes, a dramatic enhancement in density fluctuations is possible. As Fig.~\ref{fig:delta} shows, both the maximum boost in structure and the wavenumber at which this boost occurs, are monotonically increasing with decreasing $\pi-|\Theta_0|$ and thus $f/f_{\pi/2}$. 

\begin{center} \emph{Parametric resonance} \end{center}
The dramatic growth of $\theta_{\vect{k}}$---and thus $\delta_{\vect{k}}$---perturbations for $\tilde{k}\sim 1$~modes can be understood in terms of a parametric resonance instability.  
After the onset of oscillation, we can expand to subleading order in the amplitude of the zero mode, $\bar{\Theta}$, which itself is decreasing slowly, but on a time scale much slower than the oscillatory time scale. This turns the zero mode cosmological evolution equation into one for a damped non-linear harmonic oscillator. Using the Poincar{\'e}-Lindstedt method~\cite{poincare1893methodes}, the zero mode itself can be found to behave according to:
\begin{align}
\Theta = \bar{\Theta} \cos (\bar{\omega} t_m) + \frac{\bar{\Theta}^3}{192}\left[\cos(\bar{\omega} t_m) - \cos(3 \bar{\omega} t_m)\right], \label{eq:ThetaPL}
\end{align}
where $\bar{\omega} = 1-\bar{\Theta}^2/16$. 

We can recast Eq.~\ref{eq:evoltheta} in terms of a damped Mathieu equation, i.e.~a damped harmonic oscillator with a periodically modulated fundamental frequency:
\begin{align}
\frac{\dd^2 \theta_{\vect{k}}}{\dd \tau^2} + c \frac{\dd \theta_{\vect{k}}}{\dd \tau} + \left[ \delta + \epsilon \cos(\tau) \right] \theta_{\vect{k}} = 0, \label{eq:Mathieueq}
\end{align}
where we have defined $\tau \equiv 2 \bar{\omega} t_m$. Above, we have ignored the forcing term from Eq.~\ref{eq:thetaforcing}, and identified the perturbatively small quantities:
\begin{align}
c = \frac{3}{2 \tau}, ~ \delta - \frac{1}{4} = \left[- \frac{ \bar{\Theta}^2}{32} + \frac{\tilde{k}^2}{2 \tau}\right], ~ \epsilon = -\frac{\bar{\Theta}^2}{16}. \label{eq:Mathieuvars}
\end{align}
Eq.~\ref{eq:Mathieueq} has several instability regions; the primary one at small $|\epsilon|$, and the one of interest to us, is the region $|\epsilon| > c + 4(\delta - 1/4)^2$ corresponding to a parametric variation of the natural frequency at approximately twice the natural frequency. The parametric resonance instability can be understood as a process where the quartic interaction converts two zero-mode particles into two finite-momentum particles with $\tilde{k} \neq 0$.

The two exponential growth rate eigenvalues for the amplitudes of $\theta_{\vect{k}}$, expressed in the original $t_m$ coordinates, are:
\begin{align}
\Gamma_\mathrm{PR}^\pm(\tilde{k},t_m) = -\frac{3}{4 t_m} \pm \frac{\bar{\Theta}^2}{16} \sqrt{1-\left(1-\frac{8 \tilde{k}^2}{t_m \bar{\Theta}^2} \right)^2 }.\label{eq:GammaPR}
\end{align}
We see that in the limit $\bar{\Theta} \to 0$ or $\tilde{k}\to 0$, the $\theta_{\vect{k}}$ amplitude decays as $t_m^{-3/4}$, commensurate with the redshifting of the zero mode's energy density redshifting as $ \bar{\Theta}^2 \propto t_m^{-3/2}$. 
For $\tilde{k}\gg 1$, the second term becomes purely imaginary and produces an additional oscillatory behavior with frequency $\tilde{k}^2 / 2 t_m$ that redshifts with time; there is no parametric resonance growth, just as expected for relativistic modes. 

Axion density perturbations will exhibit exponential growth when $\tilde{k}^2 \simeq t_m \bar{\Theta}^2/8$, i.e.~when the root in Eq.~\ref{eq:GammaPR} is real. At least one mode will undergo a substantial growth phase as long as the inequality $\bar{\Theta}^2 \gtrsim 8/t_m$ is satisfied at some point.
Because the amplitude growth is exponential in time (with a rate given in Eq.~\ref{eq:GammaPR}), much of the parametric resonance amplification is dominated by the period in which $\bar{\Theta} < 1$.\footnote{As we will show later in the top panel of Fig.~\ref{fig:quarticcollapsecond}, some amplification also occurs in the nonperturbative regime of $\bar{\Theta}>1$.} For simplicity, we integrate the growth term of Eq.~\ref{eq:GammaPR} starting from $t_{m,0}$, defined as the time at which $\bar{\Theta}= 1$ (or the energy density is $\rho \simeq m^2 f^2 /2$), and take $\bar{\Theta}^2 =  (t_m/t_{m,0})^{-3/2}$. For axions starting near the top of the cosine potential, a good approximation is 
\begin{align}
t_{m,0} \approx 0.596 \big[t_m^\mathrm{osc} + 4 \ln t_m^\mathrm{osc} \big]^{4/3} \label{eq:tm0}
\end{align}
with $t_m^\mathrm{osc}$ as in Eq.~\ref{eq:tC}.
The boost in axion power from parametric resonance is 
\begin{align}
\mathcal{G}(\tilde{k},t_m) \simeq \zeta \exp\left\lbrace 2 \int_{t_{m,0}}^{t_m} \dd t_m' \, \mathrm{Re}\left[ \Gamma_\mathrm{PR}^+(\tilde{k},t_m') + \frac{3}{4 t_m'}\right]\right\rbrace. \label{eq:GPR}
\end{align}
Curvature fluctuations at high $\tilde{k}$ have already partially decayed away to a value that is $\mathcal{O}(1/\tilde{k}^2 t_{m,0})$ smaller than their maximum by the time the axion starts oscillating at $t_{m,0}$ (see Eq.~\ref{eq:curvpert}), leading to a suppression of the initial curvature forcing in Eq.~\ref{eq:thetaforcing}. We account for this effect (that is unrelated to parametric resonance) by the multiplicative suppression factor $\zeta = [1+\tilde{k}^2 t_{m,0} / \pi^2]^{-2}$. 

In the top panel of Fig.~\ref{fig:PR}, we plot the exact numerical results for the relative matter power spectra of axions vs CDM, at a time $t_m = 10^3$.\footnote{The axion transfer function $|\delta_{\vect{k}}/\Phi_{0,\vect{k}}|^2$ is as calculated in Fig.~\ref{fig:delta}, while the CDM perturbation obeys $\delta_{\vect{k}}/\Phi_{0,\vect{k}} = -9 \big[t_k^{-1}  \sin t_k + t_k^{-2} \cos t_k - t_k^{-3} \sin t_k + \ln t_k - \mathrm{Ci}(t_k) + \gamma_\mathrm{E} - 1/2 \big]$ in this notation, where $\mathrm{Ci}$ is the cosine integral function and $\gamma_\mathrm{E}$ is the Euler-Mascheroni constant~\cite{zhang2017evolution}.} The bottom panel shows the function $\mathcal{G}(\tilde{k}, t_m)$ evaluated at $t_m = 10^3$, displaying qualitative agreement with $|\delta_{\vect{k}}/\delta_{\vect{k}}^\mathrm{CDM}|^2$ of the top panel, and justifying the identification of structure growth as due to a parametric resonance effect. We note that the $\mathcal{G}$ function gives an overestimate to the boost in power at low $\tilde{k}$; this difference is due to the forcing of long-wavelength modes after $t_{m,0}$, an effect that is also responsible for the nodes and oscillatory behavior which are present in the top panel (but not the bottom panel) of Fig.~\ref{fig:PR}.

With the above assumptions and simplifications, the asymptotic boost in power relative to that in a CDM scenario, namely $\mathcal{G}(\tilde{k}) \equiv \mathcal{G}(\tilde{k},t_m \to \infty)$, can be expressed in closed form:
\begin{align}
\mathcal{G}(\tilde{k}) 
= \frac{\exp\bigg\lbrace 2\tilde{k}\sqrt{t_{m,0}-4 \tilde{k}^2}-4 \tilde{k}^2 \mathrm{arccos} \Big[\frac{2 \tilde{k}}{\sqrt{t_{m,0}}}\Big]  \bigg\rbrace}{\left(1 + \frac{\tilde{k}^2 t_{m,0} }{ \pi^2}\right)^2}. \label{eq:Gasymptote}
\end{align}

The parametric resonance shuts off entirely at a time $t_m = t_{m,0}^3 / (16 \tilde{k}^4)$ or when the perturbation becomes nonlinear; in practice, this asymptotic form is thus reached rather quickly.

The numerator of Eq.~\ref{eq:Gasymptote} is maximized at $k_*$, with:
\begin{equation*}
\tilde{k}_*=C_k \sqrt{t_{m,0}} \approx 0.2 \sqrt{t_{m,0}}
\end{equation*}
\begin{equation}
\label{eq:Gstar}
\mathcal{G}(\tilde{k}_*)= \zeta_* \exp \left\lbrace \xi' t_{m,0}  \right\rbrace \approx \frac{e^{0.18 t_{m_0}}}{1+0.2 t_{m,0}^2/\pi^2},
\end{equation}

As we will discuss below, the parametric form of the expressions in Eq.~\ref{eq:Gstar} holds for other (time-independent) potentials as well, with different values for the constants $C_k$ and $\xi'$.\footnote{The constant $C_k \approx 0.2$ is a solution to the transcendental equation $2 C_k = \cos \sqrt{1/(16 C_k^2) - 1/4}$, and the constant $\xi' = C_k \sqrt{1-4 C_k^2} \approx 0.18$.} Finally, we note that the boost in halo scale density $\mathcal{B}$ is proportional to the boost in $|\delta_{\vect{k}}|^3 \propto \mathcal{G}^{3/2}$, justifying our claim from Eq.~\ref{eq:Bgeneral} up to polynomial correction factors.

We have so far focused on the case of a cosine potential. However, the parametric resonance instability is quite generic: there is always an unstable wavenumber $\tilde{k}$, as long as the nonlinearities in the potential are large enough to overcome Hubble friction. For a Lagrangian parametrized as $\mathcal{L} = f^2 (\partial \theta)^2/2 - m^2 f^2 (\theta^2 / 2 - \tilde{\lambda} \theta^4/4! + \dots)$, the condition for parametric resonance is
\begin{align}
\tilde{\lambda} \bar{\Theta}^2 \gtrsim \frac{8}{t_m}. \label{eq:paramrescond}
\end{align}
For the cosine potential of Eq.~\ref{eq:cosinepot}, $\tilde{\lambda} = 1$, so given the scaling of $\bar{\Theta}^2 \simeq (t_m/t_{m,0})^{-3/2}$, all that is required is a delay in the onset of axion oscillations from its natural time scale of $t_{m,0} \sim 1$.  For a cosine potential---including for the QCD axion potential in Sec.~\ref{sec:QCD}---this is achieved by having the initial misalignment angle close to the top of the potential, cfr.~Eqs.~\ref{eq:tm0} and \ref{eq:tC}. We postpone a discussion of these peculiar initial conditions to Sec.~\ref{sec:flat}.

Parametric-resonance-fueled growth of density perturbations happens more naturally for ``flatter'' potentials, those for which $t_{m,0}$ can be much larger than unity even for generic initial conditions. We work out two such cases in Sec.~\ref{sec:flat} for two axion potentials given by Eqs.~\ref{eq:ratiopot} and \ref{eq:linwingspot}, which have $\tilde{\lambda} = 6$ and $\tilde{\lambda} = 3$, respectively. For general potentials, all appearances of $\bar{\Theta}^2$ in Eqs.~\ref{eq:ThetaPL}, \ref{eq:Mathieuvars}, and \ref{eq:GammaPR} need to be substituted by $\tilde{\lambda} \bar{\Theta}^2$.
The asymptotic boost factor in the power spectrum, analogous to Eq.~\ref{eq:Gasymptote}, can then be found by performing the integral of Eq.~\ref{eq:GPR}.
The results in Eq.~\ref{eq:Gstar} remain valid, provided one makes the replacements $C_k \to \sqrt{\tilde{\lambda}} C_k$ and $\xi' \to \tilde{\lambda} \xi'$.
Note that the temporal scaling of $\bar{\Theta}^2$ is in general different for time-dependent potentials, such as that of the QCD axion in Sec.~\ref{sec:QCD}, in which case the integral of Eq.~\ref{eq:GPR} does not yield Eq.~\ref{eq:Gasymptote}.

If one extrapolates the nearly scale-invariant primordial curvature perturbation spectrum measured by \textit{Planck}~\cite{Aghanim:2018eyx} all the way to small scales, one can expect fluctuations on the order of $\Phi_{\vect{k},0} \sim \mathcal{O}(10^{-4.5})$. The extreme growth of density perturbations, illustrated by transfer functions $|\delta_{\vect{k}}/\Phi_{\vect{k},0}|^2$ as large as $\gtrsim 10^{10}$ in the top right of Fig.~\ref{fig:delta}, can thus lead to early nonlinearities in the axion perturbations and the subsequent possibility of collapsed structures, which we discuss in Sec.~\ref{sec:nonlinear}. In Sec.~\ref{sec:linNewton}, we will first work out the evolution of perturbations that remain linear long after parametric resonance effects cease. In this case, Newtonian linear perturbation theory is a good approximation at late times, when numerical integration of the equations of motion (Eqs.~\ref{eq:evolTheta} and \ref{eq:evoltheta}) is computationally expensive.

\subsubsection{Newtonian treatment} \label{sec:linNewton}

In the subhorizon, nonrelativistic limit, we can study the evolution of density perturbations using a Newtonian fluid approach.\footnote{See Ref.~\cite{khlopov1985gravitational} for an equation-of-motion treatment of the gravitational instability of a self-interacting scalar field.} This approximation amounts to integrating out the harmonic oscillations of the axion,  and makes it feasible to study the evolution over many $e$-folds of the Universe's expansion. We can then stitch our early-time solution from Sec.~\ref{sec:linGR} onto the Newtonian equations to get the late-time behavior.

At sufficiently late times, namely
\begin{align}
t_m \gg \max\left\lbrace t_{m,0}, \frac{1}{\tilde{k}^2} \right\rbrace, \label{eq:newtoncond}
\end{align}
a Newtonian fluid approximation becomes appropriate. Well beyond the onset of axion oscillations $t_m \gg t_{m,0}$, we can average over the effects during one period of the axion oscillation, as the natural axion frequency is much larger than the Hubble rate, and we can also treat the nonlinearities in the axion potential perturbatively (i.e.~only include effects from the quartic). The inequality $t_m \gg 1/\tilde{k}^2$ ensures that the perturbation is well within the horizon, as well as nonrelativistic ($k/ma \ll 1$). Both the axion background density $\rho$ and its fractional perturbations $\delta_{\vect{k}}$ should then obey standard Newtonian fluid equations.

The zero mode energy density will redshift as $\rho \propto a^{-3(1+w)}$ where $w = P/\rho$ is the equation of state. For an axion with a cosine potential, the pressure equals $P = - \rho^2/16m^2 f^2$~\cite{turner1983coherent}. The fractional density perturbation obeys the differential equation~\cite{marsh2016axion,refId0,
PhysRevD.92.023510}:
\begin{equation} \label{eq:newtonianfluid}
\ddot{\delta}_{\vect{k}} + 2 H \dot{\delta}_{\vect{k}} - \left[ 4 \pi G \rho - \frac{c_s^2 \vect{k}^2}{a^2} \right] \delta_{\vect{k}} = 0
\end{equation}
where $c_s \equiv \sqrt{\delta P / \delta \rho}$ is the sound speed of perturbations.  It receives a $k$-dependent kinetic pressure contribution~\cite{hwang2009axion,park2012axion} as well as an adiabatic contribution $\dd P / \dd \rho$ from the quartic nonlinearity: 
\begin{equation} \label{eq:newtonfluid0}
c_s^2 \simeq \frac{\vect{k}^2/a^2}{4 m^2} - \frac{\rho}{8 m^2 f^2} = \frac{\tilde{k}^2}{4 t_m} - \frac{\langle \bar{\Theta}^2 \rangle}{16}.
\end{equation}
For generalized axion potentials with a different quartic interaction $\tilde{\lambda}$ (cfr.~the discussion around Eq.~\ref{eq:paramrescond} and in Sec.~\ref{sec:flat}), the quartic contribution to the sound speed is to multiplied by $\tilde{\lambda}$.

It is convenient to rewrite Eq.~\ref{eq:newtonfluid0} as a differential equation in the variable $y \equiv a/a_{\mathrm{eq}} = 2^{1/4} \sqrt{t_m H_{\mathrm{eq}}/m}$:
\begin{multline} \label{eq:newtonfluid}
(1+ y) \frac{\mathrm{d}^2 \delta_{\vect{k}}}{\mathrm{d} y^2} + \left( \frac{1}{y} + \frac{3}{2} \right) \frac{\mathrm{d} \delta_{\vect{k}}}{\mathrm{d} y} \\
=  \left[ \frac{3}{2 y} - \frac{\tilde{k}^4}{y^2} + \frac{3}{4 \sqrt{2}} \tilde{k}^2 \frac{M_{\mathrm{Pl}}^2}{f^2} \frac{H_{\mathrm{eq}}}{m} \frac{1}{y^3} \right] \delta_{\vect{k}}
\end{multline}
which also takes into account the transition of the Universe from radiation-domination ($y < 1$) into matter-domination ($y>1$).  The initial conditions for this equation must be found by patching to the solutions from Sec.~\ref{sec:linGR} at some intermediate time $t_m^p$ which satisfies both Eq.~\ref{eq:newtoncond} and $(y^p)^2 = 2 t_m^p H_{\mathrm{eq}} / m \ll 1$.   In other words, we choose a patch time long after the field has started oscillating nonrelativistically but long before matter-radiation equality. The matching conditions for the perturbations are then:
\begin{equation} \label{eq:patchcond}
\delta_{\vect{k}} \bigg\vert_{y^p} = \delta_{\vect{k}} \bigg\vert_{t_m^p} \; ; \qquad \frac{\mathrm{d}\delta_{\vect{k}}}{\mathrm{d} y} \bigg\vert_{y^p} = 2 t_m^p \delta_{\vect{k}}' \bigg\vert_{t_m^p}.
\end{equation}
Patching our solutions from Sec.~\ref{sec:linGR} allows us to evolve them out of radiation-domination to the present day, which we use for many of the observables discussed in Sec.~\ref{sec:signatures}.  

\begin{figure}[!ht]
\hspace*{-0.03\textwidth}\includegraphics[width=0.54\textwidth]{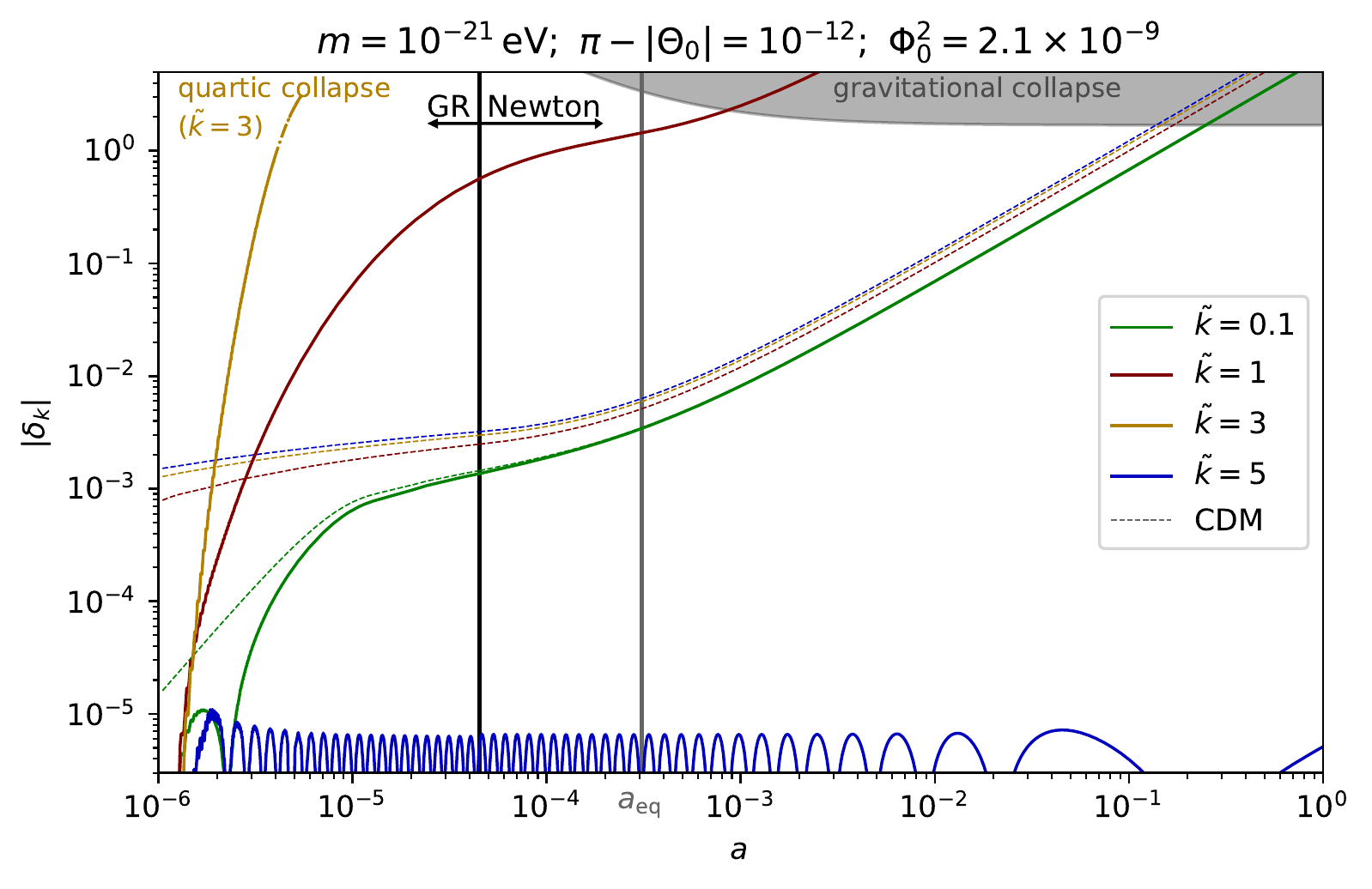}
\caption{Evolution of fractional axion energy density fluctuations $\delta_{\vect{k}}$ as a function of the scale factor for four rescaled wavenumbers $\tilde{k} = \lbrace 0.1, 1, 3, 5\rbrace$, corresponding to comoving wavenumbers of $k = \lbrace 9, 29, 50, 65\rbrace\, \mathrm{Mpc}^{-1}$ for $m = 10^{-21}\,\mathrm{eV}$. The general-relativistic evolution is patched onto the Newtonian one at $t_m = 4\times 10^4$, at the black vertical line. The $\tilde{k}=5$ mode is suppressed and oscillates due to kinetic pressure, while long-wavelength fluctuations (e.g.~$\tilde{k}=0.1$) match onto the CDM predictions (dashed lines). For an axion misalignment angle of $|\Theta_0| = \pi - 10^{-12}$, the $\tilde{k}=1$ mode receives a boost in structure, causing it to collapse gravitationally earlier during matter domination, while modes around $\tilde{k}=3$ collapse due to self-interactions and will lead to oscillon production.}\label{fig:time}
\end{figure}

We demonstrate this full, patched evolution of a few representative $\tilde{k}$-modes in Fig.~\ref{fig:time}. As long as the patching procedure satisfies Eq.~\ref{eq:newtoncond}, there is no dependence of $|\delta_{\vect{k}}|$ on the patching time. Indeed, the qualitative behavior of the modes is the same in the Newtonian regime of Fig.~\ref{fig:time}: the $\tilde{k} = 5$ density perturbation keeps oscillating with the same amplitude and a period that steadily increases (stays constant in $\log a$ time), while the $\tilde{k} = 1$ mode continues to grow in amplitude (with non-negligible contributions from the third term in Eq.~\ref{eq:newtonfluid}). Modes with $\tilde{k} \gtrsim 1$ have too much kinetic pressure at matter-radiation equality to experience this gravitational Jeans instability, and commence linear growth $\delta_{\vect{k}} \propto a$ only after $a \sim \aeq \tilde{k}^4$. After matter-radiation equality, all modes with $\tilde{k} \lesssim 1$ exhibit a gravitational instability, and will undergo linear growth $\delta_{\vect{k}} \propto a$. These modes will eventually become nonlinear---the topic of discussion in Sec.~\ref{sec:nonlinear}.

\subsection{Nonlinear regime} \label{sec:nonlinear}

In the linear regime of Sec.~\ref{sec:linear}, we have seen that the amplitude $\delta_{\vect{k}}$ of density perturbations with $\tilde{k} \sim 1$ can experience a rapid burst of growth during radiation domination, shortly after the field starts oscillating.  Provided the transfer function $|\delta_{\vect{k}}/\Phi_{\vect{k},0}|^2$ is less than the inverse of dimensionless primordial power $\mathcal{P}_\Phi(k)$ at the relevant wavenumber, the perturbations remain linear during radiation domination but have much larger values of $|\delta_{\vect{k}}|$ at matter-radiation equality than predicted in a $\Lambda$CDM universe.  They will thus undergo gravitational collapse---with slight modifications due to kinetic pressure of the scalar field---much earlier than they would have in $\Lambda$CDM, and will form correspondingly denser halos~(Sec.~\ref{sec:gravcollapse}). If the halos exceed a threshold density, they will undergo gravothermal collapse, resulting in a central profile consisting of a steep density cusp cut off by a soliton in the core (Sec.~\ref{sec:gravcooling}). In even more extreme cases (e.g.~the top-right portion of Fig.~\ref{fig:delta}), a density perturbation may even go nonlinear and collapse during radiation domination due to the attractive axion self-interactions.  We devote Sec.~\ref{sec:quarticcollapse} to the conditions for such ``quartic collapse''. Finally, in Sec.~\ref{sec:tidalstripping}, we discuss tidal stripping of halos, relevant for late-time observables discussed in Sec.~\ref{sec:signatures}.

\subsubsection{Gravitational collapse; halos and solitons} \label{sec:gravcollapse}

During matter domination, linear axion density perturbations grow with the scale factor, $\delta_{\vect{k}} \propto a$ as long as $a \gtrsim \aeq \max\lbrace 1,\tilde{k}^4\rbrace$. Thus for standard primordial power spectra, subhorizon fluctuations will become nonlinear before the present day ($a = 1$) unless $\tilde{k} \gtrsim 5$. For axions with large misalignment angles, fluctuations with $\tilde{k} \sim 1$ will go nonlinear \emph{earlier} than in a $\Lambda$CDM universe. $\Lambda$CDM simulations show that overdensities with solely gravitational interactions form gravitationally self-bound objects---halos---with a density profile well-fitted by a Navarro-Frenk-White (NFW) profile~$\rho(r) = 4 \rho_s / [(r/r_s) (1+r/r_s)^2]$~\cite{navarro1997universal}.\footnote{We note that the NFW fit has been thoroughly verified only for nearly scale-invariant power spectra within $\Lambda$CDM contexts, where one expects many mergers. In light of Sec.~\ref{sec:gravcooling}, it should definitely not be trusted at radii $r \lesssim 1/m v_s$ for axion DM. 
A \emph{spike} in the power spectrum---a shape more similar to what is generated by the large-misalignment mechanism---produces \emph{cuspier} halos, with an inner density profile $\rho(r) \propto r^{-3/2}$~\cite{delos2018density}.\label{fn:profile}} 
The scale radius $r_s$, scale density $\rho_s = \rho(r_s)$, and scale mass $M_s = 4\pi \int_0^{r_s} \dd r\,r^2 \rho(r) = 8 \pi \rho_s r_s^3 (\ln 4 - 1)$ remain approximately constant for times subsequent to the formation of the halo~\cite{ludlow2013mass,correa2015accretion}, and are relatively robust against moderate tidal stripping (see Sec.~\ref{sec:tidalstripping}).\footnote{This is in contrast to the oft-used quantities $r_{200}$, the radius within which the mean halo density is 200 times the Universe's, and $M_{200} = \int_0^{r_{200}} \dd^3 r\, \rho(r)$, the mass inside that radius. Both these quantities increase with scale factor, but can drastically decrease with tidal stripping (even if the halo is not completely disrupted).} We will therefore describe axion compact halos, the nonlinear structures resulting from axion overdensities, in terms of their scale quantities $M_s$ and $\rho_s$, the latter enhanced relative to a typical CDM halo due to the boost in $\delta_{\vect{k}}$ over a small range in $k$ and thus scale mass $M_s$. We define the scale potential as the gravitational potential at the scale radius, namely $\Phi_s \equiv \Phi(r_s) = - 16 \pi \ln(2) G_N \rho_s r_s^2$, and use the scale velocity $v_s \equiv \sqrt{-\Phi_s}$ as a measure of internal velocity dispersion. 

Gravitational collapse dynamics can be understood analytically within the Press-Schechter formalism \cite{Press:1973iz}, where a spherical tophat perturbation decouples from the ambient Hubble flow to form a virialized object at $a_\mathrm{coll}$, the scale factor at which linear perturbation theory would have predicted the fractional overdensity to have equaled $\delta_c \approx 1.686$ in a matter-dominated Universe. The virial density of the resulting halo is approximately $178$ times the mean density of the Universe at $a_\mathrm{coll}$.
A question still remains about the precise conditions for collapse, because axion density fluctuations $\delta(\vect{r}) = (2\pi)^{-3} \int  \dd^3 k \, \tilde{\delta}(\vect{k}) e^{i \vect{k}\cdot \vect{r}}$ are a (initially Gaussian) random field, with overdensities that are neither spherically symmetric nor even of similar shape and amplitude.  In practical terms, 
to explore fluctuations at different scales, $\delta(\vect{r})$ is smoothed to a density field $\delta(\vect{r},R_\mathrm{S})$ over a size $R_\mathrm{S}$ using an appropriate window function $W(\vect{r}-\vect{r}',R_\mathrm{S})$:
\begin{align}
\delta(\vect{r},R_\mathrm{S}) = \int \dd^3 r' \, W(\vect{r}-\vect{r}',R_\mathrm{S}) \delta(\vect{r}').\label{eq:smoothdens}
\end{align}
Inspired by the spherical collapse model, the window function is commonly taken to be a spherical tophat $W(\vect{r},R_\mathrm{S}) = \Theta(R_\mathrm{S} - r)(3/4\pi R_\mathrm{S}^3)$. One then posits that a point $\vect{r}$ is part of a halo of mass $M_s \geq M_\mathrm{S} \equiv (4\pi/3)\rho^0_\mathrm{DM}R_\mathrm{S}^3$ when $\delta(\vect{r},R_{\mathrm{S}}) \gtrsim  \delta_c$.

\begin{figure}[tbp]
\includegraphics[width=0.48\textwidth]{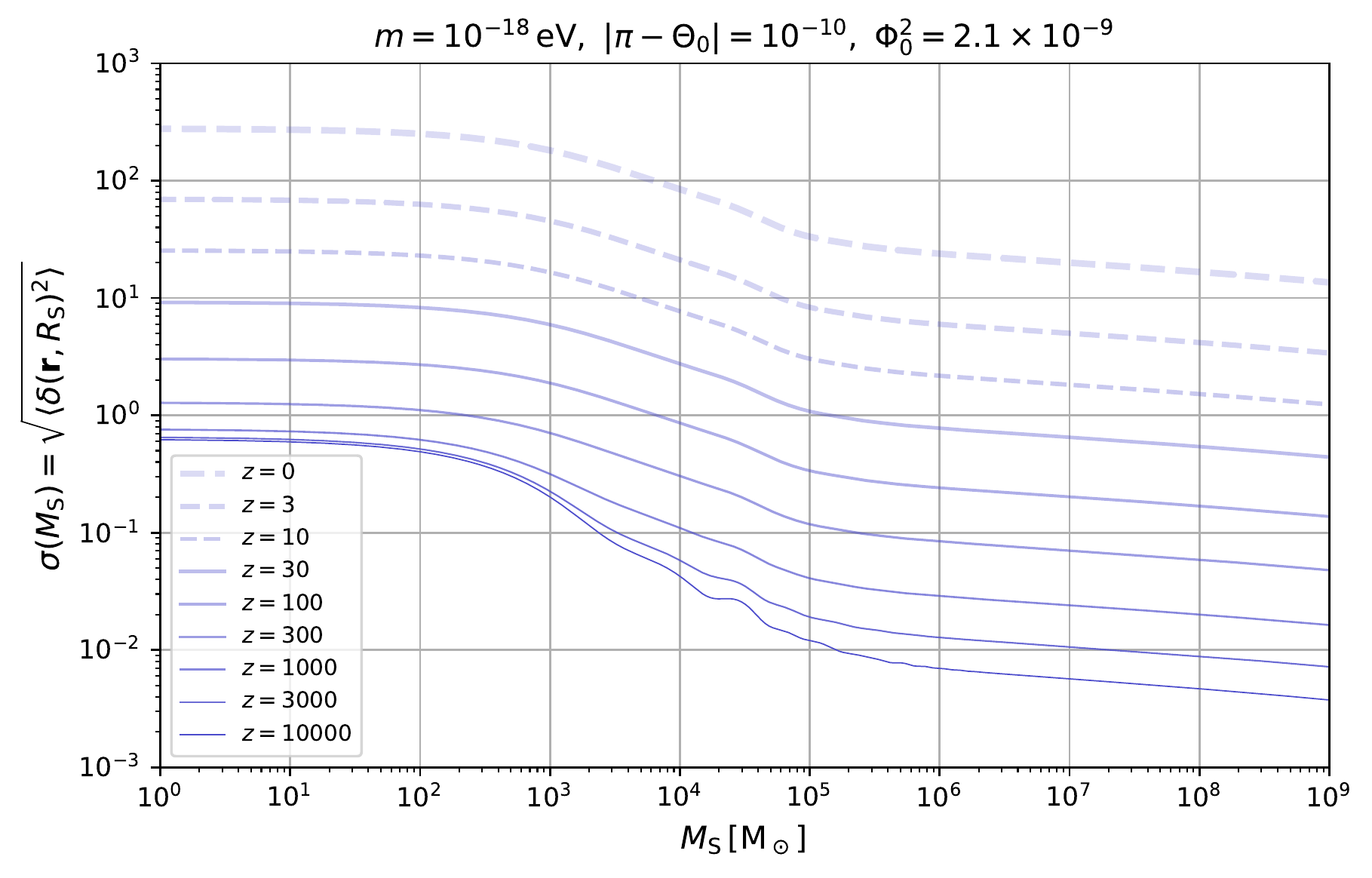}
\includegraphics[width=0.48\textwidth]{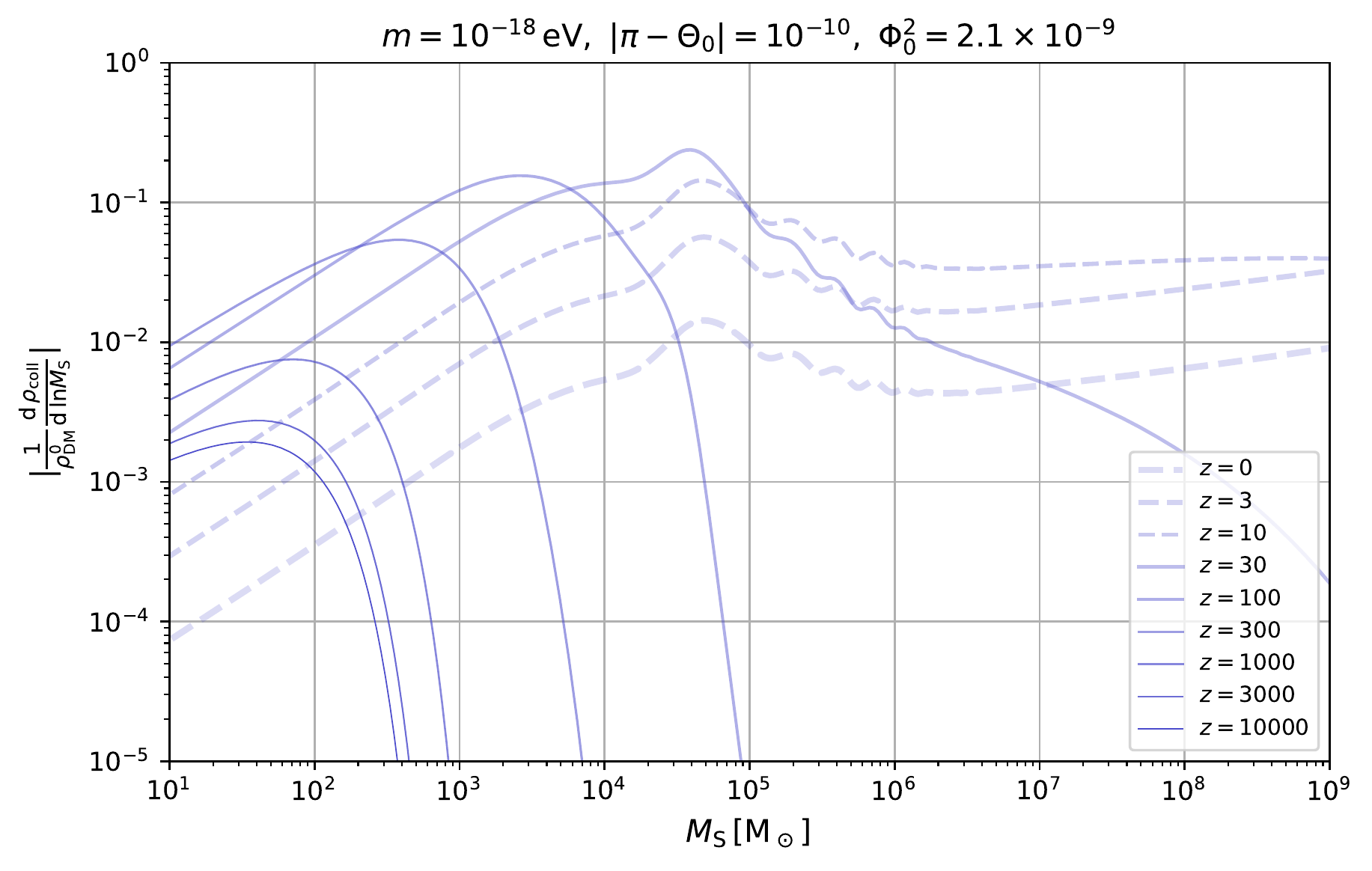}
\caption{Standard deviation of the smoothed axion density field (top panel) and the resulting differential energy density fraction in collapsed halos per logarithmic mass bin (bottom panel), as a function of the smoothing mass scale $M_\mathrm{S} = \frac{4\pi}{3} \rho^0_\mathrm{DM} R_\mathrm{S}^3$ of the spherical tophat window function with radius $R_\mathrm{S}$. Our results are plotted for the benchmark case of $m = 10^{-18}\,\mathrm{eV}$ and $\pi - |\Theta_0| = 10^{-10}$ also plotted in Fig.~\ref{fig:halospec}, at different redshifts $z$. Despite our input of a standard scale-invariant spectrum of curvature fluctuations, $\mathcal{O}(1)$ density perturbations at small scales are already common by matter-radiation equality. Further growth at these scales occurs during matter domination, albeit slightly delayed relative to large scales due to effects of kinetic pressure, leading to a collapsed halo fraction of 56\% (82\%) by redshift $z=100$ ($z=30$) entirely in dense halos lighter than $10^5\,\mathrm{M_\odot}$. After $z \sim 30$, these halos are assimilated into larger CDM-like halos.
}\label{fig:haloPS}
\end{figure}

The variance $\sigma^2(M_\mathrm{S}) \equiv \langle \delta(\vect{r},R_\mathrm{S})^2\rangle$ of the density field at the mass scale of $M_\mathrm{S}$ can be written as
\begin{align}
\sigma^2(M_\mathrm{S}) = \int \dd \ln(k) \, \mathcal{P}_\Phi(k) \left|\frac{\delta_{\vect{k}} }{\Phi_{\vect{k},0}}\right|^2 \left|W(\vect{k},R_\mathrm{S})\right|^2
\end{align}
where $W(\vect{k},R_{\mathrm{S}}) = \int \dd^3 r \, W(\vect{r},R_\mathrm{S}) e^{-i \vect{k}\cdot \vect{r}}$ is the Fourier transform of the window function. In the top panel of Fig.~\ref{fig:haloPS}, we show the standard deviation $\sigma(M_\mathrm{S})$ as a function of the smoothing mass scale $M_\mathrm{S}$ for an axion mass $m = 10^{-18}\,\mathrm{eV}$ and misalignment $\pi - |\Theta_0| = 10^{-10}$.
Assuming the fluctuations are Gaussian-distributed, the collapsed fraction of structures with a smoothing mass larger than $M_\mathrm{S}$ is $F(M_\mathrm{S}) = \mathrm{erfc}[\delta_c/\sqrt{2}\sigma(M_\mathrm{S})]$ in the extended Press-Schechter formalism. We can then construct a differential collapsed energy density per logarithmic smoothing mass $\frac{\dd \rho_\mathrm{coll}}{\dd \ln M_\mathrm{S}} \equiv \rho_\mathrm{DM}^0 \frac{\dd F(M_\mathrm{S})}{\dd \ln M_\mathrm{S}}$, and a differential collapsed fraction that evaluates to:
\begin{align}
\frac{1}{\rho_\mathrm{DM}^0} \frac{\dd \rho_\mathrm{coll}}{\dd \ln M_\mathrm{S}} = \sqrt{\frac{2}{\pi}} \frac{\delta_c}{\sigma(M_\mathrm{S})} \left|\frac{ \dd \ln \sigma(M_\mathrm{S})}{\dd \ln M_\mathrm{S}} \right| e^{\frac{-\delta_c^2}{2\sigma^2(M_\mathrm{S})}}.\label{eq:diffcollfrac}
\end{align}
We plot this function in the bottom panel of Fig.~\ref{fig:haloPS} for the same axion parameters as in the top panel. Already at $z = 3000$, $F(M_\mathrm{S}) \approx 1\%$ of perturbations exceed the critical threshold of $\delta_c$. The majority of points in space are in a dense, gravitationally-collapsed halos before redshift $z = 100$. Over time, the differential collapsed fraction at small smoothing masses $M_\mathrm{S}$ decreases as halos at these mass scales become part (i.e.~subhalos) of larger halos. 

One drawback of the Press-Schechter procedure with a spherical tophat window function is that it largely fails to account for halo substructure. For example, $\delta(\vect{r},R_\mathrm{S})$ can be large even when there is no structure at scales of order $R_\mathrm{S}$, as long as there is structure on scales bigger than $R_\mathrm{S}$. Likewise, the differential collapsed fraction of Eq.~\ref{eq:diffcollfrac} does \emph{not} include structures of mass $M_\mathrm{S}$ that are already assimilated into more massive halos. So while the above procedure and the results of Fig.~\ref{fig:haloPS} are useful to track parts of the density field's statistics, they are crude instruments for extracting the halo spectrum.

The two issues pointed out above---\emph{non-isolation} and \emph{undercounting} of substructure at the scale $R_\mathrm{S}$---stem from the fact that the Fourier transform of the spherical tophat window $W(\vect{k}, R_\mathrm{S}) = 3[\sin (k R_\mathrm{S}) - k R_\mathrm{S} \cos (k R_\mathrm{S})]/ (k R_\mathrm{S})^3$ has nonzero support even for $k \ll R_\mathrm{S}^{-1}$. Therefore, rather than summing the cumulative structure above $R_\mathrm{S}$, which is effectively what the spherical tophat smoothing procedure does, one can also use a window function that \emph{isolates} the structure at a length scale $R$:
\begin{align}
W(\vect{k},R) = N \exp\Bigg\lbrace - \frac{\big[\ln (k R/\pi)\big]^2}{4 \tilde{\sigma}^2} \Bigg\rbrace
\end{align}
with $\tilde{\sigma} = 1/2$ and a normalization constant $N$ such that $\int\dd\ln(k) \, |W(\vect{k},R)|^2 = 1$. The disadvantage of this window function is that its volume in real space formally diverges, and therefore cannot be interpreted as a smoothing kernel as in Eq.~\ref{eq:smoothdens}. Nevertheless, we find this window function useful to construct a halo spectrum, i.e.~a typical mass-density relation $\lbrace M_s, \rho_s \rbrace$:
\begin{align}
M_s &\equiv C_M \frac{4\pi}{3} \rho_\mathrm{DM}^0 R^3 \label{eq:halospec1}\\
\rho_s &\equiv  C_\rho \rho_\mathrm{DM}^0 a_\mathrm{coll}^{-3}; \qquad a_\mathrm{coll} = \left\lbrace a \big| \sigma(R_\mathrm{S}) = \delta_c \right\rbrace \label{eq:halospec2}
\end{align}
with fiducial values of $C_M \approx 1$ and $C_\rho \approx 200$. In other words, our procedure amounts to smoothing the dimensionless linear power spectrum $\mathcal{P}(k)$ in $\ln(k)$ space, and taking a typical halo to form when a smoothed 1-sigma overdensity reaches a value of $\delta_c \approx 1.686$. Note that with our definitions, the total fraction of DM within gravitationally collapsed structures can be larger than unity, because we are counting a halo and all its subhalos (and subsubhalos etc.) separately. We expect that if linear perturbation theory predicts $\sigma^2 \gtrsim 1$ at some scale $R$ with our window function, $\mathcal{O}(1)$ of the DM is contained within structures of mass $M_s$ as in Eq.~\ref{eq:halospec1}, provided they survive tidal stripping (see Sec.~\ref{sec:tidalstripping}).

\begin{figure}[tbp]
\includegraphics[width=0.48\textwidth]{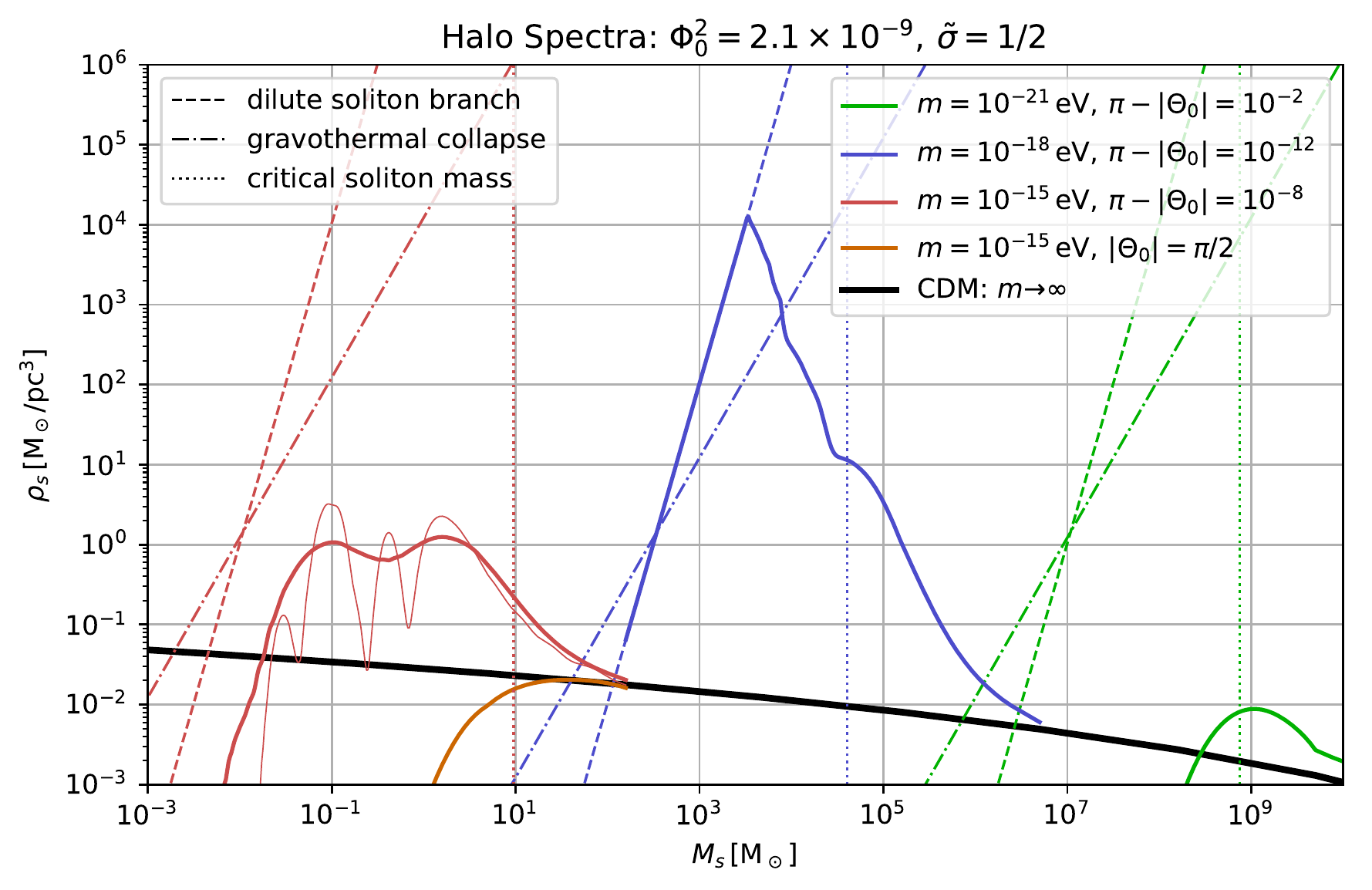}
\caption{Halo spectra in terms of scale mass $M_s$ and scale density $\rho_s$ (as in Eqs.~\ref{eq:halospec1} and \ref{eq:halospec2}) for several different axion masses $m$ and misalignment angles $\Theta_0$, as well as the reference CDM halo spectrum. The thick solid lines are computed with a dimensionless smoothing kernel of $\tilde{\sigma} = 1/2$.  For $m = 10^{-15}\,\mathrm{eV}$ and $\pi-|\Theta_0| = 10^{-8}$, we also display the halo spectrum with a narrower kernel of $\tilde{\sigma} = 1/10$ (thin red line), revealing the oscillatory behavior of the power spectrum at high wavenumber. The dashed lines depict the dilute soliton branch of Eq.~\ref{eq:solitonmassdensityrelation}, the densest possible stable axion configuration, for the same three axion masses, and the dotted vertical lines indicate the maximum (critical) soliton mass. The dot-dashed lines delineate the density above which gravothermal catastrophe occurs inside the halo, resulting in a steep internal density profile (a cusp cut off by a central soliton).}\label{fig:halospec}
\end{figure}

In Fig.~\ref{fig:halospec}, we plot the halo spectrum as defined in Eqs.~\ref{eq:halospec1} and \ref{eq:halospec2} for four different cases, assuming a scale-invariant primordial curvature power spectrum $\mathcal{P}_\Phi(k) \approx 2.1 \times 10^{-9}$. We see that the enhancement of density perturbations at scales with $\tilde{k}\sim 1$ results in halos that collapse earlier than in $\Lambda\mathrm{CDM}$ cosmological history and can be significantly denser than the $\Lambda\mathrm{CDM}$ prediction at comparable scales if $\pi - |\Theta_0| \ll 1$. The typical mass of these overdense halos is thus the one given in Eq.~\ref{eq:Msstar}.

As the halos become denser, eventually the de Broglie wavelength of the gravitationally bound axions becomes comparable to the size of the halo. At that point, the repulsive kinetic pressure of the axions becomes important for the dynamics of the halo and the halos transition to the \emph{soliton} regime, represented by the dashed line shown in Fig.~\ref{fig:halospec}.  These gravitationally-bound axion field configurations have been extensively studied in the literature \cite{Seidel:1991zh, Braaten:2015eeu, Braaten:2016dlp, Chavanis:2011zm, Chavanis:2011zi, Eby:2017azn, Eby:2017teq, Eby:2018ufi, Visinelli:2017ooc, Schiappacasse:2017ham, Mukaida:2016hwd, Salmi:2012ta}, and we devote App.~\ref{sec:boundstates} to a review of some of their properties.  There are, however, two facts that are quite relevant for the discussion here.

The first is that solitons have a well-defined relationship between mass and density.  Defining a soliton's scale radius by $r_s^{\mathrm{sol}} = \{r | \partial \ln \rho(r) / \partial \ln r = -2 \}$, we can numerically solve for the ground-state of the Schr\"odinger-Poisson equation to find:
\begin{equation}
\rho_s^{\mathrm{sol}} \approx 0.7 \, G_N^3 m^6 (M_s^{\mathrm{sol}})^4
\label{eq:solitonmassdensityrelation}
\end{equation}
where $\rho_s^{\mathrm{sol}} = \rho(r_s^{\mathrm{sol}})$ and $M_s^{\mathrm{sol}}$ is the mass enclosed within the scale radius.  For a fixed total mass of axions $M$ (with the scale mass given numerically by $M_s^{\mathrm{sol}} \approx 0.4 \, M$), this soliton state is the unique minimum-energy state, and the densest energy eigenstate.  This one-parameter family of solutions parametrized by $M_s^\mathrm{sol}$ acts as an upper bound to the scale density of a stable halo as a function of its scale radius, and we plot this bound for a few different axion masses in Fig.~\ref{fig:halospec}.  For high misalignment angles, it is possible to saturate this bound, which we also show in Fig.~\ref{fig:halospec}.

The second relevant fact is that the gravitational soliton branch described in the above paragraph has a maximum possible mass $M$ (see App.~\ref{sec:boundstates}) which corresponds to a maximum scale mass (for an axion with a cosine potential):
\begin{equation}
M_{s,\mathrm{max}}^{\mathrm{sol}} \approx 0.4 \, M_{\mathrm{max}}^{\mathrm{sol}} \approx 10 \frac{f M_{Pl}}{m}
\label{eq:solitonmaxscalemass}
\end{equation}
which we plot on Fig.~\ref{fig:halospec} for each choice of axion mass $m$ by means of a vertical dotted line.  Above this value, the attractive axion self-interactions overwhelm the repulsive kinetic pressure and no nonrelativistic, (metastable) ground state configuration exists.  Any sufficiently dense axion configuration above this mass will collapse within a dynamical time (i.e.~an infall time).  Such self-interaction-induced collapses have been studied previously in Ref.~\cite{Levkov:2016rkk}. The large-misalignment mechanism can produce dense solitons at the mass $M_s^*$ in Eq.~\ref{eq:Msstar}, which is parametrically only slightly below the critical soliton mass $M_{s,\mathrm{max}}^{\mathrm{sol}}$, by a factor of $\sim (\Heq/m)^{1/4}$. We speculate that mergers and accretion due to the gravitational cooling mechanism of Sec.~\ref{sec:gravcooling} below may tip them over the edge, thus opening up the possibility for \emph{late-time} supercritical soliton collapse into oscillon-like configurations. We leave a detailed analysis of these phenomena and their impact on detectability to future work. In Sec.~\ref{sec:quarticcollapse}, we will study the \emph{early-time}, direct production of oscillon-like states, a process that does not involve a soliton as an intermediate state.

\subsubsection{Gravitational cooling} \label{sec:gravcooling}

For the halos described above, gravitational cooling is another process, beyond mergers and accretion, that can significantly alter their structure. Compact halos not in the soliton regime can cool and form a soliton at their center, and solitons already present can accrete more mass from the cooling of their surrounding halos. The cooling timescale $\tau_{\mathrm{gr}}$ has been estimated by Ref.~\cite{Levkov:2018kau}, and in terms of the scale quantities defined in Sec.~\ref{sec:gravcollapse} their expression reads:
\begin{equation} \label{eq:gravcooltimerhom}
\tau_{\mathrm{gr}} \simeq C_{\mathrm{gr}} \frac{G m^3 M_s^2}{\rho_s} \frac{1}{\Lambda}
\end{equation}
where $C_{\mathrm{gr}}$ is an $\mathcal{O}(1)$ constant, and $M_s$ and $\rho_s$ are the halo's scale mass and density, respectively.  Here $\Lambda \sim \log (m v_s r_s )$ is a Coulomb logarithm (with $r_s$ the scale radius and $v_s$ the scale velocity), which we keep for completeness but which is $\mathcal{O}(1)$ for the whole parameter space, and so does not substantially change the results.

The cooling time scale of Eq.~\ref{eq:gravcooltimerhom} is simply the inverse rate of gravitational scattering, which is greatly increased by a bosonic enhancement factor. Indeed, Eq.~\ref{eq:gravcooltimerhom} gives the rate of gravitational scattering of quasiparticles of mass $\sim \rho_s \lambda_s^3$ and size $\lambda_s \sim 1/m v_s$; one can therefore view the gravitational cooling process as being due to the scattering of the interference fringes of the axion field~\cite{hui2017ultralight}, which cause $\mathcal{O}(1)$ density fluctuations on the scale of the de Broglie wavelength $\lambda_s$.
Ref.~\cite{Levkov:2018kau} finds that after a timescale of roughly $\tau_{\mathrm{gr}}$, a soliton will spontaneously form in the halo, and grow in mass on similar time scales, at least initially. 

For moderately enhanced halo scale densities, the soliton that forms initially is much smaller than the halo in both mass and size ($\lambda_s \ll r_s$, the ``kinetic regime'' of Ref.~\cite{Levkov:2018kau}). Nevertheless, at time $t \gg \tau_\mathrm{gr}$, the backreaction of gravitational cooling on the halo is likely to be severe. Gravitationally bound systems have a negative heat capacity, so gravitational scattering (or any form of kinetic energy exchange for that matter) generically causes a runaway instability to take place---the ``gravothermal catastrophe''. This phenomenon is known to occur in globular clusters on a time scale of $\sim 300 \, \tau_\mathrm{gr}$~\cite{lynden1980consequences, portegies2010young}, and we expect it to be operative for compact axion halos as well. 

The physical mechanism can be understood as follows: heat transfer from the dynamically warmer halo core to the colder periphery of the halo will cause the core to lose energy, and thus \emph{heat up and contract} by the negative heat capacity and the virial theorem. This process is recursive: the core will continue to collapse (heat up but shrink in mass $M_\mathrm{core}$ while its density $\rho_\mathrm{core}$ increases) by using its immediate outskirts as a heat sink. Ref.~\cite{lynden1980consequences} showed that for the case of gravitational scattering, there is an attractor solution for this process, with the collapsing core expected to leave behind a cuspy halo density profile of $\rho(r) \sim \rho_s (r/r_s)^{-\alpha}$ for $r \ll r_s$. Ref.~\cite{lynden1980consequences} argues that $\alpha$ takes values between 2 and 2.5, with numerical simulations favoring $\alpha \approx2.21$. (We expect the halo scale radius and density to be only moderately increased and decreased, respectively, by the gravitational cooling process.) 

In the case of axion dark matter, the core collapse should be halted when the core reaches a size where repulsive kinetic pressure becomes important, i.e.~when the line $\lbrace M_\mathrm{core},\rho_\mathrm{core}\rbrace$ intersects the soliton branch of Eq.~\ref{eq:solitonmassdensityrelation}, depicted also in Fig.~\ref{fig:halospec} for some benchmark axion parameters. The assumption of self-similar collapse combined with the above reasoning thus allows us to derive a relation between the solitonic core mass and the host halo mass. The core density and a function of its mass is $\rho_\text{core} \propto M_\text{core}^{-\alpha/(3-\alpha)}$, resulting in a core soliton of mass:
\begin{align}
M^\text{sol}_\text{core} = \left( \frac{4\pi}{3} \frac{\rho_s M_s^{\frac{\alpha}{3-\alpha}}}{G_N^3 m^6} \right)^{\frac{3-\alpha}{3(4-\alpha)}}.
\end{align}
For $\alpha=2.21$, this gives $M_\text{core}\propto M_s^{0.41}$, which is to be contrasted with the expectation of $M_\text{core}\propto M_s^{1/3}$ for an isothermal profile, where $\alpha=2$. The latter relation appears to arise in fuzzy DM simulations~\cite{PhysRevLett.113.261302}. We do not believe this to be in conflict with what we are describing here. In our mechanism with self-interactions, $\rho_s$ is drastically enhanced and gravitational cooling is more efficient than for a free scalar field minimally coupled to gravity. We point out that a transition from an NFW to an isothermal profile is expected as the first step in the gravothermal collapse process.\footnote{The scaling relation of $M_\text{core} \propto M_s^{1/3}$ has been extrapolated to halos heavier than those simulated to place constraints on axions above $10^{-22}\,\mathrm{eV}$~\cite{Bar:2018acw, Safarzadeh:2019sre} in mass. We do not believe these constraints should be trusted; the above scaling applies to isothermal profiles when the average velocity inside the solitonic core is equated with the velocity right outside. This core-halo mass relation should then break down in NFW halos for which the thermalization radius (the radius within which $\tau_\mathrm{gr}\sim H_0^{-1}$ and out to which the halo profile now becomes isothermal) is smaller than the scale radius $r_s$. For particle masses of $10^{-19}\,\mathrm{eV}$, this happens in halos heavier than $10^{7}\,\mathrm{M_\odot}$, and this cutoff scales as $m^{-3/2}$ for other axion masses. Above this halo mass cutoff, calculating the radius for which $\tau_\mathrm{gr}\sim H_0^{-1}$ and relating this radius to the halo mass suggests that $M_\text{core}\propto M_s^{2/15}$ and the extrapolation used in the above references clearly does not apply.}

In Fig. \ref{fig:halospec}, we show the minimum halo scale density at which gravothermal core collapse is expected to occur. Specifically, the dot-dashed lines are contours at which $\tau_\mathrm{gr}^{-1} = 300 \, H_0$, for the three benchmark axion masses considered. Halos above this contour, e.g.~those with $M_s \sim 10^{4}\,\mathrm{M_\odot}$ of the blue halo spectrum in Fig.~\ref{fig:halospec} with $m = 10^{-18}\,\mathrm{eV}$ and $\pi - |\Theta_0| = 10^{-12}$, will have their cores collapse to the soliton branch. Subsequent to this collapse, the central soliton is expected to accrete and therefore increase further in mass and density. For axion decay constants far below $f_\mathrm{\pi/2}$, it may be possible that this central soliton could accrete to the critical soliton mass at late times, the point at which a dramatic implosion and bosenova of the type described in Ref.~\cite{Levkov:2016rkk} and App.~\ref{sec:boundstates} would take place. For the parameters plotted in Fig.~\ref{fig:halospec}, we do not foresee this scenario to materialize, as the host halos affected by gravothermal core collapse are below the critical soliton mass of Eq.~\ref{eq:solitonmaxscalemass}, but halo mergers and accretion are possible loopholes to these arguments.  Further numerical work is needed to study this possibility; it is clear, however, that soliton formation is greatly aided by the initial enhancement of small-scale structure by our mechanism.
Finally, gravitational scattering \emph{between} compact axion subhalos may also affect the dynamics of their larger host halos. This aspect is discussed in Sec.~\ref{sec:dynconstraints}.

\subsubsection{Quartic collapse; oscillons} \label{sec:quarticcollapse}
At very large misalignment angles, namely $\pi - |\Theta_0| \lesssim 10^{-12}$ for the cosine potential, it can be deduced from Fig.~\ref{fig:delta} that the parametric resonance growth of perturbations can lead the axion field to grow nonlinear on scales $\tilde{k}\sim 1$ well before matter-radiation equality.  For the nonperiodic potentials of Sec.~\ref{sec:flat}, the same effects are obtained for $|\Theta_0|\gg 1$, as indicated in Figs.~\ref{fig:DeltaFlatPot} and \ref{fig:DeltaLinWings}. Density perturbations on these scales can potentially decouple from the expansion of the universe, leading to DM structures that collapse solely via self-interactions. In this section, we numerically examine the conditions in which this ``quartic collapse'' can occur and compare our results with a (very) simple analytic model of the collapse process. We restrict ourselves here to spherically symmetric fluctuations, but we do not expect qualitative differences in the collapse condition for $\mathcal{O}(1)$ asymmetric perturbations. 

Our numerical procedure involves taking a field configuration that consists of a zero-mode background $\theta_0$ and a spherically-symmetric Gaussian axion field wavepacket of radius $R_{m,0}$ and fractional overdensity $\delta_0$ at the center:
\begin{align}
\theta(t_{m,0},\vect{x}_m) = \theta_0 \left[ 1 + \frac{1}{2}\delta_0 \exp \left( - \frac{r_m^2}{2 R_{m,0}^2} \right) \right], \label{eq:wavepacketic}
\end{align}
where $t_{m,0}$ is the time at which we start our simulation. We also switch to a new comoving coordinate system $\lbrace t_m, \vect{x}_m \rbrace$ where the axion mass dependence drops out, and the metric is $\dd s^2 = m^{-2} (\dd t_m^2 - t_m \dd \vect{x}_m^2 )$. The dimensionless time coordinate is $t_m \equiv m/2H = mt$ as before, while $\vect{x}_m \equiv t_m^{-1/2} a m \vect{x}$ is a dimensionless spacelike coordinate in which a momentum mode characterized by $\tilde{k}$ has a wavelength of $2 \pi / \tilde{k}$. Note that, relative to Eq.~\ref{eq:RDmetric}, we are ignoring curvature perturbations and that $r_m \equiv |\vect{x}_m|$ in Eq.~\ref{eq:wavepacketic}.
Let us also assume that $\partial_{t_m}  \theta(t_{m,0},\vect{x}_m)  = 0$. We study the evolution of this wavepacket via the full nonlinear field equation (with spherical symmetry and without metric perturbations), which in this coordinate system reads
\begin{align}
\left[\partial_{t_m}^2  + \frac{3}{2t_m} \partial_{t_m} - \frac{1}{t_m} \left( \partial_{r_m}^2 + \frac{2}{r_m} \partial_{r_m} \right) \right] \theta + \sin \theta = 0, \label{eq:wavepacketpde}
\end{align}
along with the initial condition of Eq.~\ref{eq:wavepacketic}. Ignoring the forcing terms from curvature perturbations in Eq.~\ref{eq:thetaforcing} becomes an increasingly good approximation at late times, so our real-space, nonlinear simulations with Eq.~\ref{eq:wavepacketpde} capture and thus isolate the effects from the self-interactions only. They are thus complementary to the linear Fourier analysis of Sec.~\ref{sec:linGR}. We collect specifications of our numerical method in App.~\ref{sec:numerics}.

\begin{figure}[tbp]
\includegraphics[height=0.25\textwidth, trim= 0 0 0 0, clip]{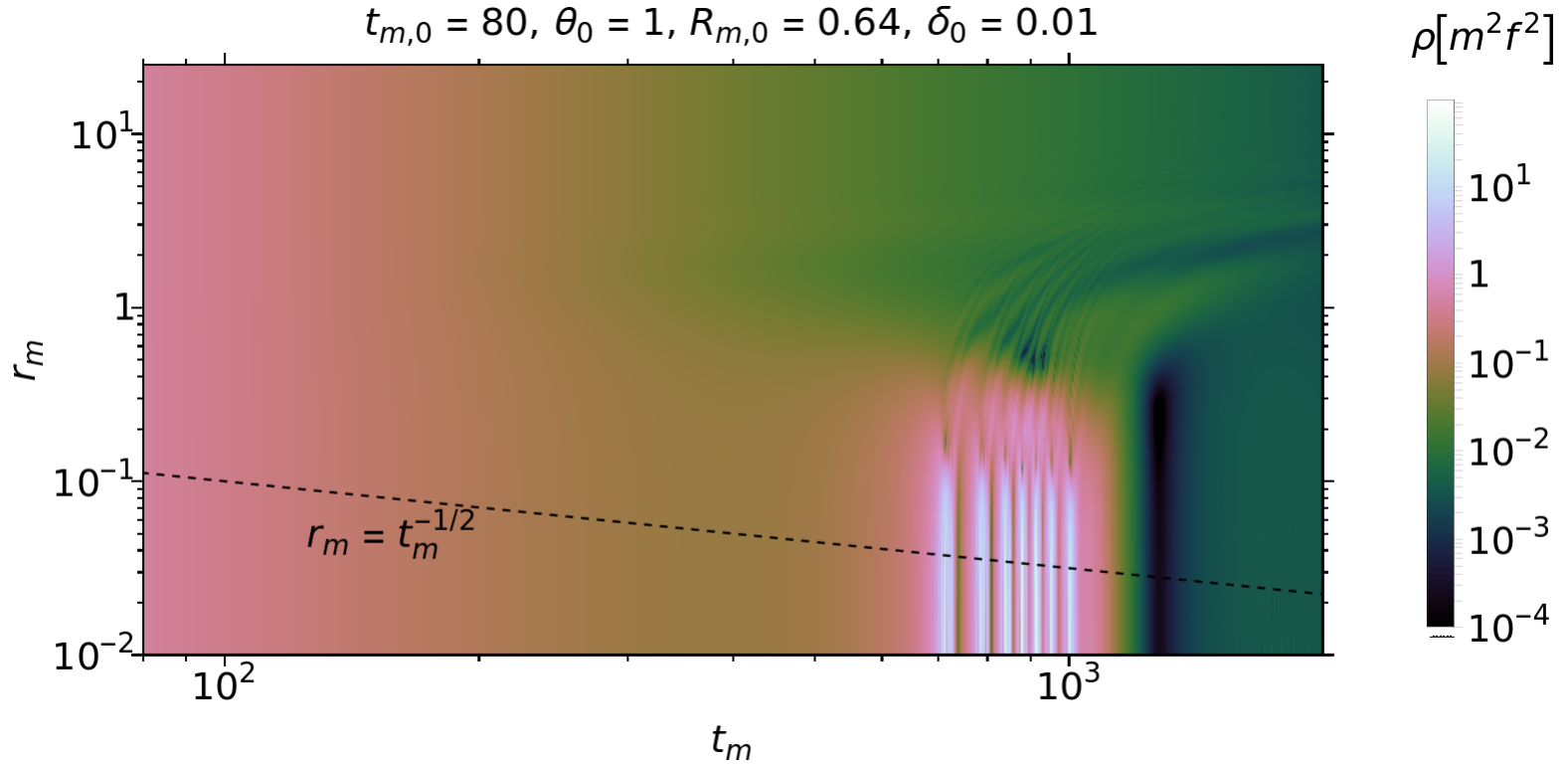}\\
\includegraphics[height=0.25\textwidth, trim= 0 0 0 0, clip]{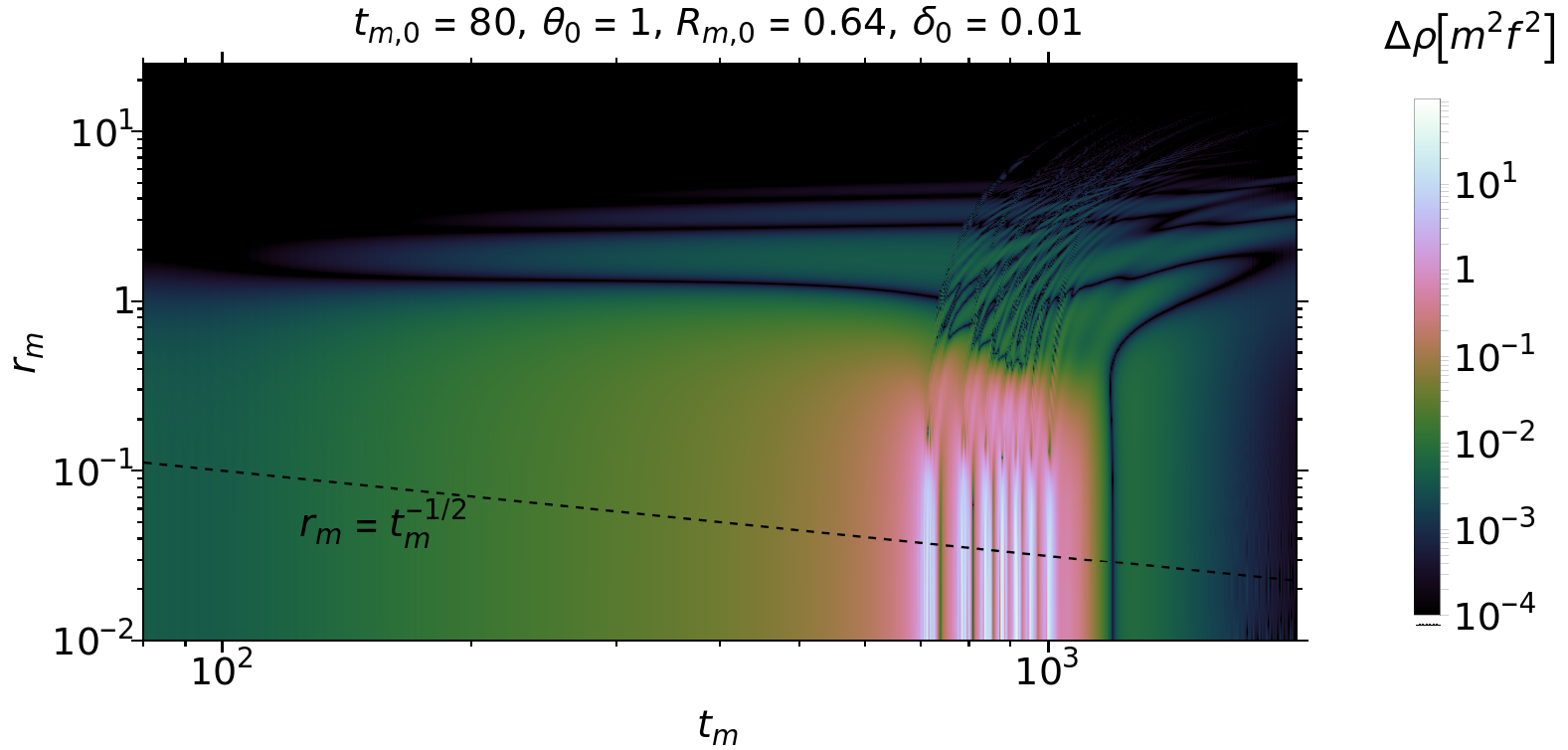}\\
\includegraphics[height=0.25\textwidth, trim= 0 0 0 0, clip]{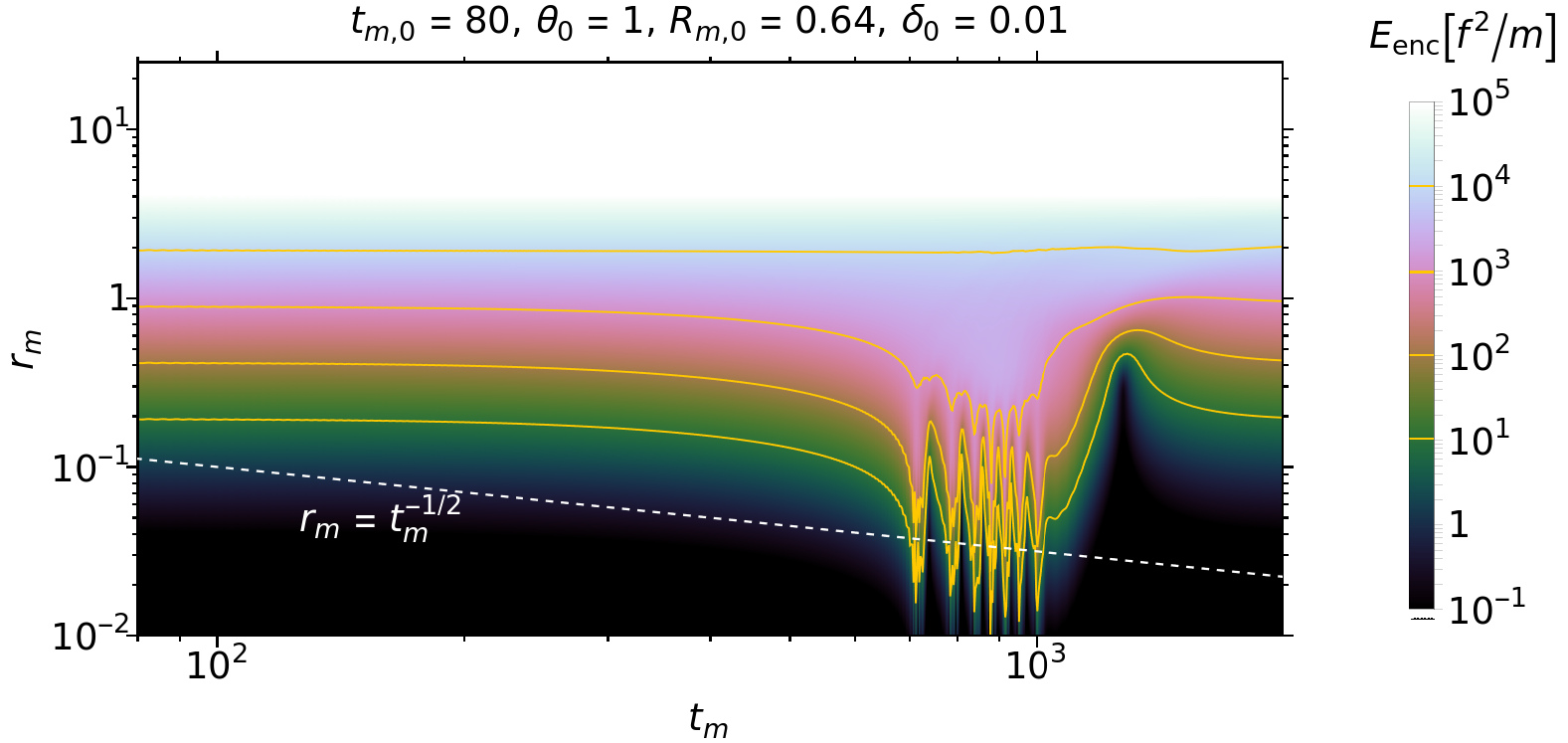}
\caption{Spherically symmetric simulation of the axion field as a function of dimensionless time $t_m$ and radius $r_m$, starting from a stationary gaussian wave packet with fractional overdensity $\delta_0 = 0.01$ and radius $R_{m,0} = 0.64$ on top of a homogeneous background with $\theta_0 = 1$ at an initial time $t_{m,0}=80$, cfr.~Eq.~\ref{eq:wavepacketic}. The evolution is governed by the differential equation of Eq.~\ref{eq:wavepacketpde}. 
The top panel shows the energy density $\rho(t_m,r_m)$ in units of $m^2 f^2$, the middle panel the density difference $\Delta \rho \equiv |\rho - \rho_\infty|$, and the bottom panel the total enclosed energy $E_\mathrm{enc}(t_m,r_m) = 4\pi \int_0^{r_m}\, r_m^2 t_m^{3/2} \rho(t_m,r_m)$ in units of $f^2/m$. The dashed line shows the scale of the physical reduced Compton wavelength $m^{-1}$.
The initially linear overdensity collapses into an oscillon by $t_m \approx 700$ and emits semi-relativistic scalar waves. 
}\label{fig:quarticcollapse}
\end{figure}

For certain values of the four parameters $\theta_0$, $t_{m,0}$, $\delta_0$, and $R_{m,0}$, the wavepacket separates from the Hubble flow and collapses into an oscillon-like object with $\rho / m^2 f^2 > 1$. In Fig.~\ref{fig:quarticcollapse}, we show the evolution of one such collapsing configuration. The initially small fractional overdensity $\delta_0 = 0.01$ deforms over the course of several e-folds, decouples from the Hubble flow expansion, and finally collapses into an oscillon-like structure by $t_m \approx 700$. The oscillon is shrinking in comoving size but is decaying more slowly in physical size $R_p = t_m^{1/2} R_m / m$. It is clearly a dynamical object, with periodic bursts of semi-relativistic scalar radiation that decrease in intensity as the central object loses energy. The semi-relativistic radiation bursts can be seen as the streaks that fan out as $r_m \propto (t_m - t_{m,\mathrm{burst}})^{1/2}$ initially but then slow down due to the expansion of the Universe. Note that the density at large comoving radius is redshifting like dark matter: $\rho_\infty \propto t_m^{-3/2}$. In Sec.~\ref{sec:GW} and App.~\ref{sec:numerics}, we study the precise characteristics of the collapse process and the outgoing radiation---both in scalar and gravitational waves---at higher resolution and without spherical symmetry but in a static (not expanding) geometry.

\begin{figure}[tbp]
\includegraphics[width=0.48\textwidth]{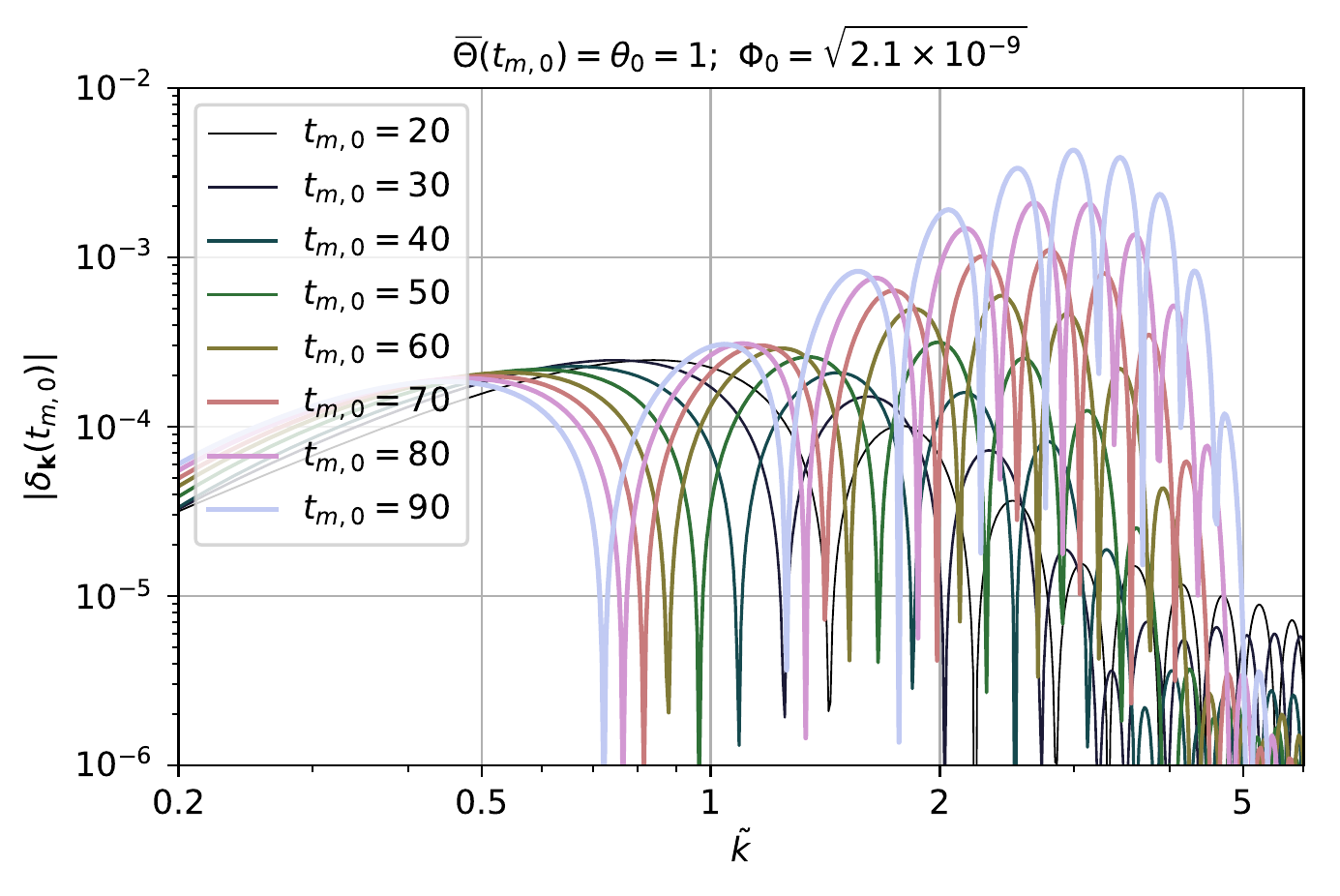}
\includegraphics[width=0.48\textwidth]{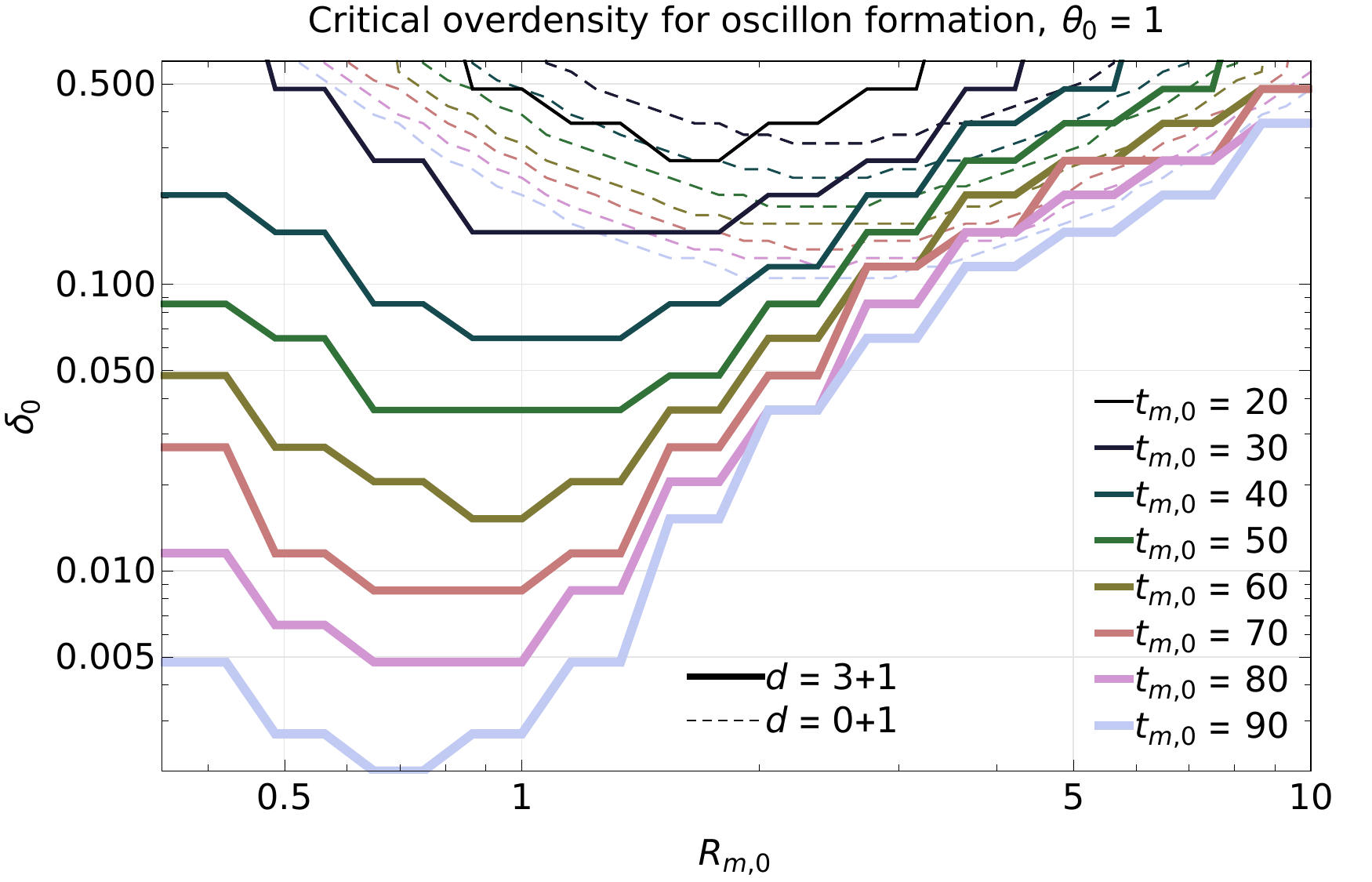}
\caption{\emph{Top panel:} Linear momentum-space analysis of axion density fluctuations $\delta_{\vect{k}}$ as a function of $\tilde{k}$ sourced by adiabatic curvature perturbations with small amplitude $\Phi_{0,\vect{k}} = \sqrt{2.1 \times 10^{-9}}$. The evolution is tracked for seven different values of misalignment angles $|\Theta_0|$ (see text) until the zero mode has amplitude $\bar{\Theta} = 1$ at seven corresponding times $t_{m,0}$. \emph{Bottom panel:} Minimum overdensity $\delta_0$ for a spherically symmetric gaussian wavepacket of radius $R_{m,0}$ (Fourier dual to $\tilde{k}$) to collapse into an oscillon, starting at the same seven start times $t_{m,0}$ at which the zero mode $\theta_0$ equals unity. Dashed lines show results based on a $(0+1)$-dimensional reduction assuming wavepacket rigidity and mass conservation, principles which break down badly for small $R_{m,0}$ due to parametric resonance and other relativistic effects.
}\label{fig:quarticcollapsecond}
\end{figure}

In the bottom panel of Fig.~\ref{fig:quarticcollapsecond}, we delineate the minimum $\delta_0$ needed to collapse into an oscillon as a function of $R_{m,0}$. We started a suite of real-space simulations all at $\theta_0 = 1$ and several benchmark starting times $t_{m,0} = \lbrace 20, 30, 40, 50, 60, 70, 80, 90 \rbrace$, which correspond to misalignment angles $\pi - |\Theta_0| = \lbrace 5.1 \times 10^{-3}, 2.3 \times 10^{-4}, 9.9 \times 10^{-6}, 3.3 \times 10^{-7}, 1.1 \times 10^{-8}, 5.1 \times 10^{-10}, 2.6 \times 10^{-11}, 9.9 \times 10^{-13} \rbrace$, respectively. In those parameter scans, ``oscillon collapse'' was operationally defined as $\rho(r_m = 0) > m^2 f^2$ before $t_m = 10^3$, i.e.~the central density exceeding double its starting value of $(1+\delta_0) m^2 f^2/2$ despite initially decreasing until the configuration becomes nonlinear. 
In the top panel of Fig.~\ref{fig:quarticcollapsecond}, we show the results of a linear Fourier analysis, using the methods of Sec.~\ref{sec:linGR} to evolve axion density perturbations $\delta_{\vect{k}}$ from $t_m = 0$ to $t_{m,0}$ for different $\tilde{k}$, the Fourier dual of $R_{m,0}$. We took the axion fluctuations to be sourced by adiabatic curvature perturbations of standard size: $\Phi_{0,\vect{k}}^2 = 2.1 \times 10^{-9}$.
The linear evolution was performed for the same parameters as in the bottom panel, i.e.~with initial misalignment angles such that the amplitude of the zero mode, $\bar{\Theta}$, equals unity at $t_{m,0}$.
With a misalignment of $\pi - |\Theta_0| \lesssim 2.6 \times 10^{-11}$, $\bar{\Theta} = 1$ is reached at $t_{m,0} \gtrsim 80$, when  one-sigma axion overdensities between $1 \lesssim \tilde{k} \lesssim 5$ will reach values $\delta_{\vect{k}} \gtrsim 0.002$ and are rapidly growing. Comparison against the real-space results of the bottom panel reveals that these perturbations are destined to collapse. For these supercritical parameters, the collapse time $t_{m,\mathrm{coll}}$ is shortly after the fluctuation becomes nonlinear with only a weak dependence on $\delta_0$, $R_{m,0}$, and $\pi-|\Theta_0|$. It is always several e-folds after the zero mode starts oscillating, yielding the hard lower bound of $t_{m,\mathrm{coll}} \gg 10^2$.

We can attempt to capture these quartic collapse dynamics in the radiation-dominated era by following a variational procedure similar to that of Ref.~\cite{Chavanis:2011zi, Chavanis:2011zm}. We derive an effective equation of motion for the physical size $R_p= t_m^{1/2} R_m/m$ of the overdensity, and deduce under which conditions $R_p \to 0$ in a finite amount of time. This procedure is analogous to the standard calculation for gravitational collapse of a spherical-tophat-shaped overdensity~\cite{Press:1973iz}, which also reduces the problem from one in $d=3+1$ dimensions to one in $d=0+1$ dimension.

In order to derive the equation of motion for $R_p$, we expand the energy density of the axion field to fourth order in $\theta$:
\begin{align}
\rho \simeq m^2 f^2 \left\lbrace \frac{(\partial_{t_m} \theta)^2 + \theta^2}{2} + \frac{(\partial_{r_m} \theta)^2}{2t_m} - \frac{\theta^4}{24}\right\rbrace \label{eq:rhoquartic}
\end{align} 
This expression can formally be expanded as a Taylor series in $\delta$: $\rho = \rho_0 + \rho_\delta + \rho_{\delta \delta} + \dots$. At every order in $\delta$, we can break down each term into a ``mass'' and ``interaction'' piece, $\rho = \rho^{M} + \rho^\mathrm{int}$, corresponding to the first and last two terms of Eq.~\ref{eq:rhoquartic}, respectively. The mass of the initial state wavepacket (cfr. Eq.~\ref{eq:wavepacketic}) is then: 
\begin{align}
M = \int \dd^3V \, \rho^{M}_\delta  \simeq \frac{f^2}{m} \frac{\delta}{2} \theta_0^2 t_{m,0}^{3/2}  R_m^3. \label{eq:wavepacketM}
\end{align}
The combination $\theta^2 t_m^{3/2}$ is approximately a constant to zeroth order in $\delta$, and in the absence of any dynamics, $\delta$ and $R_m$ are constant as a function time as well, such that the physical radius of the wavepacket $R_p =  t_m^{1/2} R_m/m$ is expanding with the Hubble flow. However, the wavepacket does have nontrivial dynamics due to its interaction energy, which can be estimated as:
\begin{align}
E_\mathrm{int} = \int \dd^3V \, \rho^\mathrm{int}_{\delta \delta} \simeq M  \frac{\delta}{2^{9/2}} \left[ \frac{3}{t_m R_m^2} - \theta_0^2 \right].
\label{eq:ChavanisE}
\end{align}
In the subhorizon, nonrelativistic limit, and assuming wavepacket rigidity\footnote{A ``rigid'' wavepacket is one whose (in this case Gaussian) shape is preserved. Wavepacket rigidity assumes that the variational ansatz that we have used to convert the $d=3+1$ Schr\"odinger equation to a $d=0+1$ equation for the wavepacket size $R_p$ is a good solution to the original equations of motion for a stationary state. The middle panel of Fig.~\ref{fig:quarticcollapse} clearly shows wavepacket deformation before collapse.} and mass conservation, the physical radius of the wavepacket should then obey a Newtonian ODE:
\begin{align}
\ddot{R}_p &= - \frac{\dd}{\dd R_p} \left[ \Phi_\mathrm{FRW} + \frac{E^\mathrm{int}}{M} \right]\\
&\simeq - \frac{R_p}{4t^2} + \frac{\delta}{2^{9/2}} \left[\frac{6}{m^2 R_p^3} - \frac{3 \theta_0^2}{R_p}\right].\label{eq:Rpode}
\end{align}
The first term is the leading correction that takes into account the deceleration of the Universe's expansion~\cite{baumann2012cosmological}, with $\Phi_\mathrm{FRW} = -(\dot{H} + H^2) R_p^2 / 2 = R_p^2 / 8 t^2$ during radiation domination. The second term is the leading self-interaction force. The initial conditions corresponding to those of Eq.~\ref{eq:wavepacketic} are:
\begin{align}
R_p(t_0) = t_{m,0}^{1/2} r_m; \quad  \dot{R}_p(t_0) = \frac{R_p(t_0)}{2 t_0} + \frac{\pi}{m R_p(t_0)}, \label{eq:Rpic}
\end{align}
where in the latter equation, the first term is due to the Hubble flow velocity $H R_p$ and the second term takes into account the ``spreading'' of the wavepacket. Again, we define a collapsing wavepacket as one for which $R_p \to 0$ in finite time.

In Fig.~\ref{fig:quarticcollapsecond}, we depict the critical parameters for collapse using the $R_p$ equation with dashed lines. One can observe that the dichotomy between collapsing and comoving configurations of Eqs.~\ref{eq:wavepacketic} and \ref{eq:wavepacketpde} is captured by the simplified dynamics of Eqs.~\ref{eq:Rpode} and \ref{eq:Rpic} only at large wavepacket sizes $R_{m,0}\gtrsim 3$, and then only approximately. For smaller wavepacket sizes, the $0+1$-dimensional reduction breaks down spectacularly. As evident from Fig.~\ref{fig:quarticcollapse}, the assumption of wavepacket rigidity (constant shape) is badly violated even in the linear regime. Likewise, the assumption of mass conservation is also not a good principle at small $R_m$, as parametric resonance (see Sec.~\ref{sec:linGR}) can be understood as a process wherein two axions with zero momentum (the background) are converted into two axions with finite momentum (part of the perturbation).
 
Our numerical simulations further show (see App.~\ref{sec:numerics} for details) that the collapsing structures eventually settle into evaporating oscillons, scalar field configurations whose dynamics are dominated entirely by the dynamics of the axion potential, with little influence from gravity. This relaxation happens mainly through scalar wave emission, some of which can be seen in Fig.~\ref{fig:quarticcollapse}. Oscillons have been known to exist generically for potentials containing attractive self-interactions, and they can be relatively long-lived for some axion potentials, although there is no simple quantitative or qualitative understanding for their longevity. Our high-resolution simulations show that the oscillon lifetime in physical units is $\lesssim \mathcal{O}(10^3)~m^{-1}$ for the cosine potential, not long enough to be cosmologically relevant.\footnote{As we will discuss in Sec.~\ref{sec:flat}, the oscillon lifetime can be significantly longer than $\mathcal{O}(10^3)~m^{-1}$ for potentials other than a cosine and/or for very large oscillons whose evaporation rate is suppressed by a form factor. This raises the possibility of DM being comprised of oscillons; some of the potential signatures of oscillon DM are discussed in Sec.~\ref{sec:signatures}.} 
Since the actual structures collapsing via these self-interactions are $\mathcal{O}(1)$ asymmetric, they can also emit gravitational waves during their infall and collapse, which we discuss in Sec.~\ref{sec:GW}.

The violent dynamics of the oscillons' implosion and evaporation leaves behind regions of axion debris with $\mathcal{O}(1)$ density fluctuations. This is quite analogous to the case of dissipating oscillons which form or become part  of QCD axion miniclusters, if the Peccei-Quinn phase transition occurs after inflation (see e.g. Ref.~\cite{Buschmann:2019icd}). We expect that these regions are slightly larger in comoving scale than the original density perturbations, and that they will \emph{gravitationally} collapse into ultra-dense halos and solitons at around matter-radiation equality, cfr.~Sec.~\ref{sec:gravcollapse}. We still expect $\mathcal{O}(1)$ fraction of DM to be in these structures; the debris of the oscillons' decay will be the bulk of the dense DM matter substructure, and their signatures will be discussed in~Sec.~\ref{sec:signatures}.

\subsubsection{Tidal stripping} \label{sec:tidalstripping}
The halos that result from the parametric-resonance-fueled growth of axion overdensities are the densest objects in the Universe upon their initial formation. They are therefore robust against tidal stripping effects even as they are assembled into larger DM halos such as those of galaxies and clusters. However, present-day baryonic structures such as stars, globular clusters, and the Milky Way (MW) disk are of course much denser than typical ambient DM densities. Most of the observational and experimental signatures of Secs.~\ref{sec:GRinteractions} and \ref{sec:directdetection} rely on the survival of the halos in our Galaxy, so one needs to address the possibility that they are tidally disrupted by the MW disk or its stellar constituents. We divide our discussion into two distinct cases, depending on whether the halo scale radius $r_s$ is either much smaller ($r_s \ll \Delta r_\mathrm{star}$) or much larger ($r_s \gg \Delta r_\mathrm{star}$) than the average interstellar separation in the MW disk: $\Delta r_\mathrm{star} \sim \mathrm{pc}$. For the intermediate regime $r_s \sim \Delta r_\mathrm{star}$, there is no separation of scales, but it should be approximately correct to interpolate between the constraints of the two limiting regimes.

First, we discuss the case of halo scale radii much smaller than the interstellar separation, the case of interest in particular for the femtohalos of Sec.~\ref{sec:directdetection}. In this regime, stellar encounters are brief compared to the (internal) dynamical time of the halo, so the relevant quantity is the differential velocity kick imparted on axions on opposite sides of the halo in the impulse approximation: 
\begin{align}
\Delta v(b) &\simeq \frac{4G_N M_{\mathrm{star}} r_s}{b^2 v_\mathrm{rel}} \label{eq:vb}\\
&\approx 8\times 10^{-15}\left( \frac{b_\mathrm{typ}}{b}\right)^2 \left( \frac{M_{s}}{10^{-18} M_{\odot}}  \frac{10^3}{\mathcal{B}_\odot} \right)^{1/3} \nonumber 
\end{align}
In the above estimate, we assumed a relative velocity of $v_\mathrm{rel} \approx 10^{-3}$ and a solar-mass perturber $M_\mathrm{star} \approx \mathrm{M_\odot}$. We also defined a typical impact parameter $b$ as $b_\mathrm{typ} = (M_\mathrm{star}/\pi \Sigma_\odot)^{1/2} \approx 0.07\,\mathrm{pc}$, with the surface mass density of the MW disk at the Sun's position equaling $\Sigma_\odot \approx 60\,\mathrm{M_\odot}\,\mathrm{pc}^{-2}$. The local density boost factor is $\mathcal{B}_\odot \equiv \rho_s / \rho_\mathrm{DM}^\odot$. By contrast, the scale velocity of a halo is $v_s = \sqrt{16 \pi \ln(2) G_N \rho_s r_s^2 }$, or numerically:
\begin{align}
v_s \approx 5 \times 10^{-13} \left( \frac{M_{s}}{10^{-18} M_{\odot}}\right)^{1/3} \left( \frac{\mathcal{B}_\odot}{10^3} \right)^{1/6}. \label{eq:vs}
\end{align}
Comparison of Eqs.~\ref{eq:vb} and \ref{eq:vs} shows that a single disk crossing has little effect on the interior structure of a moderately overdense halo. 

Of course, the halo may experience $N$ disk crossings over the course of its lifetime, with a minimum expected impact parameter of $b_\mathrm{min} = b_\mathrm{typ}/\sqrt{N}$. The requirement that $\Delta v(b_\mathrm{min}) < v_s$ is equivalent to a mass-independent lower bound on the scale density, or equivalently the boost factor:
\begin{align} \label{eq:minBoost}
\mathcal{B}_\odot \gtrsim \frac{\pi G_N}{\ln 2} \frac{\Sigma_\odot^2 \rho_\mathrm{DM}^\odot}{v_\mathrm{rel}^2} N^2 \approx 740 \left( \frac{N}{100} \right)^2. 
\end{align}
We regard Eq.~\ref{eq:minBoost} as a conservative lower bound on the minimum overdensity necessary to prevent a catastrophic tidal disruption event for a halo that crosses the disk $N$ times. Typical halos will have $N$ at most $\sim 150$, while those on eccentric orbits or recently accreted onto the MW could have substantially lower values of $N$.
Instead, one could consider the process wherein the internal binding energy per unit mass ($-v_s^2/4$) 
of the halo is gradually reduced by dynamical heating of $N$ tidal encounters,  each interaction dumping kinetic energy per unit mass of $v_s \Delta v(b)$, under the assumption of mass conservation. One then arrives at a bound similar to that of Eq.~\ref{eq:minBoost}, except stronger by a factor of $(4 \ln N)^2$ on the RHS. However, tidal interactions do cause partial mass loss---preferentially of particles on more weakly-bound orbits, leaving behind more deeply bound particles and a denser halo. Ref.~\cite{van2017disruption} indicates that even Eq.~\ref{eq:minBoost} may be overly restrictive: a tidal shock energy far exceeding the halo's original binding energy can result in a surviving halo fragment. We therefore expect halos with $r_s \ll \Delta r_\mathrm{star}$ to survive tidal interactions inside the Milky Way if they are only moderately overdense.

In the case of larger subhalos with $r_s \gg \Delta r_\mathrm{star}$, tidal survival constraints are relaxed because the subhalos are effectively probing a lower-density medium; the tidal forces from individual stars are only strong on scales much smaller than the subhalo itself, and cannot cause its entire disruption. In the commonly-adopted simplified model of Ref.~\cite{king1962structure}, one posits that all mass of subhalo outside the tidal radius $r_t$ is tidally stripped by a spherically symmetric perturber with enclosed mass function $M_p(R)$. If the subhalo is on a circular orbit at radius $R$ from the center of the host halo, the tidal radius is implicitly given by:
\begin{align}
\frac{M(r_t)}{r_t^3} = \left(3 - \left. \frac{\dd \ln M_p(R)}{\dd \ln R} \right|_R \right) \frac{M_p(R)}{R^3}. \label{eq:rt}
\end{align}
Above, $M(r)$ is taken to be the enclosed mass function of the subhalo. If we require that $r_t > r_s$ on a circular orbit at the Sun's radius $R \approx 8.3\,\mathrm{kpc}$ from the MW with scale radius $r_s^\mathrm{MW} \approx 18\,\mathrm{kpc}$ and scale density $\rho_s^\mathrm{MW} \approx 2.6 \times 10^{-3}\,\mathrm{M_\odot}\,\mathrm{pc}^{-3}$~\cite{mcmillan2011mass}, we arrive at the weak constraint $\mathcal{B}_\odot \gtrsim 1.2$. 
Tidal fields from density variations in the Galactic disk on scales of order the subhalo size can be significantly larger, as one can generally expect $\mathcal{O}(1)$ overdensities in the disk with mean local density $\rho_{d,\odot} \approx 0.087\,\mathrm{M_\odot}\,\mathrm{pc}^{-3}$~\cite{mcmillan2011mass}. Still applying Eq.~\ref{eq:rt} and conservatively taking the RHS to be $4\pi \rho_{d,\odot}$, we find that $r_t > r_s$ requires that $\mathcal{B}_\odot \gtrsim 11$.
Most of the mass is located outside the scale radius of an NFW-shaped halo, so if these inequalities are only barely satisfied, one can expect survival but with substantial mass loss from tidal stripping.

\section{Observational prospects} \label{sec:signatures}
In Sec.~\ref{sec:evolution}, we described how the attractive self-interactions of axion DM at large initial misalignment give rise to compact halos much denser than the $\Lambda$CDM expectation at similar scales. In Secs.~\ref{sec:QCD} and \ref{sec:flat}, we will repeat this analysis for the QCD axion and for generalized axion potentials, respectively, with similarly boosted DM power spectra and thus denser halos. When formed, these halos constitute $\mathcal{O}$(1) fraction of the DM, and their spatial distribution will trace the ambient DM density.

In this section, we describe how we expect DM phenomenology to change in our scenario. We divide the observable signatures of compact axion halos into four categories.  In Sec.~\ref{sec:GRinteractions}, we consider direct gravitational interactions between these halos and astrophysical objects such as stars.  These include perturbations in stellar phase space distributions, various gravitational lensing signatures, and potentially-observable dynamical friction effects. The rough region of affected parameter space is shaded in blue in Fig.~\ref{fig:summary}, and the reader interested in the key results of this section should focus first on Fig.~\ref{fig:GRinteractions}.

We then move in Sec.~\ref{sec:directdetection} to a discussion of how such compact halos affect DM direct detection experiments that search for nonminimal axion couplings to the SM.  This is relevant for high axion masses (shown by the green region in Fig.~\ref{fig:summary}), and the key results are summarized in the final two paragraphs of Sec.~\ref{sec:directdetection} as well as Fig.~\ref{fig:clumpreachplot}.  In particular, we point out the importance of these effects for the QCD axion (see also Sec.~\ref{sec:QCD}).

We next consider indirect gravitational effects on baryonic structures and early star formation in Sec.~\ref{sec:baryons}.  These are relevant only for the lightest axions (with masses less than $\mathcal{O}(10^{-18})\,\mathrm{eV}$), a region shaded in brown in Fig.~\ref{fig:summary}, and we report the key findings on star formation in Fig.~\ref{fig:starformation}.  In the final paragraph of this section we also discuss effects observable in Lyman-$\alpha$ forests, and why current constraints on ultralight dark matter do not apply and must be reanalyzed in our case.

Finally, in Sec.~\ref{sec:GW}, we study the extreme case when collapse happens well before matter-radiation equality and oscillons are formed.  The collapsing structures will emit gravitational waves and form a stochastic GW background, and for light axions (masses less than $\mathcal{O}(10^{-14})\,\mathrm{eV}$), this background may be detectable in the future.  We shade the affected region of parameter space in orange in Fig.~\ref{fig:summary}, and Fig.~\ref{fig:GWest} contains our estimates of power in the stochastic background as well as the potential reach of upcoming experiments. 

\subsection{Direct gravitational interactions} \label{sec:GRinteractions}

\begin{figure*}
\includegraphics[width = 0.95\textwidth]{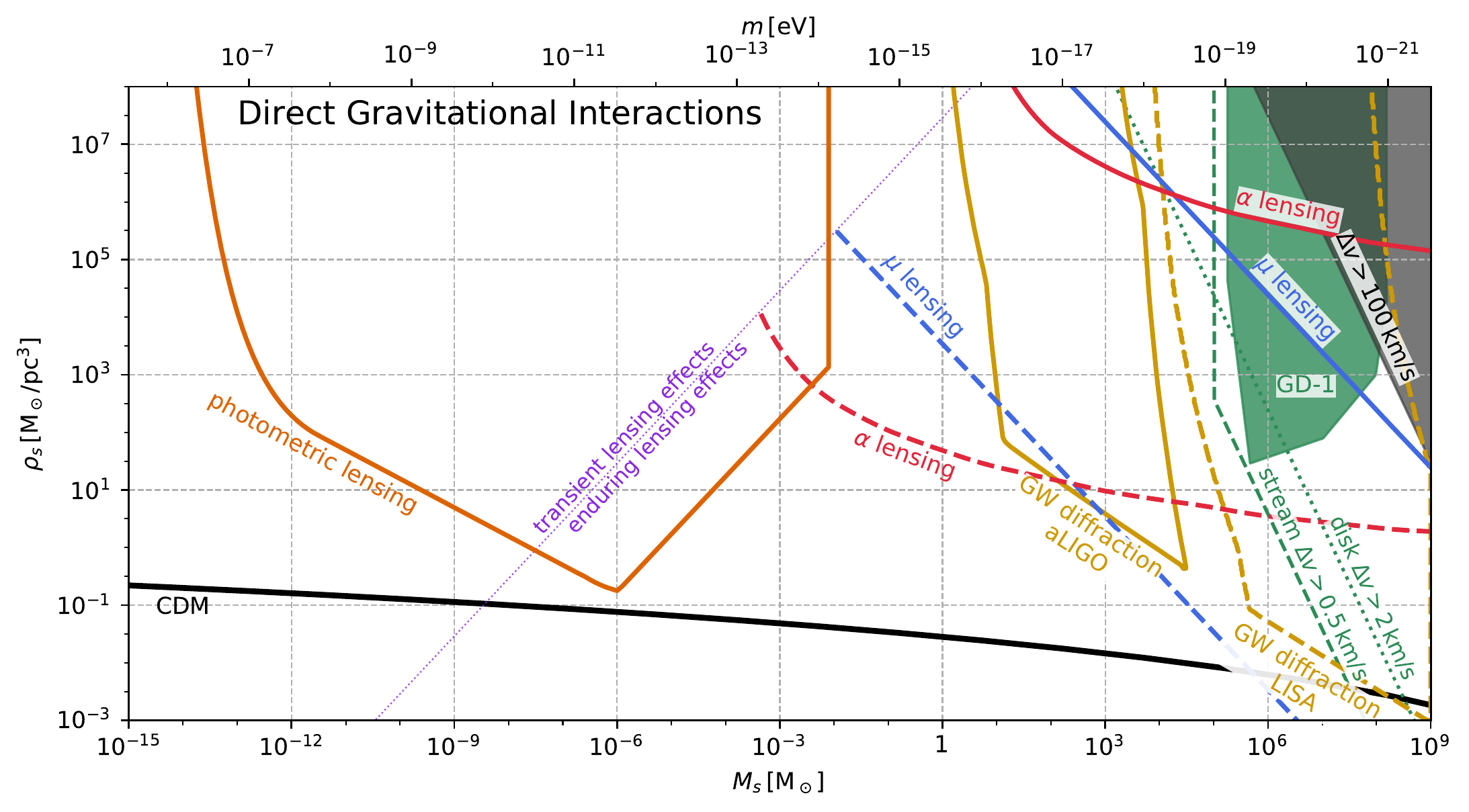}
\caption{Astrophysical probes of direct gravitational effects from compact halos, parametrized in terms of their sensitivity to halo scale mass $M_s$ and scale density $\rho_s$. Above the dashed (dotted) green line, compact subhalos would produce observable velocity kicks in stellar streams (the Galactic disk). The green region outlines the best-fit parameters of one such tentative impact on the GD-1 stream. In the dark gray region, these kicks can be strong enough to eject stars from the Galactic disk or even halo. Above the solid (dashed) blue line, astrometric lensing by compact halo induces localized distortions in the proper motion $\mu$ of background sources that are observable by \emph{Gaia} (SKA). Likewise, correlations in stellar proper accelerations $\alpha$ induced by astrometric weak lensing are detectable by \emph{Gaia} (\emph{Theia}) above the solid (dashed) red line. On the purple line, the halo scale radius equals the typical distance traveled over a 9-year observation time, demarcating the transition between transient and enduring lensing effects for unmagnified sources. Inside the gold-colored solid (dashed) line, an observable fraction of GW events at aLIGO (LISA) will be diffracted. Photometric irregularities in the microlensing light curve of highly magnified, caustic-transiting stars may be observable above the orange line.}
\label{fig:GRinteractions}
\end{figure*}

The compact halos formed through the large-misalignment mechanism can be large enough to gravitationally bend or magnify the light emitted by astrophysical objects as they move in front of them, or to gravitationally affect the motion of nearby stars as they move through the Galactic halo. Here we analyze these effects in detail, and Fig.~\ref{fig:GRinteractions} summarizes the parameter space that each effect probes as a function of the halo scale mass $M_s$ and the halo scale density $\rho_s$. Purely from the minimal coupling to gravity, there are discovery prospects for halos seeded by large-misalignment axions with masses as high as $m \sim 10^{-5}\,\mathrm{eV}$. We note that most of the effects in Fig.~\ref{fig:GRinteractions} do not rely on subhalos that transit the MW disk or can only probe relatively dense subhalos, and are thus robust to tidal stripping.

We begin in Sec.~\ref{sec:localgreffects} by discussing how compact subhalos perturb local stars. In Sec.~\ref{sec:astrometry}, we show that the most powerful probe in a large part of the parameter space is astrometric weak lensing. DM subhalos' lensing of stellar light can appear as a distortion of the apparent motion of stars. We consider two types of observables, one based on the apparent velocity of background luminous sources such as distant stars or quasars (blue curves in Fig.~\ref{fig:GRinteractions}), the other based on apparent stellar accelerations (red curves in Fig.~\ref{fig:GRinteractions}).

In Secs.~\ref{sec:microlensing}, \ref{sec:stronglensing}, and~\ref{sec:LIGOlensing}, we discuss signatures of DM subhalos that rely mainly on strong gravitational lensing, where lensing produces significant magnification and multiple images of the lensed object. We find that DM subhalos within our galaxy are generically too diffuse to satisfy the strong lensing criterion, but that for some rare extragalactic stars, located behind critical-lensing caustics of galactic clusters, can lead to observable signatures in a very wide range of parameter space (Sec.~\ref{sec:stronglensing}).   
For extragalactic halos that almost but not quite satisfy the strong lensing criterion, we describe possibly detectable anomalous dispersion in LIGO events, although more analysis is required to firmly establish the reach of such techniques (Sec.~\ref{sec:LIGOlensing}).

At the end of Secs.~\ref{sec:astrometry} and~\ref{sec:microlensing}, we also contemplate the possibility that \emph{oscillons} survive to the present day and constitute a significant component of DM.  In this case, we outline their corresponding lensing signatures and constraints.  This scenario does not apply to the cosine potential we have considered thus far because it does not support cosmologically long-lived oscillons, but could be relevant for the generalized axion potentials we will consider in Sec.~\ref{sec:flat}.  As we discuss there, oscillon configurations in other axion potentials can be significantly longer lived, although we do not yet know whether these or other potentials support oscillons that survive to the present day.

Finally, in Sec.~\ref{sec:dynconstraints}, we discuss dynamical friction effects coming from massive DM subhalos, but deem current observations not sufficiently robust to constrain our scenario. 

\subsubsection{Local gravitational perturbations}\label{sec:localgreffects}
As DM subhalos traverse the Galaxy, they will gravitationally attract nearby stars and perturb their 6D phase space distribution. A star that passes near a compact subhalo with impact parameter $\vect{b}$, which we assume to be spherical for simplicity, will receive a velocity kick of:
\begin{align}
\Delta \vect{v} &= - \hat{\vect{b}} \frac{2 G_N M(b)}{b V}  \label{eq:velocitykick} \\
&\approx - \hat{\vect{b}} ~ 0.5 \,\mathrm{km\,s^{-1}} \left[\frac{M(b)}{10^7\,\mathrm{M_\odot}}\right]\left[\frac{166\,\mathrm{km\,s^{-1}} }{V} \right]\left[\frac{\mathrm{kpc}}{b}\right], \nonumber
\end{align}
where $V$ is the relative velocity between the subhalo and the star and $M(b)$ is the subhalo mass enclosed within the impact distance $b$. As the subhalo moves through the interstellar medium, it causes a correlated distortion in the real-space distribution depending on the time elapsed since impact. It has been proposed to search for these perturbations in the 6D phase space distribution of stars in the MW's disk~\cite{feldmann2014detecting} and stellar halo~\cite{buschmann2018stellar}, with purported sensitivities to subhalos with masses down to perhaps as low as $10^7\,\mathrm{M_\odot}$ in the CDM paradigm. The effect of Eq.~\ref{eq:velocitykick} is too small to be seen on any one star for all but the most massive and/or densest subhalos, since the velocity dispersions in the Galactic disk and stellar halo are $\sim 25\,\mathrm{km\, s^{-1}}$ and $\sim 166\,\mathrm{km\, s^{-1}}$, respectively. Since the effect of Eq.~\ref{eq:velocitykick} is coherent for all stars along the subhalo's trajectory, one can in principle average down this intrinsic dispersion noise, as well as any additional instrumental uncertainties. However, to what extent this averaging procedure can beat down the noise remains an open question, as it depends on the degree of pre-existing departures from kinetic equilibrium, which have recently been found in both the disk~\cite{antoja2018dynamically} and stellar halo~\cite{myeong2018sausage,helmi2018merger,fiorentino2014weak,deason2018apocenter,belokurov2018unmixing,necib2018dark}. In Fig.~\ref{fig:GRinteractions}, we mark by the green dashed line as potentially detectable those subhalos for which the velocity kick produced by a compact subhalo's passage is larger than $2\,\mathrm{km\,s^{-1}}$.

Promising alternative targets for local gravitational perturbations caused by DM substructure are stellar streams~\cite{ibata2002uncovering,johnston2002lumpy,siegal2008signatures,bovy2016detecting,carlberg2016modeling,erkal2016number}, the tidal debris tails originating from disrupted globular clusters or dwarf galaxies. They can be thought of as low-noise ``antennae" of gravitational effects, as they are inherently dynamically cold, out-of-equilibrium systems. This is because their velocity dispersion is bounded from above by the dispersion of the progenitor, and their morphology delineates their orbit, i.e.~the velocity vectors of their stellar constituents are approximately tangential to the stream. For example, the GD-1 stream has a dispersion of about $2.3\,\mathrm{km\,s^{-1}}$, a length of $\gtrsim 100\,\mathrm{kpc}$, and a width of about $30\, \mathrm{pc}$~\cite{malhan2018constraining}. Close encounters with a dense subhalo would kick stars out of the stream, creating a local underdensity near the point of impact (a ``gap'') and a potentially observable secondary stream (a ``spur'') emanating from the gap~\cite{carlberg2012dark,yoon2011clumpy}. Interestingly, such features have recently been discovered in the GD-1 stream~\cite{bonaca2018spur}. Tantalizingly for our purposes, if these features are due to a subhalo puncturing the stream, they appear to have been caused by one that is \emph{denser} than predicted in the standard CDM framework. Unfortunately, it is challenging to unambiguously attribute the disruption features to a dark subhalo, as they become apparent only after about a MW orbital time, so it is hard to exclude close encounters with known or unknown globular clusters. In Fig.~\ref{fig:GRinteractions}, we recast the posterior best-fit parameters from the potential DM subhalo impact of Ref.~\cite{bonaca2018spur} in green. We also outline the parameter space for which one can generate velocity kicks large enough to disrupt a very cold stream: $\Delta v \gtrsim 0.5\,\mathrm{km\,s^{-1}}$ for $\mathrm{max}\lbrace b, r_s \rbrace \gtrsim 10\,\mathrm{pc}$.

\subsubsection{Astrometric weak gravitational lensing}\label{sec:astrometry}
Compact subhalos in the Milky Way can also induce apparent motions of stars and other luminous sources through gravitational lensing whenever they are near the line of sight to the background light source, without producing multiple images or magnification. Astrometric weak lensing was first considered for point-like objects in Refs.~\cite{boden1998astrometric,dominik2000astrometric,belokurov2002astrometric} and for cuspy minihalos in Refs.~\cite{erickcek2011astrometric,li2012new}. A program of searches with optimal observables for both compact object and extended subhalos was outlined in Ref.~\cite{van2018halometry}, in light of ongoing (\textit{Gaia}~\cite{brown2018gaia}, HSTPromo~\cite{bellini2014hubble}) and future astrometric surveys (WFirst~\cite{spergel2015wide,sanderson2017astrometry,gould2014wfirst}, \textit{Theia}~\cite{boehm2017theia}, SKA~\cite{fomalont2004microarcsecond}, TMT~\cite{schock2014thirty}, etc.) with much improved precision and/or catalogue size. 

Time-domain astrometric lensing signatures can be usefully divided into two categories: \emph{transient} and \emph{enduring} effects, depending on whether the minimum impact parameter $b$ to the line-of sight is smaller or larger, respectively, than the change in impact parameter over a typical multi-year astrometric survey $v \tau \sim \mathcal{O}(10^{-3}\,\mathrm{pc})$. Unless the subhalo is extremely cuspy (e.g.~$\rho(r) \propto r^{\gamma}$ with $\gamma < -2$), the lensing deflection angle is maximized for impact parameters near the scale radius, i.e.~$b \sim r_s$. A subhalo can thus produce a gravitational lensing transient only if
\begin{align}
\rho_s \gtrsim \frac{3}{4\pi} \frac{M_s}{(v \tau)^3} \approx 2 \times 10^{10}\,\rho_\mathrm{DM}^\odot \left[\frac{M_s}{\mathrm{M_\odot}} \right]\left[\frac{10^{-3}\,\mathrm{pc}}{v \tau} \right]^3.
\end{align}
An object that forms via gravitational collapse of a linear density perturbation cannot have a density that parametrically exceeds the density at matter-radiation equality (see Sec.~\ref{sec:gravcollapse}):
\begin{align}
\rho_s\big|_\mathrm{gr-coll} \lesssim 200 \rho_\mathrm{eq} \approx 2 \times 10^7 \rho_\mathrm{DM}^\odot. \label{eq:rhogrmax}
\end{align}
Therefore, only very light ($M_s \lesssim 10^{-3}\,\mathrm{M_\odot}$) gravitationally collapsed subhalos can produce transients, but at densities bounded by Eq.~\ref{eq:rhogrmax}, they yield too small an angular deflection ($4G_N M_s/ b \approx 0.04\,\mathrm{\mu as}$ for $M_s = 10^{-3}\,\mathrm{M_\odot}$ and $b = 10^{-3}\,\mathrm{pc}$) to be detectable by current state-of-the-art astrometric observatories, which reach at best $\mathrm{\mu as}$-level precision for bright sources. We outline the boundary of this transient regime by the purple line in Fig.~\ref{fig:GRinteractions}. In this transient regime, pulsar timing arrays may shed light on compact substructures via the Shapiro time delays and Doppler effects that they induce~\cite{dror2019pulsar}. 
 
 Enduring gravitational lensing effects arise for impact parameters (and subhalo radii) larger than about $10^{-3}\,\mathrm{pc}$. The instantaneous angular deflection is in practice unobservable because the true celestial positions of luminous sources is not known, and the lensing-induced number density changes are much smaller than intrinsic and shot-noise density fluctuations over angular scales that a subhalo subtends over the sky. However \emph{time derivatives} of the angular deflections, specifically lensing-induced proper motions and accelerations, are observable in practice. Ref.~\cite{van2018halometry} proposed to look for local (using templates) and global (using correlations) evidence of these distortions. In Fig.~\ref{fig:GRinteractions}, we show their projections for the reach of local proper motion templates (blue) with \emph{Gaia} (solid) and SKA (dashed), and of global acceleration correlations (red) with \emph{Gaia} (solid) and \emph{Theia} (dashed), assuming a $\Omega_\mathrm{sub}/\Omega_\mathrm{DM} \approx 0.30$ DM fraction in subhalos of mass $M_s$ and density $\rho_s$. For other DM fractions and at fixed $M_s$ and signal-to-noise ratio, one can employ the approximate scalings $\rho_s \propto \Omega_\mathrm{sub}^{-1}$~\cite{van2018halometry}.
 
\paragraph*{Astrometric weak lensing from oscillons.} Observable astrometric lensing transients can be produced by oscillons, as their internal density is parametrically equal to:
\begin{align}
\rho_s^\mathrm{osc} &= C^\mathrm{osc}_\rho m^2 f^2 \simeq \frac{C^\mathrm{osc}_\rho}{2^{5/2} C_{\pi/2}} \rho_\mathrm{eq} \left[\frac{m}{\Heq}\right]^{3/2}\left[\frac{f}{f_{\pi/2}} \right]^2 \nonumber \\
&\approx 2 \times 10^{23}\, \rho_\mathrm{DM}^\odot \, \frac{C^\mathrm{osc}_\rho}{C_{\pi/2}} \left[\frac{m}{10^{-15}\,\mathrm{eV}}\right]^{3/2}\left[\frac{f}{f_{\pi/2}} \right]^2, \label{eq:rhoosc}
\end{align}
where $C^\mathrm{osc}_\rho$ is a model-dependent constant of order unity. The typical mass of oscillons forming through our mechanism of Sec.~\ref{sec:quarticcollapse} is:
\begin{align}
M_s^\mathrm{osc} &= C^\mathrm{osc}_M \frac{f^2}{m} \simeq \frac{C^\mathrm{osc}_M}{2^{5/2} C_{\pi/2}} \frac{\rho_\mathrm{eq}}{m^{3/2} \Heq^{3/2} } \left[\frac{f}{f_{\pi/2}} \right]^2 \nonumber \\
&\approx 6 \times 10^{-4} \, \mathrm{M_\odot} \, \frac{C^\mathrm{osc}_M}{C_{\pi/2}} \left[\frac{10^{-15}\,\mathrm{eV}}{m}\right]^{3/2}\left[\frac{f}{f_{\pi/2}} \right]^2
\end{align}
with $C^\mathrm{osc}_M$ a model-dependent constant that is $\mathcal{O}(10^3)$ for a cosine potential at $t_{m,0} \sim 90$ but can be larger for other potentials and very small values of $f/f_{\pi/2}$.
The density of Eq.~\ref{eq:rhoosc} is so high that oscillons are effectively point-like when it pertains to their lensing signatures. 
Ref.~\cite{van2018halometry} projected that the ongoing \emph{Gaia} survey has the potential to discover point-like objects making up at least a percent of dark matter down for masses greater than $10^{-4}\,\mathrm{M_\odot}$ by the end of its mission. Time-domain, astrometric, weak lensing is thus a powerful probe of axion models with ``flat'' potentials (such that oscillons are cosmologically long lived), low $f/f_{\pi/2}$ (such that they form at high abundance), and axion masses less than $ 10^{-15}\,\mathrm{eV}$.

\subsubsection{Photometric microlensing}\label{sec:microlensing}

One of the most promising purely-gravitational probes of our scenario is photometric microlensing~\cite{paczynski1986gravitational}.  Historically, this is a program which has set tight constraints on sub-unity DM fraction in compact objects down to $10^{-10}\,\mathrm{M_\odot}$~\cite{alcock2000macho,tisserand2007limits,niikura2017microlensing, griest2014experimental,zumalacarregui2018limits}, but such constraints are limited to extremely dense objects.  Microlensing surveys search for the transient order-unity increase in brightness of a background luminous source caused by the passage of a lens near the Einstein radius $\theta_\mathrm{E} = \sqrt{ 4 G_N M D_{LS}/(D_L D_S) }$
where $D_L$, $D_S$, and $D_{LS}$ are the angular diameter distances to the lens, to the source, and from the lens to the source respectively.
This expression is only valid when the entire mass $M$ is enclosed within $\theta_\mathrm{E}$, but with the exception of potentially long-lived oscillons, the axion minihalos discussed here are not dense enough to strongly lens, and so prior constraints do not apply.  We can, however, employ a technique first discussed in Ref.~\cite{Dai:2019lud}.

The basic idea is to exploit single stars at $z \gtrsim 1$ that are located near gravitational lensing caustics of intervening galaxy clusters and are thus highly magnified (with magnification $\mu \sim 10^2 - 10^3$, see e.g. Refs.~\cite{Kelly:2017fps, Chen:2019ncy, Kaurov:2019alr}).  Very small changes in the mass distribution of the lensing cluster can shift the location of the image closer to or further away from the caustic and result in large changes in measured brightness, so tracking the brightness of such stars over time can provide information about the cluster subhalo distribution.  In particular, Ref.~\cite{Dai:2019lud} suggests using stellar microlensing events (when one of these source stars is additionally magnified due to microlensing by a star in the lensing cluster), and finds that with reasonable observing parameters they should be able to detect variances in the lensing convergence $\kappa$ down to one part in $10^4$ at length scales $\ell \sim 10$--$10^4\,\mathrm{AU}/h$ in the lensing cluster.  Here we repeat an abbreviated analysis for our case using slightly more conservative values: We assume only that one can detect variances in $\kappa$ of $\mathcal{O}(10^{-3})$ at length scales of $\ell \sim 30-10^4\,\mathrm{AU}$.

The lensing convergence $\kappa$ is defined as the ratio of the surface density of the lens to the critical surface density $\Sigma_{\mathrm{crit}} = 1/(4 \pi G D_{\mathrm{eff}})$ where $D_{\mathrm{eff}}$ is an effective distance given by $D_{\mathrm{eff}} = D_L D_{LS} / D_S$.  In the event that a lens halo is composed of several subhalos (and our line of sight through the halo passes through several such subhalos), the power spectrum of the convergence due to halo substructure is given by~\cite{Dai:2019lud}:
\begin{equation}
    P_{\kappa}(q) = \frac{\Sigma_\mathrm{cl}}{\Sigma_{\mathrm{crit}}^2} \int \frac{\dd M_s}{M_s^2} \frac{\dd f(M_s)}{\dd \ln M_s} | \tilde{\rho}(q ; M_s) |^2
\end{equation}
where $\Sigma_{\mathrm{cl}}$ is the surface density of the cluster, $f(M_s)$ is the subhalo mass distribution, and $\tilde{\rho}(q ; M_s)$ is the Fourier transform of the subhalo density distribution $\rho(r ; M_s)$.  In the case of spherical symmetry this is simply:
\begin{equation}
    \tilde{\rho}(q ; M_s) \equiv 4 \pi \int_0^{\infty} r^2 \dd r \frac{\sin ( q r )}{q r} \rho(r ; M_s).
\end{equation}
The relevant measure of fluctuations in $\kappa$ is then given in terms of the power spectrum above by:
\begin{equation}
    \Delta_{\kappa}(q) \equiv \sqrt{\frac{q^2 P_{\kappa}(q)}{2 \pi}},
\end{equation}
where here and above $q$ can be mapped onto a specific length scale $\ell$ by $\ell = 2 \pi / q$.  We can now estimate how sensitive this technique will be for our case.  We take $D_{\mathrm{eff}} \sim 1\,\mathrm{Gpc}$, $\rho(r ; M_s)$ to be an NFW profile of given scale mass and density, and $f(M_s)$ to be a delta function with 30\% of the DM concentrated in subhalos of a fixed mass.  Because we select for stars located on strong lensing caustics, we take $\Sigma_{\mathrm{cl}} \simeq 0.8\, \Sigma_{\mathrm{crit}}$, the factor of $0.8$ allowing for a star that is nearby but does not exactly lie on a caustic.  The lens model for the star of Ref.~\cite{Kelly:2017fps}, for example, predicts that for that star, $\Sigma_{\mathrm{cl}} = 0.83 \, \Sigma_{\mathrm{crit}}$~\cite{liangdaiprivate}.

Finally, we must check that the assumption of many subhalos along our line of sight is valid, and that the timescale of the fluctuations is shorter than the timescale of a typical intracluster-star-microlensing event $\tau_{\mathrm{microlens}} = \mathcal{O}(10^6)\,\mathrm{s}$.  During such an event, if the lensing star and the source star have a relative velocity $v_{\mathrm{rel}}$, then the image of the source star moves an approximate distance $d_{\mathrm{microlens}} \sim v_{\mathrm{rel}} \mu \tau_{\mathrm{microlens}}$ where $\mu \sim 10^{2}$--$10^{3}$ is the magnification.  Typical cluster velocities are of the order $10^{-2}$--$10^{-3}$, so we have $d_{\mathrm{microlens}} \sim 10^6\,\mathrm{s} \sim 10^3\,\mathrm{AU}$.  To ensure that there are many subhalos along our line of sight, we require that $\Sigma_{\mathrm{cl}} \pi d_{\mathrm{microlens}}^2 \gtrsim 10 M_s$, and $d_{\mathrm{microlens}} \gtrsim r_s$ is required for the timescale of fluctuations to be shorter than a typical microlensing event duration.

Assuming these requirements are satisfied, we calculate $\Delta_{\kappa}(2 \pi / \ell)$. We mark as potentially detectable parameter space wherein $\Delta_{\kappa}(2 \pi / \ell) > 10^{-3}$ for at least one length scale in the range $30\,\mathrm{AU} < \ell < 10^4\,\mathrm{AU}$, and we delineate the lower boundary of this region by the orange line in Fig.~\ref{fig:GRinteractions}.  Because this technique can probe even relatively low boost factors (and thus relatively weakly bound structures), simulations of subhalo mergers and accretion are needed to refine our estimate here.

\paragraph*{Microlensing from oscillons.}
As mentioned above, inducing a substantial change in brightness during a usual microlensing event requires the lens halo to lie entirely within its Einstein radius on the sky.  This can be translated to a requirement on internal density:
\begin{align}
\rho_s \gtrsim\frac{1}{(4\pi G_N D_l)^{3/2} M_s^{1/2}} \sim 10^{16} \rho_{\mathrm{DM}}^{\odot} \left[\frac{\mathrm{M_\odot}}{M_s}\right]^{1/2} \left[\frac{\mathrm{kpc}}{D_L}\right]^{3/2}. \label{eq:rhostrong}
\end{align}
Comparing Eqs.~\ref{eq:rhoosc} and \ref{eq:rhostrong} shows that oscillons within the MW (with $D_L \lesssim 10\,\mathrm{kpc}$) can satisfy this, meaning the photometric microlensing surveys of Refs.~\cite{alcock2000macho,tisserand2007limits,niikura2017microlensing,griest2014experimental} are sensitive to oscillons that are cosmologically long-lived and produced at high fractional abundance. They can thus test axion models wherein oscillons are produced at $\gtrsim 10\%$ fractional abundance and the axion mass is in the range $ 10^{-11}\,\mathrm{eV} \lesssim m \lesssim  10^{-19}\,\mathrm{eV}$ (such that $ 10^{-10}\,\mathrm{M_\odot} \lesssim  M_s^\mathrm{osc} \lesssim 10^2 \,\mathrm{M_\odot}$).

\subsubsection{Extragalactic strong gravitational lensing}\label{sec:stronglensing}
Flux ratio anomalies in multiply-imaged background sources can provide indirect windows into the substructure of strongly lensing galaxies~\cite{mao1998evidence,metcalf2001compound,chiba2002probing,
dalal2002direct,metcalf2002flux,kochanek2004tests}. DM substructure can also perturb the position~\cite{koopmans20022016,chen2007astrometric,williams2008lensed,more2009role} and relative time delays~\cite{keeton2009new,congdon2010identifying} of the lensed images, and many studies~\cite{inoue2005three,inoue2005extended,koopmans2005gravitational,vegetti2009bayesian,vegetti2009statistics,
vegetti2014density,hezaveh2016measuring,hezaveh2013dark} have explored the potential to pin down the subhalo spectrum of strong gravitational lenses. Ref.~\cite{hezaveh2016detection} claims a detection of a subhalo of $M_s \sim 10^9\,\mathrm{M_\odot}$, and also derived limits on the abundance of subhalos down to $M_s \sim 2 \times 10^7 \, \mathrm{M_\odot}$. The interpretation of the upper limits on subhalo abundance depend strongly on poorly determined quantities such as the host galaxy's mass and concentration, so it would be interesting to characterize these uncertainties more quantitatively and recast the observations of Ref.~\cite{hezaveh2016detection} to constrain axion subhalo mass functions such as those depicted in Fig.~\ref{fig:halospec}. 

\subsubsection{Diffraction of gravitational waves}\label{sec:LIGOlensing}

Gravitational waves emitted from BH-BH merger events will be lensed by the intervening mass distribution and can potentially provide another probe of dark matter substructure.  Even if the lens is not massive enough to lead to multiple images (detectable as multiple copies of the same merger event at different time delays), it can imprint characteristic distortions in both the waveform's amplitude and phase \cite{dai2018detecting}.  The strength of these distortions is characterized by a dimensionless parameter $w$:
\begin{equation}
w \simeq 1.3 (1 + z_L ) \left[ \frac{f_\mathrm{GW}}{10^2 \, \mathrm{Hz}} \right] \left[ \frac{M_{\mathrm{enc}}}{100 \, M_{\odot} } \right]
\end{equation}
where $z_L$ is the redshift of the lens, $f_\mathrm{GW}$ is the GW frequency, and $M_{\mathrm{enc}}$ is the mass enclosed within the impact parameter of the lens.  Distortion effects are maximized when $w \sim \mathcal{O}(1)$.  The detection potential for such distortions has been studied by Ref.~\cite{dai2018detecting}, who claim that high-signal-to-noise-ratio events (SNR $\gtrsim 20$--$30$) at advanced LIGO (aLIGO) will be able to probe BH-BH merger events with $w \sim \mathcal{O}(1)$ out to $\gtrsim 1 \, \mathrm{Gpc}$.  Since aLIGO operates at frequencies of $\mathcal{O}(10^1 \text{--} 10^3) \, \mathrm{Hz}$, it will thus be sensitive to DM substructure with mass of order $\mathcal{O}(10 \text{--} 1000) \, M_{\odot}$ enclosed within the impact parameter.

As Ref.~\cite{dai2018detecting} points out, the GW diffraction effect can change significantly based on the lens mass profile. Compact axion halos produced from the large-misalignment mechanism have a different internal density profile than CDM halos (see footnote~\ref{fn:profile}) so a reanalysis is necessary for a precise appreciation of the sensitivity. We can make conservative estimates for this GW diffraction technique by using an NFW profile down to a smoothing scale of $2 \pi / (m v_s)$. We do this as follows:

In the case of strong self-interactions, we expect that a large fraction $f_{s}$ of the DM is bound up in minihalos of a characteristic mass $M_s$ and density $\rho_s$.  The probability of any given BH-BH merger passing by such a minihalo with an impact parameter at most $b$ is roughly \cite{dai2018detecting}:
\begin{equation}
\mathcal{P}(b) \sim 0.045 f_s \left[\frac{1 + z_L}{2} \right]^3 \left[\frac{D_\mathrm{BH}}{5\,\mathrm{Gpc}}\right] \left[\frac{10^5\,M_\odot}{M_s}\right]\left[\frac{b}{1\,\mathrm{pc}}\right]^2
\end{equation}
where $D_\mathrm{BH}$ is the proper distance from us of the BH--BH merger event and $z_L$ is the redshift of the lens.  Taking $D_\mathrm{BH} \sim 5~\mathrm{Gpc}$, $z_L \sim 0.3$, and $f_s \sim 0.3$, we compute the smallest impact parameter $b_{\text{min}}$ such that at least 1\% of the BH--BH events will be lensed with $b < b_\mathrm{min}$. If $b_{\text{min}}$ is less than the smoothing scale $2 \pi / (m v_s)$, then we take $b_{\text{min}}$ to be the smoothing scale instead). We then require that there exists a $b > b_{\text{min}}$ such that the lens mass enclosed within a cylinder of radius $b$ leads to $0.5 < w < 5$ for some GW frequency $10^1\,\mathrm{Hz} < f_\mathrm{GW} < 10^3\,\mathrm{Hz}$. In addition, we check that this $b$ is no larger than ten times the Einstein radius for this mass, as suggested by the discussion in Ref.~\cite{dai2018detecting}.  If these requirements are satisfied, we mark the parameters $M_s$ and $\rho_s$ as potentially detectable in Fig.~\ref{fig:GRinteractions} by aLIGO.  Finally, we repeat the same analysis for LISA~\cite{Audley:2017drz} but for the frequency window $10^{-4}\,\mathrm{Hz} < f_\mathrm{GW} < 10^{-1}\,\mathrm{Hz}$.

We find that this technique is a second promising probe of regions of parameter space also covered by present or future astrometric lensing surveys, but we caution that these results are schematic estimates and a full reanalysis is necessary to be more precise.

\subsubsection{Dynamical constraints}\label{sec:dynconstraints}
Massive subhalos will experience a dynamical friction force from their collective gravitational scattering of the surrounding medium~\cite{chandrasekhar1943dynamical}, and will thus gradually lose angular momentum and sink to the center of their host halo. Following Ref.~\cite{binney2011galactic}, a subhalo on a circular orbit of initial radius $r_i$ and speed $v_c$, embedded in an isothermal halo with density profile $\rho(r) = v_c^2 / 4\pi G_N r^2$ made up of constituents much less massive than $M_s$, will sink to the center in a time:
\begin{align}
t_\mathrm{DF} \simeq \frac{1.17}{F} \frac{r_i^2 v_c}{G_N M_s} \approx \frac{4.0 \times 10^{10}\,\mathrm{y}}{F} \left[ \frac{10^8\,\mathrm{M_\odot}}{M_s}\right] \left[\frac{r_i}{8\,\mathrm{kpc}}\right]^2,\label{eq:tDF}
\end{align}
with $v_c \approx 235\,\mathrm{km\,s^{-1}}$ appropriate for the MW halo at the Sun's location. The form factor $F$ is an effective Coulomb logarithm $F = [\ln(1+\Lambda^2) - \ln(1+\Lambda_s^2)]/2$ with $\Lambda \equiv b_\mathrm{max} v_c^2 / G_N M_s$ and $\Lambda_s \equiv \Lambda \sqrt{R_s^2/b_\mathrm{max}^2 + 2R_s / \Lambda b_\mathrm{max}}$, that depends on the maximum impact parameter, $b_\mathrm{max} \approx 200\, \mathrm{kpc}$ for the MW, and the minimum impact parameter, which we take to be the scale radius of the subhalo $R_s$. For reference, $F\simeq \ln \Lambda \approx 10 (15)$ for $M_s = 10^8\,\mathrm{M_\odot} (10^6\,\mathrm{M_\odot})$, as long as $R_s \ll 3\,\mathrm{pc} (0.03\,\mathrm{pc})$. For larger sizes $R_s \gtrsim G_N M_s / v_c^2$, the Coulomb logarithm is suppressed and tends to $F \simeq \ln(b_\mathrm{max}/R_s)$ regardless of $M_s$ and $v_c$.

Eq.~\ref{eq:tDF} does not take into account backreaction, subhalo-subhalo scattering, baryonic components, mass loss from tidal disruption, orbit eccentricity, nor the more complicated density profile of the MW halo, but we nevertheless presume it to be a reasonable approximation. We expect the MW's evolution to be drastically altered if a significant fraction of its constituents have a dynamical friction timescale shorter than a Hubble time. It is evident from Eq.~\ref{eq:tDF} that MW subhalos as light as $10^7\,\mathrm{M_\odot}$ are significantly affected by dynamical friction, but until galaxy-scale simulations are performed and compared to data, we refrain from extracting constraints pertaining to dynamical friction effects on the evolution of the MW. 

The flipside to the above dynamical friction effects is that subhalos also have the capacity to dynamically heat their surrounding medium, including star clusters or compact ultra-faint dwarf galaxies. Ref.~\cite{brandt2016constraints} has employed this effect on a star cluster in Eridanus II and ten compact dwarfs to set constraints on point-like dark matter objects of masses $\gtrsim 5\,\mathrm{M_\odot}$. For extremely compact objects such as long-lived oscillons, those constraints would likely apply without change. It would be interesting to repeat the analysis of Ref.~\cite{brandt2016constraints} and investigate the phenomenology for compact subhalos: in this scenario, the stars can also dynamically \emph{cool} by gravitational scattering on the internal structure of the subhalos, so the limits will likely weaken.
A related effect, namely the catastrophic tidal disruption of wide stellar binaries (as opposed to the diffusive dynamical heating from tidal forces), is in principle also sensitive to sub-pc dark matter objects heavier than a few tens of solar masses~\cite{chaname2004disk,yoo2004end}, although current observations are not yet sufficiently robust to exclude an order unity dark matter fraction in such objects~\cite{quinn2009reported}.

To conclude, dynamical friction or heating effects from compact subhalos are a promising probe of DM substructure, but we believe more work is required in order to consider them robust.


\subsection{Femto-halo effects in direct detection}\label{sec:directdetection}

For heavier axion masses, the large misalignment mechanism enhances power at scales too small to be relevant cosmologically or even astrophysically. Still, if the axion has nonzero interactions with the SM, these changes to the power spectrum can affect the prospects for direct detection. In this section, we will focus on axion halos with masses at or below $10^{-15}\,\mathrm{M_\odot}$, which we will refer to as {\it{femto-halos}} (FHs). As we will see in Sec.~\ref{sec:QCD}, this part of the parameter space is also relevant for QCD axion DM searches. These FHs have a large number density and can potentially be observed by Earth-bound direct DM detection experiments, as the FH incidence rate on Earth is:
\begin{align}
\gamma \approx \frac{0.3}{\mathrm{year}}\left[ \frac{10^3}{\mathcal{B}_\odot}\right]^{2/3} \left[ \frac{10^{-18} \,\mathrm{M_\odot}}{M_{s}}\right]^{1/3} 
\end{align}
where $\mathcal{B}_\odot\equiv \rho_s/\rho^\odot_{\mathrm{DM}}$ is the femto-halo's density boost relative to the local DM density.

\begin{figure}[tbp]
\includegraphics[width=0.48\textwidth]{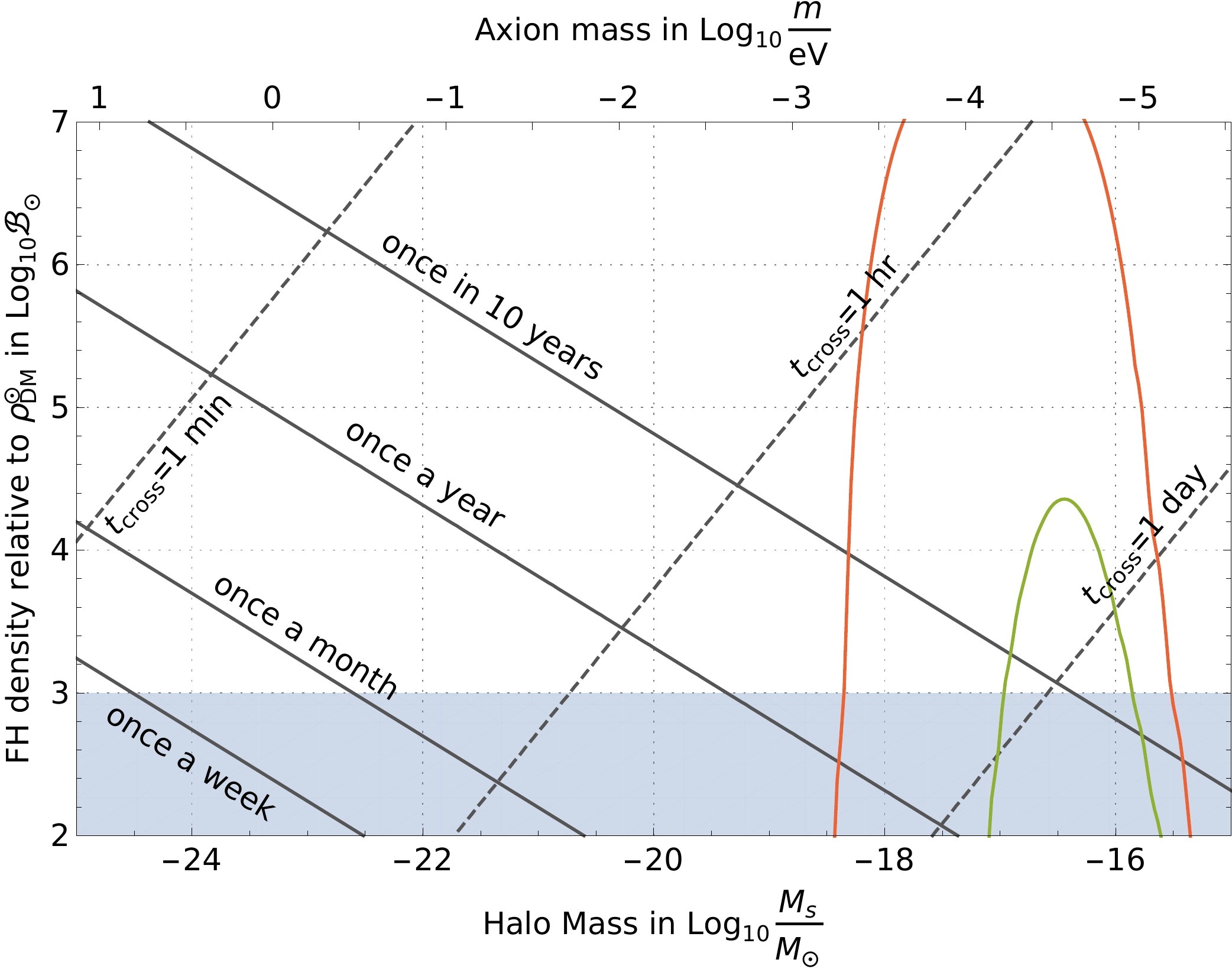}
\caption{Contours of constant incidence rate (solid lines) and detector crossing time (dashed lines), as a function of the FH mass and density boost factor $\mathcal{B}_\odot$ relative to the local DM density. The shaded region gives a conservative estimate of the tidal disruption constraint from disk crossings, as estimated in Sec.~\ref{sec:tidalstripping}. We also show halo spectra for the QCD axion with decay constants of $f_a = 10^{10}\,\mathrm{GeV}$ and $f_a = 2 \times 10^{10}\,\mathrm{GeV}$ in red and green, respectively, derived in Sec.~\ref{sec:QCD} and also shown in Fig.~\ref{fig:halospecQCD}.  For reference, we display on the upper horizontal axis the axion mass $m$ that yields the value of $M_s^*$ (see Eq.~\ref{eq:Msstar}) on the lower horizontal axis, but as the QCD axion halo spectra demonstrate, any fixed value of $m$ leads to FHs with a couple orders of magnitude variation in mass.}\label{fig:directdetection}
\end{figure}

Current direct axion DM searches look for a monochromatic signal at frequency $f \simeq m / 2\pi$ that is coherent for roughly $v_\mathrm{vir}^{-2}\approx 10^6$ periods. The amplitude of the signal is set by the local DM density and is typically assumed to be stationary. Axion searches are mostly resonant and, since the axion frequency is unknown, the resonant frequency is scanned.\footnote{The most notable axion experiment that falls in this category is ADMX~\cite{ADMX}. We also refer the reader to the Particle Data Group review of axions \cite{PDG} for a summary of other proposed experiments that are relevant for our discussion.} As we have seen in Sec.~\ref{sec:nonlinear}, the large misalignment mechanism may result in only a fraction of DM being in the form described above. With most of the axion DM in FHs, the DM signal becomes \emph{transient}, lasting for the FH's crossing time:
\begin{align}
t_\mathrm{cross} = \frac{r_s}{v_\mathrm{rel}} \approx 0.3\,\mathrm{day} \left[ \frac{10^3}{\mathcal{B}_\odot}\right]^{1/3} \left[ \frac{M_{s}}{10^{-18} \,\mathrm{M_\odot}}\right]^{1/3},
\end{align}
where we have taken $v_\mathrm{rel} = 10^{-3}$ for definiteness. For completeness, we note that this corresponds to a FH scale radius:
\begin{align}
r_s\approx 2\times 10^{-7}\,\mathrm{pc}\left[ \frac{10^3}{\mathcal{B}_\odot}\right]^{1/3} \left[ \frac{M_{s}}{10^{-18} \,\mathrm{M_\odot}}\right]^{1/3}.
\end{align}
During an encounter with a FH, the expected signal power is a factor of $\mathcal{B}_\odot$ higher than expected from a smooth DM component. Fig.~\ref{fig:directdetection} shows contours of constant incidence rate and crossing time as a function of the FH mass and the overdensity relative to the local  DM density.

FH axions have a much lower velocity dispersion relative to galactic axions, greatly increasing the effective coherence time of a DM signal in any axion experiment while a FH goes through the detector. The corresponding ratio between the scale velocity inside the FH $v_{s}$, and the virial velocity outside is:
\begin{align}
\frac{v_{s}}{v_\mathrm{vir}}\approx 4\times 10^{-11}\left[ \frac{\mathcal{B}_\odot}{10^3}\right]^{1/6} \left[ \frac{M_{s}}{10^{-18} \, \mathrm{M_\odot}}\right]^{1/3}.
\end{align}
The effective fractional spread in the frequency of the FH's DM signal is then
\begin{align}
\hspace{-0.5em}\frac{\delta f}{f}= v_\mathrm{rel} v_{s}\approx 2\times10^{-17}\left[ \frac{\mathcal{B}_\odot}{10^3}\right]^{1/6} \left[ \frac{M_{s}}{10^{-18} \, \mathrm{M_\odot}}\right]^{1/3} ,
\end{align}
with $v_\mathrm{rel} \sim 10^{-3}$ the relative velocity between the DM FH and the detector.

A natural question is to what degree these dynamically ultra-cold structures are distorted by tidal effects upon their entry into the Solar System. The tidal force from the Sun is practically always much greater than the self-gravity of the FH as it approaches Earth. Nevertheless, the FH does not get completely torn apart before it reaches our planet, due to the limited time it spends traversing the Solar System. We estimate the fractional change in the FH's size to be:
\begin{align}
\frac{\Delta r_s}{r_s} \sim \frac{G_N \mathrm{M_\odot}}{\mathrm{AU} v_\mathrm{rel}^2}\sim \mathcal{O}(10^{-2}),
\end{align}
where $\mathrm{AU}$ is the Earth-Sun distance. We thus expect the shape and the size of the FH to be essentially unaltered from their prior values.

Tidal effects will primarily affect DM searches via the differential velocity they impart across the FH. This differential velocity is typically much larger than the FH's internal scale velocity, and will appear as a frequency drift in the laboratory's rest frame, drastically reducing the effective coherence time in a practical axion DM search.
(In principle, one can construct frequency-drifting signal templates, but these are computationally costly to implement, as shown by searches for monochromatic gravitational waves in LIGO~\cite{LIGOCWsearch}.)  
We estimate the total differential velocity across the FH to have a magnitude of $\delta v_\mathrm{tidal} \sim G_N M_\odot r_s / (\mathrm{AU}^2 v_\mathrm{rel})$ upon its arrival at Earth. 
The resulting frequency drift is then determined by how much of this differential velocity is experienced during a ``shot'' time $t_\mathrm{shot}$, which we take to be a small fraction of $t_\mathrm{cross}$: 
\begin{align}
\delta f_\mathrm{drift}\sim \frac{m}{2 \pi}v_\text{rel} \delta v_\mathrm{tidal}  \frac{t_\mathrm{shot}}{t_\mathrm{cross}} \sim\frac{m}{2 \pi} v_\mathrm{rel} \frac{G_N \mathrm{M_\odot}}{\mathrm{AU}^2}t_\mathrm{shot}.
\end{align}
Requiring that the frequency drift be small enough that it may be ignored during any one shot, one gets an upper bound on the shot time as a function of the FH mass, density, and axion mass, i.e.~by requiring $t_\mathrm{shot}<\delta f_\mathrm{drift}^{-1}$. Breaking up the total integration time into shots of duration $t_\mathrm{shot}$ that saturates this inequality constitutes an axion DM search with effective fractional frequency resolution of:
\begin{align}
\label{eq:taucoherence}
\frac{\delta f}{f}\sim 10^{-12} \left(\frac{10^{-4}\,\mathrm{eV}}{m}\right)^{1/2}.
\end{align}
There is thus a parametric gain in effective coherence time---$10^{12}$ periods or more instead of the usual $v_\mathrm{vir}^{-2} \sim 10^{6}$ periods---even though the effects Solar System's tidal forces are substantial.\footnote{Since the FH size is much larger than the size of the earth for nearly all of the parameter space discussed here, we believe the tidal effects of the earth to be subdominant.}

Based on the above considerations, we can outline a new strategy for axion DM in the form of FHs. First of all, the intermittent nature of the signal favors a broadband data-recording approach: looking at a more extended range of frequencies increases the probability that the experiment is operating at the right frequency when a FH is going through the detector. Since most axion experiments are based on resonant antennae, a few comments are in order. Any experiment, resonant or non-resonant by design, can be run in a broadband mode. The problem is that for some resonant experiments such as ADMX, many of the components are optimized over an extremely narrow bandwidth, which makes running the experiment off-resonance suboptimal. This can be ameliorated by redesigning this hardware to respond to a wider range of frequencies.

This brings us to the second point: the reduced sensitivity off resonance can be compensated by the long coherence time and the boost in power relative to a search for a diffuse Galactic axion DM component. In fact, the sensitivity in axion coupling for FH DM searches can ultimately be improved relative to a search for a standard axion signal, provided an optimized broadband data-taking protocol is implemented. The signal power is not stationary: it is expected to spike at the incidence rate $\gamma$ for a duration $t_\mathrm{cross}$ by the local axion density boost factor $\mathcal{B}_\odot$. Such intermittent signals will be missed more often than not in most currently implemented experimental protocols, and sometimes even downright rejected if they are confused with a systematic background transient. Instead of slowly scanning a narrow frequency bandwidth over the total running time of the experiment, a better strategy is to coherently integrate the data stream and record a broader frequency bandwidth over a the longest possible shot satisfying $t_\mathrm{shot}<\delta f_\mathrm{drift}^{-1}$, and then incoherently adding the Fourier signal power of the shots. A more optimized search strategy could involve frequency-drifting matched filters over longer shot times (perhaps up to $t_\mathrm{cross}$), at the cost of considerable additional computational complexity and data volume. Our suggested protocol entails a data volume of $\mathcal{O}(t_\mathrm{shot} f)$ bytes every $t_\mathrm{cross}$, the result of taking the average Fourier signal power of $t_\mathrm{cross}/t_\mathrm{shot}$ number of shots.

\begin{figure}[tbp]
\includegraphics[width=0.48\textwidth]{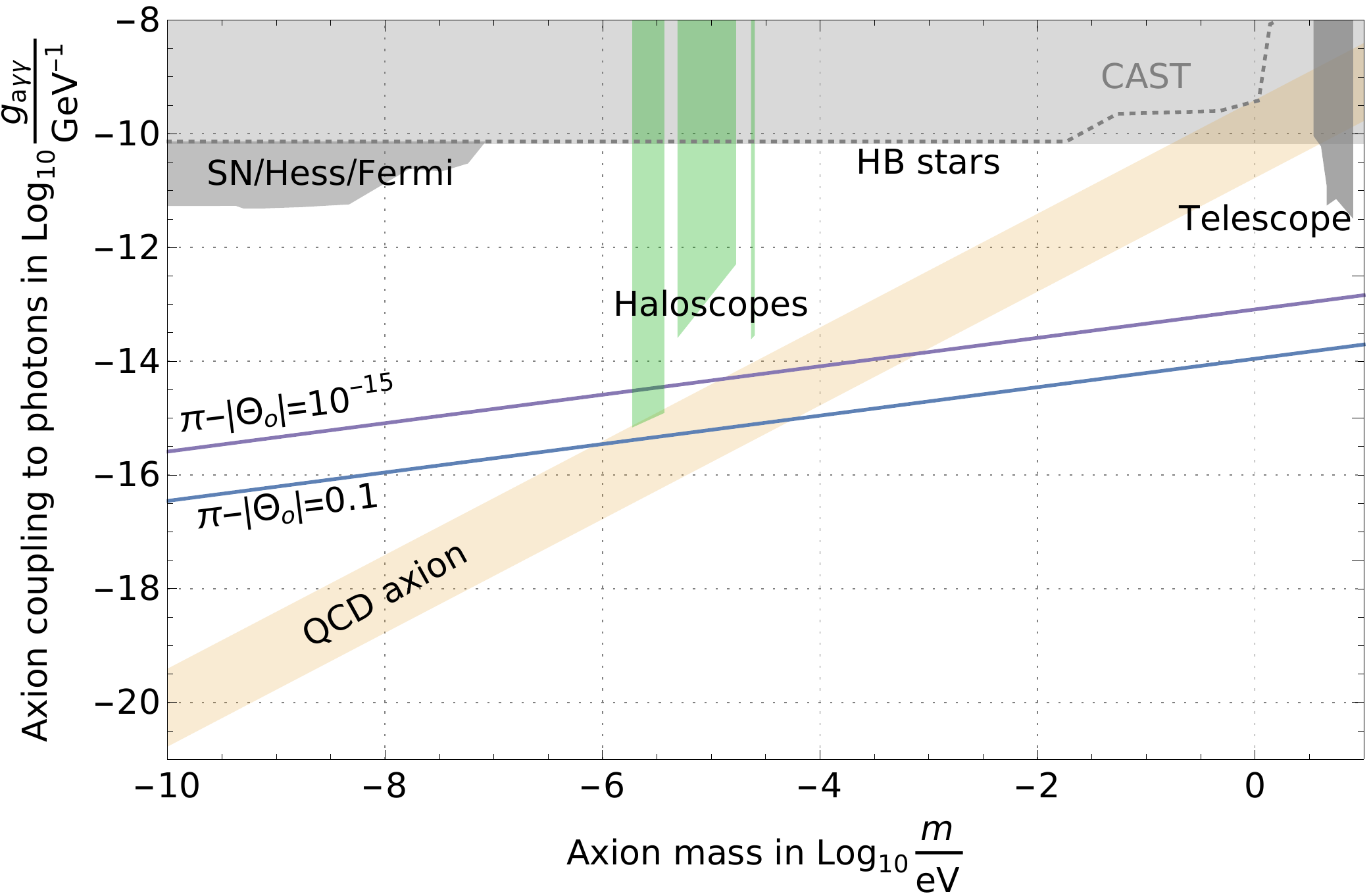}
\caption{Axion-photon coupling vs axion mass plot, adapted from Ref.~\cite{PDG}. Assuming that the axion-photon coupling is given by $g_{a \gamma \gamma}\sim\frac{\alpha}{4 \pi f}$, where $\alpha$ is the fine structure constant, we show the value of the coupling as a function of the axion mass for which the axion displacement is $\pi-|\Theta_0|=0.1$ (blue line) and $\pi-|\Theta_0|=10^{-15}$ (purple line). The blue line thus defines the region above which the large-misalignment mechanism can be responsible for DM production, and a large fraction of the axion DM is in FHs. The DM constraints on this plot, namely the haloscope and telescope searches (as well as any prospective discovery reach curves) should be recasted in terms of their sensitivity in this region.}\label{fig:clumpreachplot}
\end{figure}

Besides being temporally intermittent and more coherent, the signal from axion FHs is distinguishable from the standard Galactic axion signal also in terms of its spatial properties. For a standard axion, the mean velocity $\bar{v} = \langle v \rangle$ and the spread in velocity, e.g.~quantified by the standard deviation $\sigma_v = \langle (v-\bar{v})^2 \rangle^{1/2}$, are of the same order, so the spatial coherence length $\lambda_\mathrm{coh} \sim 1/m \sigma_v$ roughly equals the typical reduced de Broglie wavelength $\lambda_\mathrm{dB} =  1/m \bar{v}$. For FH axions, the typical reduced de Broglie wavelength is $\lambda_\mathrm{dB} = 1/m v_\mathrm{rel}$ where $v_\mathrm{rel}$ is the speed of the FH in the lab frame, but the coherence length is much larger: $\lambda_\mathrm{coh} \sim 1 / m v_s$. Two or more detectors with separations larger than $\lambda_\mathrm{dB}$ but shorter than $\lambda_\mathrm{coh}$ will therefore still pick up spatially phase-correlated signals, unlike for standard Galactic axions not bound in ultra-cold FHs. Such an array of detectors can even reconstruct the FH's velocity from these phase correlations, and would aid rejection of systematic transient backgrounds.

To summarize, these are the following key points to keep in mind when designing an experimental search for FH axion DM:
\begin{itemize}
\item{As a FH crosses the detector, the signal is a factor of $\mathcal{B}_\odot$ bigger in power and at least $\sim10^6$~times more coherent than a DM signal coming from a diffuse Galactic component.}
\item{Given that most searches are based on resonant antennae, it is imperative to also look for signals off resonance. The loss in sensitivity off resonance can be often be recovered by the boost in power and longer coherence time.}
\item{The experiment needs to record data for an extended period of time over large bandwidths, to ensure that a FH has an $\mathcal{O}(1)$ probability to cross the detector at least once during the experimental running time, at each frequency. Special care needs to be taken to handle the large data volume.}
\end{itemize}

Ultimately, the exact data analysis strategy would have to be independently determined  for each experimental setup, but the discussion above clearly shows that a search for an intermittent signal can be done concurrently with any search for a continuous galactic DM signal. As shown in Fig.~\ref{fig:clumpreachplot}, taking into account the possibility of axion DM in the form of FHs is of great importance for high-frequency axion DM searches. Current exclusions on the axion DM parameter space would not necessarily apply if the vast majority of DM is in the form of FHs, while some experiments may be sensitive to smaller couplings than originally envisioned. This means that numerical simulations of the large misalignment mechanism in the non-linear regime are crucial for extracting limits in axion DM searches. 


\subsection{Baryon structure and early star formation}\label{sec:baryons}

In $\Lambda$CDM, the bulk of star formation takes place in halos with a mass greater than $10^8\,\mathrm{M_\odot}$, at redshifts $z \lesssim 30$ (see e.g. Ref.~\cite{bromm2013formation} for a review).  However, when structures collapse much earlier, stars may form at much higher redshift and in lower-mass halos.  For axion masses between $10^{-22}$ and $10^{-18}$ eV, the axion self-interactions affect halo masses between $10^4$ and $10^9$ solar masses. In this section, we show that collapsed structures on these scales at high redshifts can satisfy the two main requirements for star formation: a sufficient baryon component and a cooling mechanism.  At the end, we also briefly discuss possible constraints from Lyman-$\alpha$ forests.

Baryons have a finite sound speed that inhibits their infall into perturbations on arbitrarily small scales.
Before recombination, this sound speed is close to the speed of light and the growth of baryon density perturbations is suppressed at all scales. After recombination, the baryon sound speed drops to a value set by the baryon gas temperature $T_b$ \cite{Tseliakhovich:2010bj}:
\begin{align}
c_s^2(a)=\frac{\gamma T_b}{ \mu m_H}=\frac {\gamma T_\mathrm{CMB-0}} {\mu m_H a} \left[ 1+\frac{a/a_1}{1+(a_2/a)^{3/2}} \right]^{-1}
\end{align}
where $\gamma=5/3$, $\mu=1.22$, $T_\mathrm{CMB-0} \approx 2.7\,\mathrm{K}$ is the present-day CMB temperature, and $m_\mathrm{H}$ is the hydrogen mass. The constants $a_1=1/119$ and $a_2=1/115$ in the expression above adequately capture the behavior of the baryon temperature after recombination. For redshifts larger than $\sim 100$, Compton scattering of baryons with CMB photons dominates over adiabatic cooling from the Universe's expansion, and $T_b$ tracks the photon temperature.

We na\"ively expect the effects of the finite sound speed in baryons to be captured by the Jeans scale $k_J$ above which baryons do not collapse into structures:
\begin{align}
\frac{k_J(a)}{a}=\frac{ \sqrt{4 \pi G_N \rho_m(a)}}{c_s(a)}
\end{align}
where $\rho_m(a)$ is the average matter density at a given redshift. This Jeans scale is defined by an instantaneous comparison between gravitational attraction and matter pressure in the equation governing linear perturbation growth, however even once gravity begins to dominate over pressure, the process of infall takes a finite time. This consideration leads to a more physical filtering scale, $k_f$, that accounts for the baryons' finite infall velocity. As was first shown in \cite{Gnedin:1997td}, for small co-moving momenta $k$ one can approximate the small scale structure of baryons $\delta_b(k)$ as:
\begin{align}
\delta_b(k)\simeq \delta_b-\frac{k^2}{k_f(a)^2}\delta_m
\end{align}
where $\delta_b$ and $\delta_m$ are the fluctuations in baryons and matter at very large length scales, respectively. Here $k_f$ is defined as:
\begin{align}
k_f= \delta_m^{1/2} \left[ \int_{t_\mathrm{r}}^t \frac{\dd t'}{a^2} \int_{t_\mathrm{r}}^{t'} \frac{\dd t''}{a^2} c_s^2(a) f_\mathrm{DM}\delta_b\right]^{-1/2}, \label{eq:filter1}
\end{align}
where $f_\mathrm{DM}=0.85$ is the DM fraction of the matter component. 
Unfortunately, Eq.~\ref{eq:filter1} fails to capture an effect of second order in the density perturbations that is nevertheless sizeable. Baryon acoustic oscillations produce a relative streaming velocity $v_{bm}$ between DM and baryon perturbations~\cite{Tseliakhovich:2010bj} with dispersion of $\sigma_{bm}=10^{-4}c\sim c \delta_m$ right after recombination, which subsequently redshifts adiabatically. Although of second order, this effect is thus enhanced by the large pre-recombination sound speed and can be important.

In Ref.~\cite{Naoz:2012fr}, it was shown that both this relative streaming velocity as well as the finite sound speed due to the baryon temperature can be included in a modified equation for $k_f$:
\begin{align}
k_f^{-2}= \delta_m  \int_{t_\mathrm{r}}^t \frac{\dd t'}{a^2} \int_{t_\mathrm{r}}^{t'} \frac{\dd t''}{a^2} f_\mathrm{DM} \left[c_s^2(a) \delta_b+
 (\vect{v}_{bm}\cdot \hat{\vect{k}})^2 \delta_\mathrm{DM}\right],
\end{align}
where $t_\mathrm{r}$ is the time at recombination, $v_{bm}= n \sigma_{bm}$, and $n$ quantifies the number of standard deviations of $v_{bm}$. It should be noted that the relative streaming velocity has a direction and thus the result depends on the direction of wavenumber direction $\hat{\vect{k}}$. 

\begin{figure}[tbp]
\includegraphics[width=0.48\textwidth]{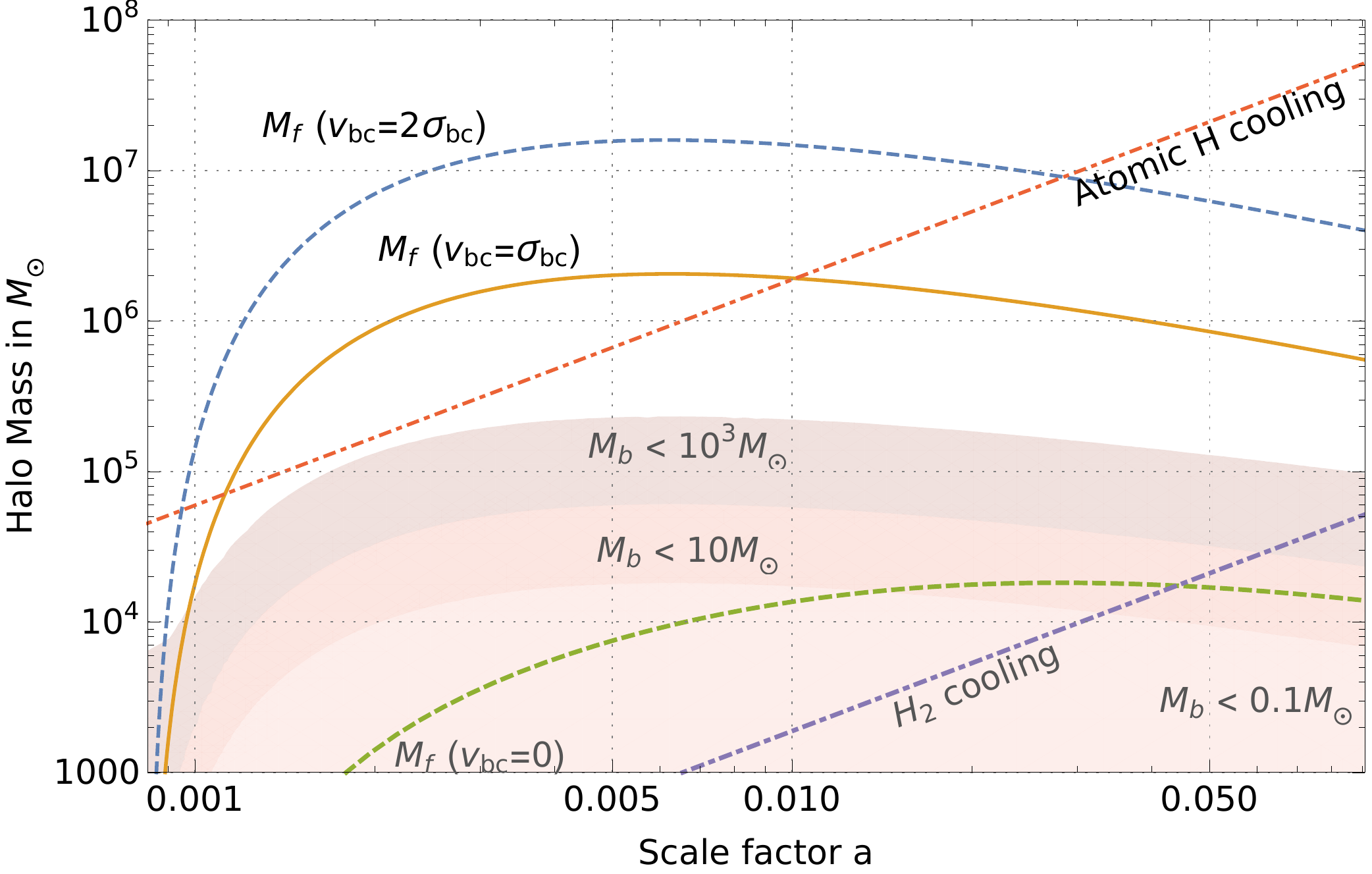}
\caption{We plot the value of the minimum mass a halo must have in order to carry a significant fraction of baryons, i.e. the filtering mass, $M_f$, for three different values of the baryon-DM relative streaming velocity $v_{bm}=0$ (dashed green line), $v_{bm}=\sigma_{bm}$ (solid yellow line), and $v_{bm}=2 \sigma_{bm}$ (dashed blue line) as a function of the halos' collapse scale factor, $a$. In the shaded regions, halos carry less than $\{10^3\,\mathrm{M_\odot},10\,\mathrm{M_\odot},0.1\,\mathrm{M_\odot}\} $ of baryonic mass $M_b$. We also show the minimum halo mass, $M_{min}$, where baryons can cool through atomic (dot-dashed red line) or molecular hydrogen cooling (dot-dashed mauve line). In order for a halo to be able to form stars, it must have at least one efficient cooling mechanism. Depending on when a compact axion halo forms, the mass of the smallest star forming halo can be as low as $10^4\,\mathrm{M_\odot}$, which corresponds to axion masses of $10^{-18}\,\mathrm{eV}$. Given that baryonic structure formation is significantly modified, this plot also shows that constraints on fuzzy DM such as those coming Lyman-$\alpha$ need to be revisited.}\label{fig:starformation}
\end{figure}

From this newly derived $k_f$, we can define a filtering mass $M_f(a)=\frac{4 \pi}{3} \rho^0_m \left({\pi}/{k_f}\right)^3$ below which we expect halos with a baryon fraction much smaller than the large-scale average. In Fig.~\ref{fig:starformation}, we plot $M_f$ for $v_{bm}=\{ 0, \sigma_{bm},2 \sigma_{bm}\}$. As discussed in Sec.~\ref{sec:nonlinear}, axion self-interactions result in collapse of DM structures at a much earlier time compared to $\Lambda$CDM cosmology, and $M_f(a_{col})$ corresponds to the minimum halo mass with a significant baryon fraction for the different possible values of the scale factor $a_{col}$ at collapse.\footnote{If collapse happens before recombination, we expect the early formed DM halos to accrete baryons post-recombination when $M_s>M_{f}(a_\mathrm{r})$. The resulting halo will consist of the earlier formed dense DM core, and the more diffuse post-recombination accreted component of matter.}
The baryon fraction scales approximately as \cite{Naoz:2012fr}:
\begin{align}
f_b=f_{b_o}\left[ 1+(2^{1/3}-1)\frac{M_f(a)}{M}\right]^{-3},
\end{align}
with $f_{b_o}=+0.15-0.005 {v_{bm}}/{\sigma_{bm}}$. Fig.~\ref{fig:starformation} shows which halos have at least $10^3 M_{\odot},~10 M_{\odot},\text{and}~0.1 M_{\odot}$ of baryonic mass, in our efforts to outline the critical condition for the formation of at least one star.

A significant baryon fraction is a necessary but not sufficient condition for star formation in a gravitationally collapsed structure. Another important requirement is that the virial temperature of the halo is large enough to allow for gas cooling. In $\Lambda$CDM, the most important form of cooling is provided by collisions of atomic hydrogen~\cite{bromm2013formation}, which only occurs in halos with virial temperatures larger than $10^4\,\mathrm{K}$. As Fig.~\ref{fig:starformation} shows, this implies that in $\Lambda$CDM, halos of mass smaller than $10^8\,\mathrm{M_{\odot}}$ have greatly suppressed star formation rates. 

For axion DM, when self-interactions are important, collapse can happen much earlier at higher densities. Atomic cooling is possible for halos of smaller mass than in $\Lambda$CDM, since $M_{\mathrm{cool,H}}(a)=2\times 10^6 \,\mathrm{M_\odot} \, (100 a)^{-3/2}$, but other cooling mechanisms could also be in effect. At the high densities of the early universe, other cooling mechanisms can also be operational. $\mathrm{H_2}$ molecular cooling, for example, is in principle active when the virial temperature is higher than $100\,\mathrm{K}$~\cite{Bromm:2003vv}, or halo masses larger than $M_{\mathrm{cool,H_2}}=2\times 10^3\,\mathrm{M_\odot} \, (100 a)^{-3/2}$. We record this minimum halo mass in Fig.~\ref{fig:starformation} as well, although the light from a few early stars can disassociate $\mathrm{H_2}$, halting further cooling and star formation~\cite{Barkana:2000fd}.  More work is thus required to understand exactly how such stellar feedback affects further star formation at high redshifts.

The discussion above clearly shows that for DM structures that collapse earlier, the minimum halo mass required to form stars can be greatly reduced from the $\Lambda$CDM prediction. In principle, as Fig.~\ref{fig:starformation} shows, the first stars could form in halos with mass as low as $10^4\,\mathrm{M_\odot}$. Unfortunately, beyond identifying that these requirements are satisfied, we cannot make further quantitative predictions. The reason is that little is understood about early star formation beyond the $\Lambda$CDM paradigm. How does gas cloud fragmentation proceed at such high densities? Does radiative feedback inhibit or help star formation at high densities? How does reionization happen?  Although it seems quite likely that stars will form in these high-redshift structures, without proper simulations to answer such questions it is impossible to be sure.

Given \textit{Planck}'s measurement of reionization, one might expect that early star formation would be highly constrained. As extensively discussed in Ref.~\cite{Loeb:2000fc}, one cannot draw such a conclusion very easily. At high densities, recombination could be much more efficient so that the ionizing radiation emitted from the first stars fails to keep the universe ionized. It is also not known from first principles how much ionizing radiation can escape a primordial halo, and thus not completely clear what the observable consequences of such early star formation would be.

In addition to changing the process of reionization, early star formation can alter the evolution history of astrophysical black holes as well as the 21-cm line history of the Universe. For astrophysical black holes, a period of star formation earlier than in the $\Lambda$CDM scenario could explain the appearance of high-redshift quasars by allowing for a longer growth period through Eddington-limited accretion. In our scenario, the black hole seed mass can be smaller by up to a factor of $\mathcal{O}(100)$. For ULAS J1342+0928~\cite{Banados:2017unc}, the most distant quasar known with an estimated mass of $8\times 10^8\,\mathrm{M_\odot}$, this would relax the seed BH mass requirement from several tens of thousands solar masses to less than $1000\,\mathrm{M_\odot}$. Given the size of the axion DM parameter space where the star formation history can be significantly altered, we thus believe our scenario deserves substantial further investigation through numerical simulations in combination with present and upcoming high-redshift data from the James Webb telescope and 21-cm probes of reionization such as EDGES~\cite{Bowman:2018yin}, HERA~\cite{DeBoer:2016tnn}, LEDA~\cite{Bernardi:2016pva}, the SKA low~frequency aperture array, and others.

Simulations are also necessary in order to understand how such shifts to the power spectrum can be probed and constrained by Lyman-$\alpha$ forests.  For ultralight masses of $\mathcal{O}(10^{-22}\,\mathrm{eV})$, axion dark matter without large-misalignment is constrained because of the matter power spectrum suppression above the wavenumber $k_*$~\cite{Irsic:2017yje}.  In our case, however, the structure enhancement discussed above will counteract some of this power suppression, and in extreme cases may provide such an enhancement that the \emph{ excess} of power will be constrained.  Ref.~\cite{Leong:2018opi} has conducted a preliminary study of this effect, but more work and simulations are necessary to understand exactly what region of the parameter space is constrained.  Lyman-$\alpha$ forests are perhaps able to probe up to masses of $\mathcal{O}(10^{-21})\,\mathrm{eV}$ at values of $f$ low enough to be in the oscillon-formation regime, which would mean that halos heavier than $10^9\,\mathrm{M_\odot}$ can be affected. This region of parameter space is also relevant for the gravitational wave signatures described below, in Sec.~\ref{sec:GW}.


\subsection{Gravitational wave emission}\label{sec:GW}
As studied in Sec.~\ref{sec:quarticcollapse}, the large-misalignment mechanism in extreme cases can lead to oscillon formation long before matter-radiation equality. The collapsing axion field structures are originally asymmetric and lose mass and angular momentum as they transition to oscillon configurations. While this process is dominated by scalar wave dynamics, the spherical asymmetry of the collapsing scalar field produces a small but potentially detectable component of stochastic gravitational waves.

We have computed this gravitational wave emission via numerical simulations which are described in App.~\ref{sec:numericgrav}, and their most relevant characteristics and implications can be estimated analytically and independently of the specific form of the potential.

We find that a good fit to the gravitational wave emission can be drawn from the standard quadrupole formula:
\begin{align}
P_{\text{GW}}\simeq\frac{G_N}{5} (\dddot{Q})^2,
\end{align}
where $\dddot{Q}$ is the third time derivative of the quadrupole moment. We assign a quadrupole moment to each oscillon of size $Q= \eta M_{\text{osc}} R_{\text{osc}}^2$, where $M_{\text{osc}}$ is the (initial) oscillon mass, $R_{\text{osc}}$ its characteristic radius and $\eta$ a factor describing its eccentricity. Since the field of axion density fluctuations is initially a Gaussian random field, eccentricity factors of $\mathcal{O}(1)$ are physically reasonable. According to our simulations, for deformations of order $25$--$50$\%, we have that $\eta \simeq 1$.

The mass, radius, and frequency of oscillation of the emitting oscillon are determined as follows. The mass of the collapsing object, which will be roughly the mass of the initial oscillon configuration, is estimated as the enclosed mass inside a volume of comoving radius $\pi/k$:
\begin{align}
    M_\text{osc}&\simeq\frac{4\pi}{3}\rho_\mathrm{DM}^0\pare{\frac{\pi}{\sqrt{2ma_\text{eq}^2H_{\text{eq}}}\tilde{k}}}^3
\end{align}
where $\rho_\mathrm{DM}^0$ is the DM density today defined in Sec.~\ref{sec:introduction}, and $\tilde{k}$ is the dimensionless comoving wavenumber from Eq.~\ref{eq:ktilde}. This mass will then collapse until the density becomes of order $m^2f^2$, which is roughly the point at which the oscillons are formed and gravitational waves are emitted. This determines the radius $R_{\text{osc}}$ that goes into the quadrupole. Finally, the angular frequency of oscillation and thus of the emitted gravitational waves, will be $\omega_{\text{GW}}\simeq \alpha R_\text{osc}^{-1}$:
\begin{align}
    \alpha\omega_{\text{GW}}^{-1}&\simeq R_\text{osc}\simeq\pare{\frac{3}{4\pi}\frac{M_\text{osc}}{m^2f^2}}^{1/3}
\end{align}
our numerical simulations imply that $\alpha\simeq\mathcal{O}(3)$.

We are considering scales $\tilde{k}$ which collapse at a time $t_m=t_{m,\text{coll}}$ well within radiation domination. The scale factor at collapse is:
\begin{align}
    a_\mathrm{coll} \simeq a_{\text{eq}}\sqrt{t_{m,\text{coll}}\frac{2H_{\text{eq}}}{m}} \, .
\end{align}

After collapse, both the energy density and frequency of the GWs will be redshifted. Assuming that $\mathcal{O}(1)$ of the DM is in these collapsing objects (an assumption supported by Ref.~\cite{Amin:2019ums}), the energy density $\Omega_{\text{GW}}$ emitted in gravitational waves relative to the DM energy density today scales as
\begin{align}
\frac{\Omega_{\text{GW}}}{\Omega_{\text{DM}}}&\simeq\frac{\omega_{\text{GW}}^{-1} P_\mathrm{GW}}{M_{\text{osc}}}a_{\text{coll}} \nonumber
\\
&\simeq 10^{-10}\eta^2\alpha^5\left[\frac{10^{-22}\,\text{eV}}{m}\right]
\frac{\sqrt{t_{m,\text{coll}}}}{ \tilde{k}^2}\left[\frac{\rho_{\pi/2}}{\rho}\right]^{1/3},
\label{eq:OmegaGW}
\end{align}
with characteristic frequency
 \begin{align}
 f_\mathrm{GW} &\simeq \frac{\omega_{\text{GW}}}{2\pi}a_{\text{coll}} \nonumber
 \\
 &\simeq 6\times10^{-15}\,\text{Hz}\, \alpha \left[\frac{m}{10^{-22}\,\text{eV}}\right]^{1/2}\tilde{k}\sqrt{t_{m,\text{coll}}}\left[\frac{\rho_{\pi/2}}{\rho}\right]^{1/3}.
 \label{eq:frequencyGW}
 \end{align}
Note that Eq.~\ref{eq:OmegaGW} and the assumptions behind it are general and hold for any potential that can give rise to configurations that collapse long before equality.

\begin{figure}
\includegraphics[width = 0.49\textwidth]{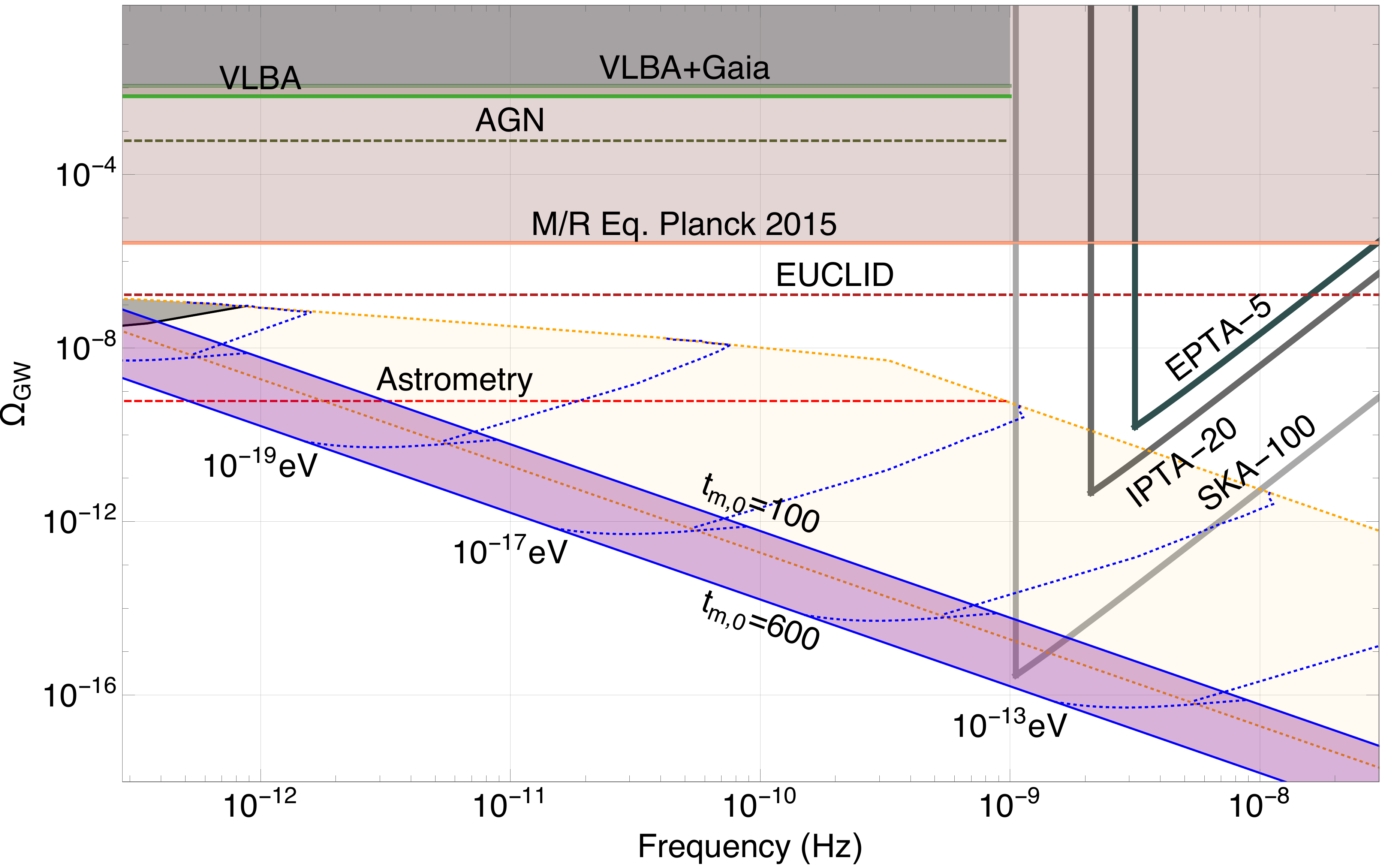}
\caption{The purple band is the expected stochastic gravitational wave background from Eq.~\ref{eq:OmegaGW} as a function of the observable frequency of  Eq.~\ref{eq:frequencyGW}. The upper end of the purple band corresponds to $t_{m,0}\simeq100$ and the lower end to $t_{m,0}\simeq600$. The black band on the lower frequency end corresponds to exclusions due to structure formation \cite{Irsic:2017yje}, since the collapsing mass is $\gtrsim 10^9\,\mathrm{M_\odot}$. The yellow band between the orange dotted lines is the prediction of the linear theory. The horizontal lines are current constraints (solid) and future reach (dashed) of different experiments. Future PTA sensitivities (solid) are also shown (see App.~\ref{sec:gwprospects} for details). The red ``Astrometry" line assumes $10^8$ quasars and $\sigma_\mu=1\,\mu\text{as y}^{-1}$ noise levels, while the SKA-100 sensitivity curve assumes $100$ pulsars observed for $30\,\text{yrs}$ with an error of $10\,\text{nsec}$ and a cadence of $14\,\text{days}$. The blue dotted lines are contours of constant axion mass. The signal from the $\sim10^{-22}\text{--}10^{-20}\,\mathrm{eV}$ and $\sim10^{-15}\text{--}10^{-14}\,\mathrm{eV}$ axion mass ranges is within expected future sensitivities.}
\label{fig:GWest}
\end{figure}

The power and frequency of the expected GW signal depend sensitively on $\tilde{k}$ and $t_{m,\text{coll}}$. Aided by numerical simulations, we observe that when the power spectrum boost $\mathcal{G}$ factor of Eq.~\ref{eq:GPR} becomes of order $10^6$ and thus $\delta\simeq 1$, the resulting nonlinear structures collapse into relativistic objects shortly afterwards (cfr.~Sec.~\ref{sec:nonlinear}). This observation allows us to estimate the collapse time from the parametric resonance formulae within the linear theory (cfr.~Sec.~\ref{sec:linear}). We furthermore check that in the $d=0+1$ rigid wavepacket approach of Sec.~\ref{sec:nonlinear}, the self-interaction term of Eq.~\ref{eq:ChavanisE} is larger in magnitude than the kinetic term. This ensures that, if a structure reaches $\delta\simeq1$ within the time that $E_\text{int}<0$, then it will collapse into an oscillon and give gravitational waves. The bottom panel of Fig.~\ref{fig:quarticcollapsecond} suggests that this approach is approximately correct at large values of $\delta$ (when parametric resonance shuts off due to the nonlinearities), especially for larger collapsing structures (or small $\tilde{k}$).

Eqs.~\ref{eq:OmegaGW} and \ref{eq:frequencyGW} suggest that the signal is dominated by the most massive collapsing structures, corresponding to the smallest possible $\tilde{k}$ that grows nonlinear. The above combination of the linear parametric resonance theory (Sec.~\ref{sec:linGR}) and the nonlinear quartic collapse analysis (Sec.~\ref{sec:quarticcollapse}) yields a set of three conditions that must be satisfied for a scale $\tilde{k}$ to collapse: (a) the power boost factor must reach a value of $\mathcal{G}(\tilde{k},t_{m,\text{coll}})=10^6$, (b) $t_{m,\mathrm{coll}}$ occurs well within radiation domination, and (c) $E_\text{int}$ of Eq.~\ref{eq:ChavanisE} is negative. Note that the time at which parametric resonance shuts off, defined below Eq.~\ref{eq:GammaPR}, is parametrically the same (with a somewhat larger numerical coefficient) as the maximum time allowed by the constraint $E_\text{int}<0$, so satisfying (a) means that (c) is automatically satisfied as well. 

From the linear treatment of perturbations, we expect a range of $\tilde{k}$ to parametrically resonate and collapse, as suggested by Fig.~\ref{fig:delta}. However, by the time the smallest $\tilde{k}$ satisfies the above conditions, higher $\tilde{k}$ have already become nonlinear, if we assume a scale-invariant spectrum of curvature perturbations. In fact, the linear theory predicts that there can be substantial time separation between the collapse of the first (and higher) $\tilde{k}$ and the collapse of the last (and smallest) one. But once the first-collapsing scales have entered the oscillon regime, the nonlinearities reduce the amount of energy available in the zero mode, essentially stunting any further growth for smaller $\tilde{k}$, and the linear regime procedure outlined above fails.  Simulations of similar systems in Ref.~\cite{Amin:2019ums} indicate that the vast majority of the axion energy density leaves the zero mode after the first scales undergo quartic collapse, rendering it unable to source the parametric resonance of the smaller wavenumbers.\footnote{One could consider a primordial power spectrum with suppressed power at these large $\tilde{k}'s$, so that the smallest collapsing $\tilde{k}$ as determined from the linear procedure is still accurate, but considering such scenarios goes beyond the scope of this paper. \label{footnote:linearGWestimate}} The same simulations indicate that this collapsing process is rapid, taking only roughly a factor of 10 in $t_m$ from the first hints of collapse to complete fragmentation into oscillons.  Thus we estimate the effects of nonlinearities on our signal by finding the smallest $\tilde{k}$ that collapses within a factor of 10 in time from the first collapsing scale, and using this $\tilde{k}$ to evaluate Eqs.~\ref{eq:OmegaGW} and \ref{eq:frequencyGW}. 

The results of the procedure outlined above are shown as the purple region in Fig.~\ref{fig:GWest} for the cosine potential. The upper blue line corresponds to $t_{m,0}\simeq100$ and the lower one to $t_{m,0}\simeq600$. As one turns up the misalignment angle and thus $t_{m,0}$, the range of collapsing $\tilde{k}$ in the linear theory gets extended on both ends, causing the \emph{first}-collapsing $\tilde{k}$ to be higher and collapse earlier, ultimately suppressing the signal. 
In the shaded region between the orange dashed lines in Fig.~\ref{fig:GWest}, we also depict the na\"ive expectation obtained by extending the linear regime description to the \emph{latest}-collapsing structures. As noted above, turning up the tuning allows for even smaller $\tilde{k}$ to parametrically resonate in the linear theory, which also collapse much later, potentially resulting in an enhancement of the signal with the tuning. In this case, the upper part of the curve (i.e.~for $f_\mathrm{GW}< 3\times 10^{-10}\,\mathrm{Hz}$) is cut off by the requirement that collapse occurs well within radiation domination, and $t_\text{coll}\leq t_\mathrm{eq}/10$, where $t_\mathrm{eq}$ is the time of matter-radiation equality. The shaded black region also corresponds to scales such that the collapsing mass is larger than $10^9\,\mathrm{M_\odot}$, and is excluded by structure formation~\cite{Irsic:2017yje}. Fig.~\ref{fig:GWest} shows how the linear description appears to overestimate by several orders of magnitude the expected GW signal. This is mainly because a much larger hierarchy in the collapse time is possible between the different collapsing $\tilde{k}$ when the nonlinear effects are neglected. As noted in footnote~\ref{footnote:linearGWestimate}, this estimate can become accurate for different primordial curvature power spectra.

There can be also GW emission from two additional regimes: (i) the interaction of two oscillons as they decay, expand, and collide; and (ii) the interaction of the scalar waves, emitted during the early collapse of a structure, with another oscillon. The power emitted from such configurations will, however, be suppressed by the usual $r^{-2}$ dilution due to propagation/expansion in 3D space as well as by the geometric cross-section of the interaction. As such, these additional contributions are subdominant with respect to the signal of Eq.~\ref{eq:OmegaGW}. These other GWs will be produced later than the ones of Eq.~\ref{eq:OmegaGW}, so both their amplitude and frequency will be less redshifted by the present day, but this is not sufficient to overcome their suppression. We thus take Eq.~\ref{eq:OmegaGW} to be an upper limit of the stochastic GW power leftover from these oscillon dynamics.

As Fig.~\ref{fig:GWest} illustrates, the gravitational wave emission can cover many orders of magnitude in frequency for different axion masses. We also show representative sensitivities of several current and upcoming experiments that promise to cover the relevant frequency window. These experiments fall in three categories: (a) looking for effects in the apparent motion of stars or quasars through astrometry, (b) pulsar timing observations, and (c) excess gravitational wave radiation manifesting as additional relativistic degrees of freedom. In App.~\ref{sec:gwprospects}, we briefly review each one of these and describe their sensitivity as shown in the figure. 

We should also note that for other potentials, such as the ones discussed in Sec.~\ref{sec:flat}, the GW signal can be enhanced in the higher mass end of the spectrum, i.e.~in the range within the PTA sensitivity curves. This is the result of the potentials being flatter at large field values, delaying the onset of the oscillations and subsequent collapse. Additionally, the larger quartic allows even smaller $\tilde{k}$ to parametrically resonate and collapse.  

For most of our parameter space, these signatures fall below existing sensitivities, but we are hopeful that advances in astrometric surveys and pulsar timing arrays will be able to probe our scenario in the near future, to constrain or detect gravitational waves from the large-misalignment mechanism in the $\sim10^{-22}\text{--}10^{-20}\,\mathrm{eV}$ and $\sim10^{-15}\text{--}10^{-14}\,\mathrm{eV}$ axion mass ranges.

\section{QCD axion}\label{sec:QCD}

We now turn away from a general analysis of ultralight scalar models to focus on the QCD axion. This proposed solution to the strong-CP problem is independently well-motivated \cite{Peccei:1977hh, axion1, axion2}, but the specifics of its potential and phenomenology mean we need to make a few major changes to the above computations.  The first is that the mass $m_a$ and decay constant $f_a$ of the field are no longer independent, and are instead related by \cite{villadoro2016qcd}:
\begin{align} \label{eq:qcdmfrelation}
m_a(T=0) = 5.70 \, \mu \mathrm{eV} \left( \frac{10^{12} \, \mathrm{GeV}}{f_a} \right)
\end{align}
This has the effect of reducing the parameter space to one dimension.  Fixing $f_a$ determines $m_a$ uniquely, and thus also determines the required initial misalignment angle $\Theta_0$ necessary to produce the proper present-day dark matter abundance (assuming dark matter is predominantly composed of QCD axions).  Because we are interested in effects that are most prominent when the field begins near the top of its potential, we will be interested in relatively smaller values of $f_a \lesssim 2 \times 10^{10} \,\mathrm{GeV}$ compared with much of the QCD axion literature.  This will correspond to masses $m_a \gtrsim 3 \times 10^{-4} \, \mathrm{eV}$.

The second major change is that the axion potential changes shape and becomes temperature-dependent.  At zero temperature, the potential is no longer a perfect cosine and depends on the masses of the light quarks \cite{villadoro2016qcd}:
\begin{align}
V(\phi ) = - m_\pi^2 f_\pi^2 \sqrt{1 - \frac{4 m_u m_d}{(m_u + m_d)^2} \sin^2 \left( \frac{\phi}{2} \right) }
\end{align}
where $m_\pi$ and $f_\pi \simeq 92 \,\mathrm{MeV}$ are the pion mass and decay constant respectively, and $m_u$ and $m_d$ are the masses of the up and down quarks.  For the measured values of the SM parameters, this potential is sharper at the top than a cosine potential, which would seem to imply a need for greater tuning in order to see the sorts of extreme boosts to structure growth that we are studying.  However, this potential is only valid at low temperatures.

At high temperatures, the dominant contribution to the potential comes from QCD instantons, and a good approximation to the potential is given by the dilute instanton gas result:
\begin{align}
V(\phi, T) = m_a^2(T) f_a^2 \left[ 1 - \cos \left( \frac{\phi}{f_a} \right) \right] \label{eq:VaT}
\end{align}
where $T$ is the temperature and $m_a^2(T)$ scales as:
\begin{align}
m_a(T)^2 \equiv \chi_\mathrm{QCD}(T) \, m_a^2(T=0) \label{eq:maT}.
\end{align}
The topological susceptibility $\chi_\mathrm{QCD}(T)$ scales as $\propto T^{-8.16}$ for temperatures $T > 1\,\mathrm{GeV}$, and can be computed numerically using lattice QCD.  For our analysis, we use the numerical results of Ref.~\cite{Borsanyi:2016ksw}.  Because we are interested in structure growth in the hot early Universe, we may approximate the full QCD potential with the form in Eqs.~\ref{eq:VaT} and \ref{eq:maT}.

\begin{figure}
\includegraphics[width = 0.5\textwidth]{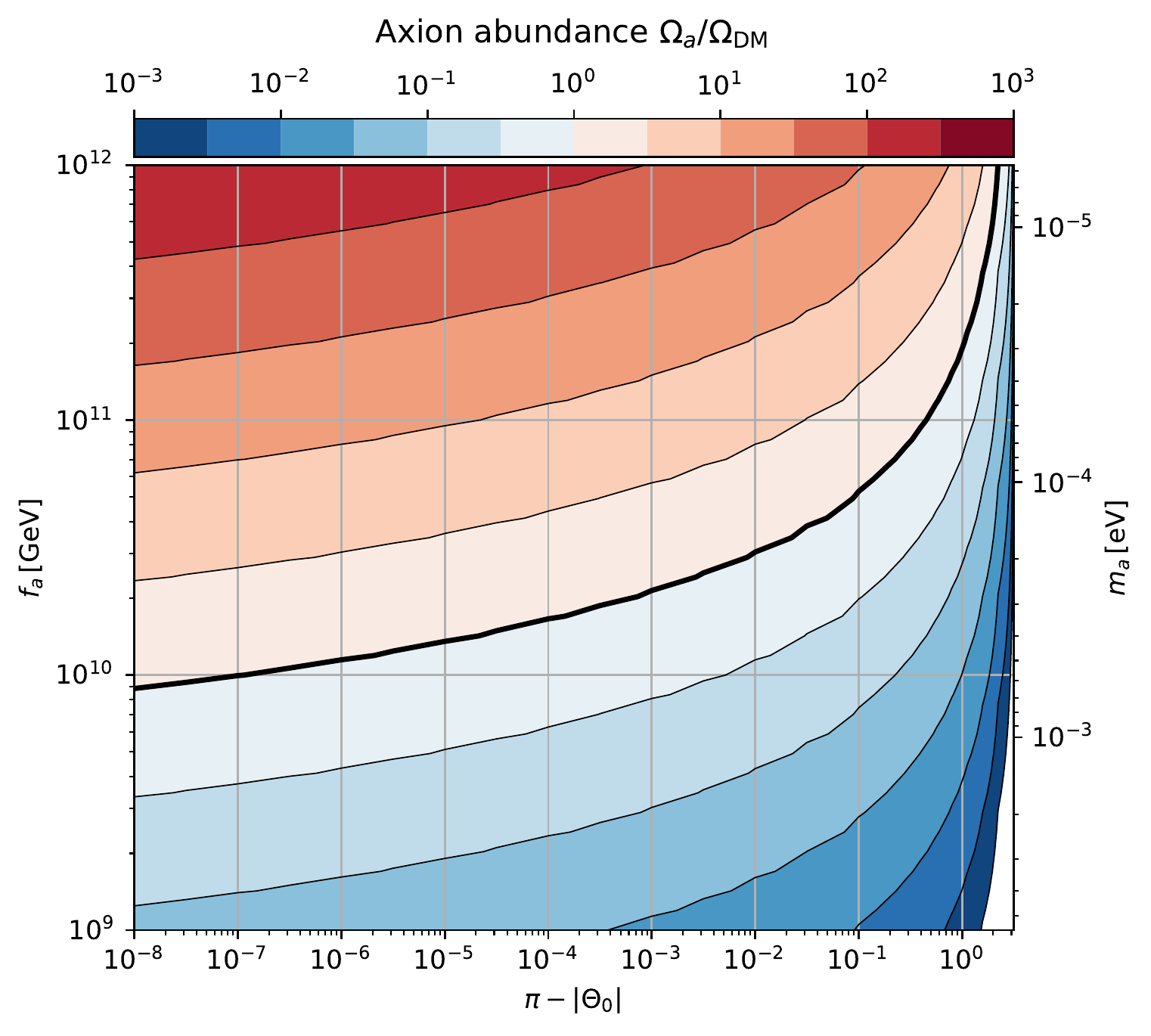}
\caption{
QCD axion abundance $\Omega_a$ as a fraction of the total DM abundance $\Omega_\mathrm{DM}$, as a function initial misalignment angle $\Theta_0$ and decay constant $f_a$. The thick black line denotes the curve where the QCD axion energy density matches the observed DM density. Parameter space above this line is therefore excluded, and a QCD axion below this line could only make up a subcomponent of the DM, but can still exhibit structure enhancement (which is only a function of $\Theta_0$). For QCD axions comprising the totality of the DM, the thick black line gives a relationship between the decay constant $f$ and the required initial misalignment angle $\Theta_0$.
}
\label{fig:QCDtuning}
\end{figure}

We can now proceed to the full analysis.  Defining $t_m$, $t_k$, and $\tilde{k}$ as in Sec.~\ref{sec:linGR} (using the zero-temperature mass $m_a(T=0)$ for $m_a$) we have that the background field evolves according to Eq.~\ref{eq:evolTheta} with $\sin(\Theta)$ replaced by $\chi_\mathrm{QCD}(T) \sin(\Theta)$.  Note that because $\chi_\mathrm{QCD}(T) \ll 1$ at high temperatures, the field may enter the horizon and begin oscillating substantially after $t_m \sim 1$, and in fact this is the case for the low-$f_a$ QCD axions under consideration here. Using this, we fix the relationship between $f_a$ and the required initial misalignment angle $\Theta_0$ for a given DM abundance, the results of which are shown in Fig.~\ref{fig:QCDtuning}.

\begin{figure}
\includegraphics[width = 0.5\textwidth]{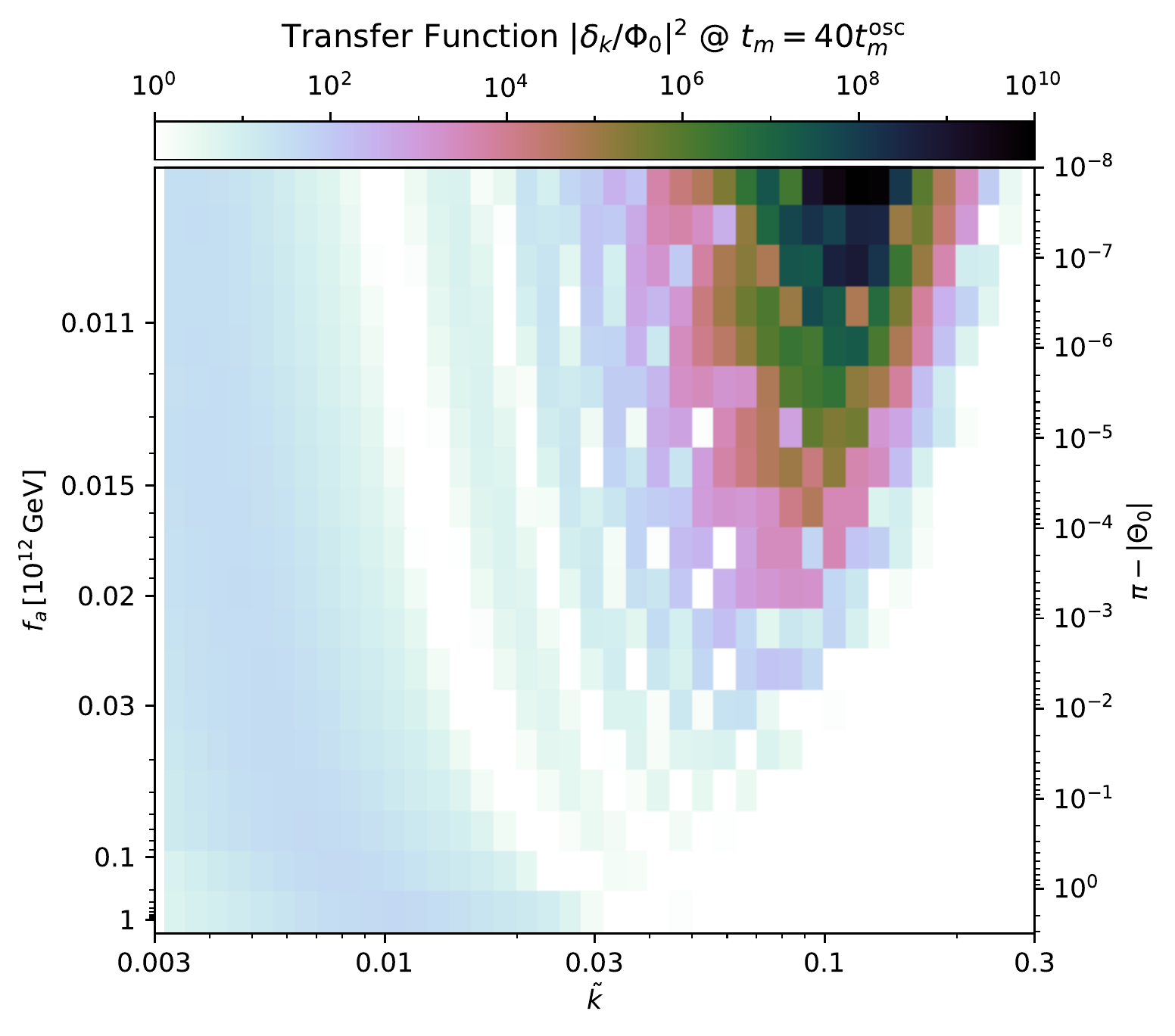}
\caption{
Transfer function $|\delta_{\vect{k}}/\Phi_{\vect{k},0}|^2$ of the QCD axion density fluctuation $\delta_{\vect{k}}$ relative to the primordial curvature fluctuation $\Phi_{\vect{k}, 0}$ evaluated at a time $t_m = 40 \, t_m^{osc}$, where $t_m^{osc}$ is the time at which the QCD axion field begins oscillating (when $m_a(T) \sim 3 H(T)$).  The results are presented as a function of comoving wavenumber $\tilde{k}$ and axion decay constant $f_a$.  In this plot we assume that the QCD axion comprises the totality of the DM, and thus the decay constant $f_a$ uniquely determines the required initial value of the misalignment angle $\Theta_0$.
}
\label{fig:DeltaQCD}
\end{figure}

Equations \ref{eq:evoltheta} and \ref{eq:thetaforcing} then describe the growth of QCD axion perturbations, and the covariant density perturbation is given by Eq.~\ref{eq:deltadef}, replacing each appearance of $\sin(\Theta)$ and $\cos(\Theta)$ with $\chi_\mathrm{QCD}(T) \sin(\Theta)$ and $\chi_\mathrm{QCD}(T) \cos(\Theta)$ respectively.  We evolve these equations numerically for a range of $\tilde{k}$ and initial misalignment angles; the results are shown in Fig.~\ref{fig:DeltaQCD}.  Note that the temperature-dependence of the QCD axion mass generally delays the onset of oscillation, so the wavenumbers that are unstable under parametric resonance are noticeably smaller than those in Fig.~\ref{fig:delta}, peaking around $\tilde{k} \sim 10^{-2}$ rather than $\tilde{k} \sim 5$.

\begin{figure}
\includegraphics[width = 0.5\textwidth]{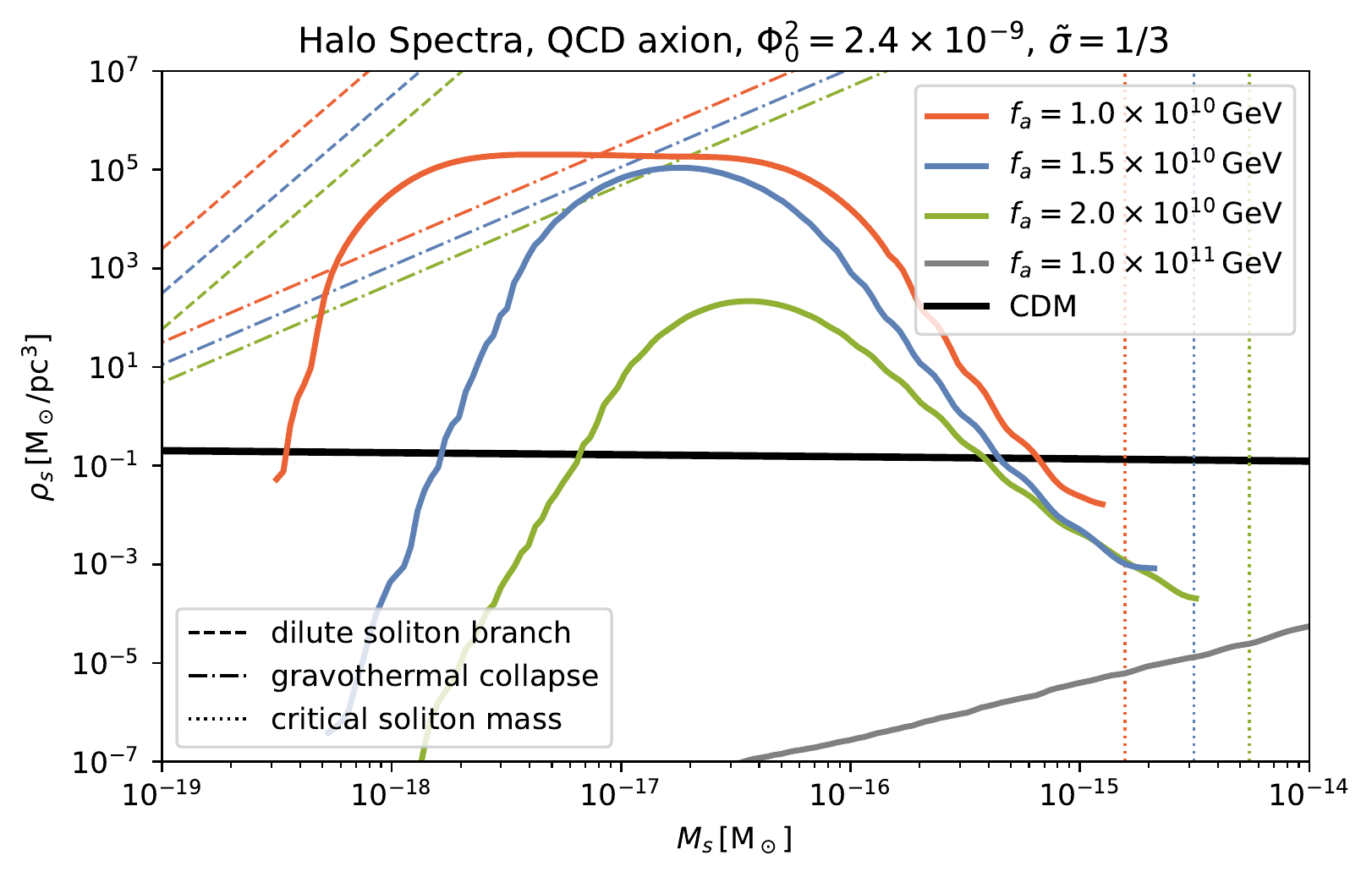}
\caption{\label{fig:halospecQCD} Halo spectrum for QCD axion DM in terms of scale mass $M_s$ and scale density $\rho_s$ for various values of decay constant $f_a$.  The predicted CDM halo spectrum is also shown, although the scales here are far smaller than any that have yet been measured.  For the red, green, and blue values of $f_a$, the dashed lines depict the dilute soliton branch of Eq.~\ref{eq:solitonmassdensityrelation}, the dotted lines depict the maximum (critical) soliton mass, and the dot-dash lines delineate the density above which gravothermal catastrophe occurs inside the halo (see Sec.~\ref{sec:gravcooling}).  Note that at scale masses larger than those that receive a boost, the predicted scale densities are actually less than those in CDM, implying a reduction of structure growth on these scales.  Although not shown on this figure, each of these colored lines will continue to decrease until they meet the gray $f_a = 10^{11}\,\mathrm{GeV}$ line at which point they will follow it back up to the CDM prediction.}
\end{figure}

A Newtonian treatment---analogous to that of Sec.~\ref{sec:linNewton}---of perturbations can be given long after they enter the horizon, and at late times ($\mathcal{O}(100)$ periods after the field begins oscillating), we stitch the exact general relativistic solution to the Newtonian solution in order to average out the oscillatory behavior.  Because all temperatures in the late-time universe are much lower than $1\,\mathrm{GeV}$, the nonlinear behavior is exactly the same as discussed in Sec.~\ref{sec:nonlinear}, and we give a present-day halo spectrum for a few representative values of $f_a$ in Fig.~\ref{fig:halospecQCD}.

Note in Fig.~\ref{fig:halospecQCD} that for scale masses larger than those that receive a boost, the predicted scale densities are actually \textit{below} the $\Lambda$CDM result.  This is due to the temperature-dependence of the QCD axion mass, which means the field does not be have like a matter fluid until the temperature drops below that of the QCD phase transition.  Scales that enter the horizon at higher temperatures have $\chi_\mathrm{QCD} < 1$ early in their history, which means the forcing term from Eq.~\ref{eq:thetaforcing} does not cause as much growth at early times as it does in CDM.  For scales that enter the horizon when $T \lesssim 100~\;\,\mathrm{MeV}$, the zero-temperature potential is a good approximation throughout their evolution and the behavior returns to the $\Lambda$CDM result.  In practice, this depression of growth at scales above the peak implies that when the structure growth is enough to cause collapse, all the power in the halo power spectrum will be confined to a smaller range of scale masses, and so the fraction of DM in structures at these scales will be higher than it would be for the axions of the previous section.

The effects discussed in this paper are most prominent for larger QCD axion masses, of the order $m_a \gtrsim 3 \times 10^{-4}\,\mathrm{eV}$, a range which will soon be probed by experiments such as MADMAX~\cite{Brun:2019lyf}, ORPHEUS~\cite{Rybka:2014cya}, HAYSTAC~\cite{HAYSTAC}, ADMX-HF~\cite{ADMX-HF}, ORGAN~\cite{McAllister:2017lkb}, QUAX~\cite{Barbieri:2016vwg}, TOORAD~\cite{Marsh:2018dlj}, dish antennae~\cite{horns2013searching}, plasma haloscopes~\cite{lawson2019tunable}, and multilayer optical haloscopes~\cite{Baryakhtar:2018doz}.  If the structure growth is enough to result in gravitationally collapsed robust against tidal stripping (i.e.~for $f_a \lesssim 2\times 10^{10}\,\mathrm{GeV}$), it is likely that most of the DM in our Galaxy will be clustered into axion femto-halos. In that case, the expected sensitivity of such experiments must be re-evaluated to take this clustering into account.  Experiments such as ARIADNE \cite{Arvanitaki:2014dfa, Geraci:2017bmq}, which are sensitive to this mass range but do not require that the QCD axion be the DM, will be unaffected.  We leave a complete reanalysis of the various constraints for future work. Although the QCD axion is the most motivated example of a light scalar with a temperature-dependent mass, it is not the only option; our results are readily modified for general time-dependent potentials.

Gravitationally bound structures in the context of the QCD axion have also been discussed in the literature under the name axion miniclusters~\cite{Hogan:1988mp, Tkachev:1986tr, Kolb:1993zz, Kolb:1993hw, Kolb:1994fi, Tkachev:2014dpa}.  Those objects are qualitatively quite different from the ones discussed here.  Axion miniclusters form during a post-inflationary Peccei-Quinn (PQ) phase transition, which leads to large density fluctuations on small scales that collapse at or slightly before matter-radiation equality.  In our case, there is no PQ symmetry present after inflation and the perturbations in the axion field are simply the primordial curvature perturbations enhanced by the axion self-interaction effects discussed in Sec.~\ref{sec:linear}.
One very important observational difference of the QCD axion miniclusters relative to the compact halos we consider here is that the former are very necessarily extremely dense. As a result, they encounter Earth only about once every $10^5$ years and cannot \emph{positively} affect axion DM searches in the laboratory.


\section{Initial conditions and general axion potentials}\label{sec:flat}

In Secs.~\ref{sec:evolution} and \ref{sec:signatures}, we have restricted ourselves to the case of the cosine potential. This is because the one instanton contribution to axion potentials is quite generic in a weakly coupled theory and it is also the case most relevant for the QCD axion. At first glance though, we seem to be faced with a serious problem of tuning. In order for the effects of self-interactions to be appreciable, the axion field has to start less than $\mathcal{O}(10^{-3}$--$10^{-2})$ from the top of the potential. The most extreme case, namely that of self-interaction-driven structure collapse during radiation domination, na\"ively requires tuning at the level of 1 part in $10^{12}$, but this figure merits a few comments.

First, there are dynamical mechanisms that can drive the field's initial value to the top of the cosine potential, in which case it is natural for it to be tuned near $\pi$.  One possible such mechanism, described in Ref.~\cite{Co:2018mho}, is to have a contribution to the axion potential during inflation that gives it a large mass (specifically $m > H_{\mathrm{inf}}$) and aligns the minimum with $\pi$ rather than 0 (both $0$ and $\pi$ are natural choices for the minimum because they are the only two values of $\Theta$ that preserve $CP$-symmetry).  The axion will then roll down to $\pi$ and remain there until the end of inflation when this potential contribution turns off.  From there, the field will evolve as discussed in Sec.~\ref{sec:evolution} with an initial value that appears to be tuned. 

The concrete model constructed in Ref.~\cite{Co:2018mho} applies specifically to the QCD axion, but similar mechanisms likely exist for other axion-like particles.  The basic ingredient necessary is a difference between the minimum of the potential during inflation and the minimum after, which should be unsurprising given that the minima of any potential are generically temperature-dependent.  During inflation the system is thermal at the Hawking temperature $T_H = {H_{\mathrm{inf}}}/{2 \pi}$, and so thermal contributions to the axion potential can easily lead to the zero-temperature maximum ($\Theta = \pi$) being a minimum during inflation. Such dynamics also have the added advantage that they suppress isocurvature fluctuations, relaxing the constraints discussed in App.~\ref{sec:isocurvature}.

Second, even if no dynamics are involved, an understanding of the tuning requires an understanding of the probability measure associated with the initial field value as well as the probability measure associated with an anthropic argument.  The latter can in principle alleviate the tuning substantially, which we investigate with a brief discussion of an anthropic argument due to Ref.~\cite{Freivogel:2008qc}.  The basic idea is that if $\rho_\mathrm{DM}$ were much less than we observe it to be, structures would not be able to collapse before the Universe entered the present era of dark-energy domination.  Since expansion would then rapidly dilute all matter, no structures would collapse and thus no observers would form.  On the other hand, if $\rho_\mathrm{DM}$ were much larger than its observed value, baryons would be proportionally rarer and thus baryonic observers would be less common.  In our case, using the technique and priors of Ref.~\cite{Freivogel:2008qc} yields an actual tuning of order the square root of the ``na\"ive'' tuning.  This analysis cannot be rigorous---the measures used are subject to significant uncertainties and disagreement in the literature---but it still serves to demonstrate that anthropic arguments can substantially alleviate the tuning necessary to observe the effects discussed in this paper.

Ultimately, we must note that the tuning depends heavily on the shape of the potential near the top.  As discussed in Sec.~\ref{sec:linGR}, the requirement for large self-interaction-induced growth in density perturbations is a ``delay'' between the time when the field starts oscillating and its na\"ive oscillation time (i.e. when $m \sim 2 H$).  For a field that begins near the top of its potential, changes in the potential's slope can lead to parametric changes in how long it takes to begin rolling.
Realistic axion potentials descending from some unknown UV completion may deviate significantly from the cosine potential of Eq.~\ref{eq:cosinepot}, and more naturally realize a delay in the onset of oscillations.

The effects that we point out in this paper are present in large classes of models with different axion potentials as long as they have attractive self-interactions.  In several of these models, including some models of axion monodromy, the potential is flatter than quadratic (that is they scale like $V(\phi) \sim \phi^p$ for some $p < 2$, or equivalently $V(\phi) < \frac{1}{2} m^2 \phi^2$ at large field values) for a large field range, which is exactly what is required for the effects described above to manifest themselves.  As discussed at the end of Sec.~\ref{sec:linGR}, the extreme growth in energy density perturbations requires a ``delay'' in the onset of oscillations from its natural timescale $t_m \sim 1$ ($m \sim 2H$).  This natural timescale is the exact result for a purely quadratic potential with mass $m$, so any delay must come from the potential being flatter than quadratic.  The precise nature of how it flattens will determine how much the field is delayed in its oscillation, but any such delay will lead to similar phenomenology: a set of wavelengths with an exponential growth instability.  To illustrate this, we consider two different toy models and then discuss how generic their behavior really is.  More discussion of these effects is also present in Ref.~\cite{Olle:2019kbo}.

The first model we consider is an axion with potential:
\begin{equation} \label{eq:ratiopot}
V( \phi ) = m^2 f^2 \frac{\phi^2}{2 f^2 + \phi^2} = m^2 f^2 \frac{ \theta^2}{2 + \theta^2},
\end{equation}
where $\theta \equiv \phi / f$.  This potential has the same mass $m$ and overall energy scale $m^2 f^2$ as the cosine potential in Eq.~\ref{eq:cosinepot}.  Such a potential can arise quite naturally for example from integrating out a heavy field in a two-scalar model.  As discussed in Ref.~\cite{Dong:2010in}, we can begin with a potential such as
\begin{equation} \label{eq:heavylightpotential}
V(\phi_L , \phi_H) = g^2 \phi_L^2 \phi_H^2 + M^2 (\phi_H - \phi_0)^2,
\end{equation}
and integrate out the heavy field $\phi_H$ to obtain the potential of Eq.~\ref{eq:ratiopot} with $m^2 = 2 g^2 \phi_0^2$ and $f^2 = M^2/(2 g^2)$.

\begin{figure}
\includegraphics[width = 0.5\textwidth]{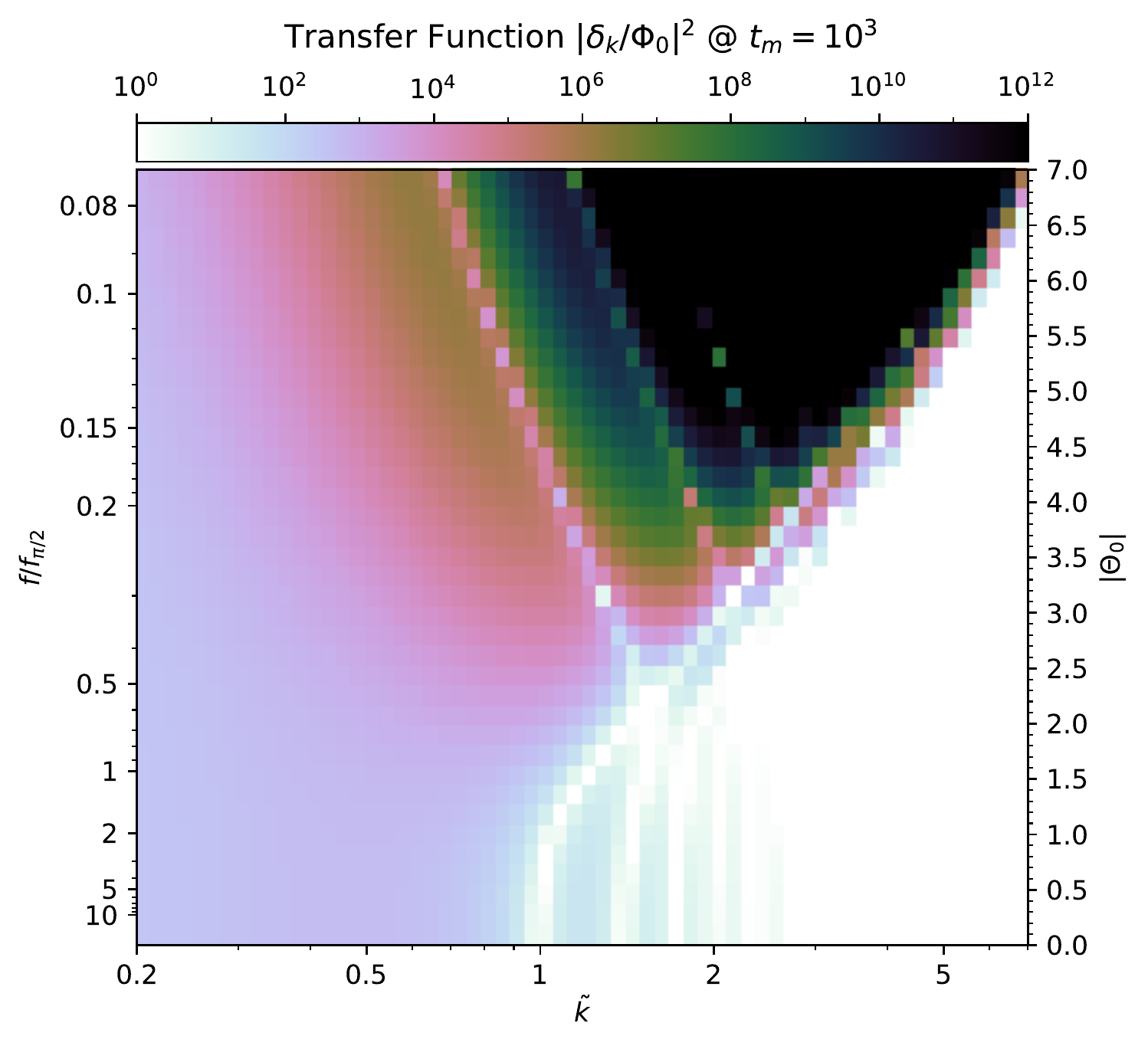}
\caption{\label{fig:DeltaFlatPot} Transfer function for the axion energy density (see Fig.~\ref{fig:delta}) for an axion with the ratio potential of Eq.~\ref{eq:ratiopot}.}
\end{figure}

We can now repeat the linear growth analysis from Sec.~\ref{sec:linGR} with this potential to obtain Fig.~\ref{fig:DeltaFlatPot}.  For $|\Theta_0| \gtrsim 2$, the field's oscillation is delayed and there are large enhancements to structure growth for a range of length scales.  For $|\Theta_0| \gtrsim 4$, some scales receive enough of a boost that they will collapse during radiation domination.  This potential will thus exhibit all of the observable phenomenology discussed in Sec.~\ref{sec:signatures}, but does not suffer from any of the tuning issues present in the cosine potential.

\begin{figure}
\includegraphics[width = 0.5\textwidth]{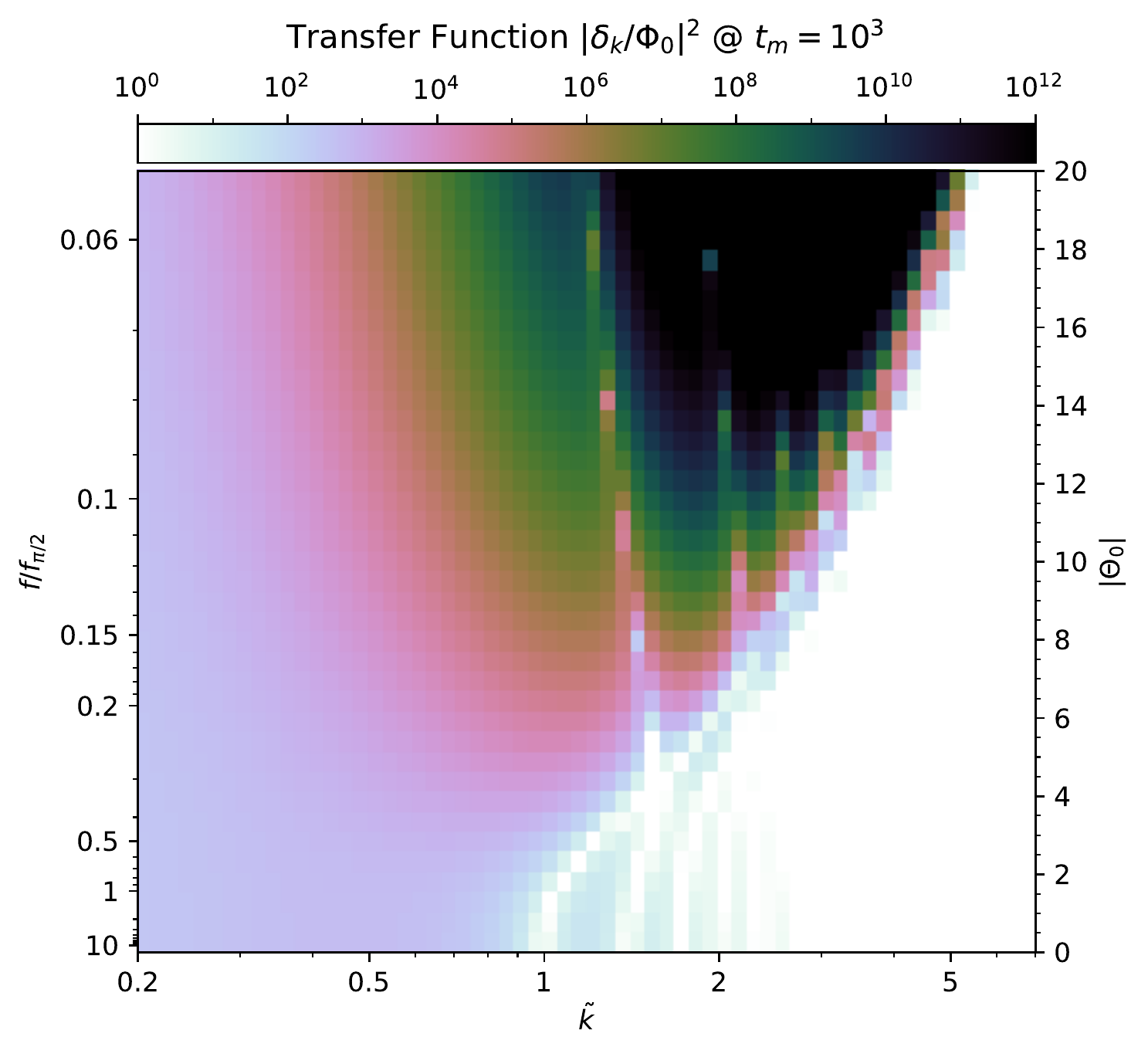}
\caption{\label{fig:DeltaLinWings} Transfer function for the axion energy density (see Fig.~\ref{fig:delta}) for an axion with the monodromy potential of Eq.~\ref{eq:linwingspot}.}
\end{figure}

Axion potentials with an unbounded field range and a flattening at large field values have also been discussed extensively in the axion monodromy literature. e.g.~Refs.~\cite{Dong:2010in, dubovsky2012axion, McAllister2014, Kaloper:2016fbr}.  As a prototypical example from this class of models we consider the case of a D4-brane stretched between two NS5-branes around an internal cylinder.  This model is discussed briefly in Ref.~\cite{Silverstein:2016ggb}, but for our purposes it is only important that the low-energy 4D-theory will include a moduli field $\theta$ corresponding to the winding of the D4-brane around the cylinder.  The potential of this field is then given by:
\begin{equation} \label{eq:linwingspot}
V( \phi \equiv f \theta ) = m^2 f^2 \left( \sqrt{1 + \theta^2} - 1 \right).
\end{equation}
This potential is quadratic near the origin and flattens to become linear at field excursions $|\theta| \gtrsim 1$.  Again we can repeat the linear growth analysis from Sec.~\ref{sec:linGR} with this potential to obtain Fig.~\ref{fig:DeltaLinWings}, where we can see that indeed there will be significant structure growth and early collapse for $|\Theta_0| \gtrsim 10$.

Both of these examples serve to demonstrate that the phenomenology and signatures discussed in this paper are not unique to the cosine potential of Eq.~\ref{eq:cosinepot} but are rather generic to any axion model with a delayed onset of oscillation relative to the natural timescale $t_m^\mathrm{osc} = 2 H_\mathrm{osc}/m \sim 1$ near any minimum where the quadratic expansion is good approximation. For models with a cosine potential, this requires an initial misalignment angle tuned quite close to the top of the potential, but for other models this is not the case. For the monodromy potential of Eq.~\ref{eq:linwingspot}, it is easy to show that $t_{m}^\mathrm{osc} \sim |\Theta_0|^{1/2}$ for very large initial field misalignments $|\Theta_0| \gg 1$. As long as $\bar{\Theta} \gg 1$, the energy density is linear in the field value, and will scale as $a^{-2}$~\cite{turner1983coherent}, so we have $\rho/m^2 f^2 \sim \bar{\Theta} \sim |\Theta_0| t_{m}^\mathrm{osc} / t_m$ during radiation domination. Hence we find that $\bar{\Theta} = 1$ at a dimensionless time $t_{m,0} \sim |\Theta_0|^{3/2}$ that can be very large indeed, leading to strong parametric resonance effects (cfr.~Eq.~\ref{eq:Gstar}). The ratio potential of Eq.~\ref{eq:ratiopot} has an even steeper dependence of $t_{m,0}$ on large initial misalignments $|\Theta_0|$. 

Intriguingly, in numeric simulations of both the above potentials, we have found metastable oscillon states with substantially longer lifetimes than similar states for the cosine potential.  We have been unable to find a precise expression for their lifetimes, but simulations confirm that both Eq.~\ref{eq:ratiopot} and Eq.~\ref{eq:linwingspot} lead to states that live at least $\mathcal{O}(10^5 / m)$ and possibly much longer (other groups have also found states living at least $\mathcal{O}(10^8 / m)$ in similar potentials \cite{Olle:2019kbo}).  If they live a few orders of magnitude longer than this, they may be cosmologically relevant and have observable signatures, some of which we have already discussed in Sec.~\ref{sec:GRinteractions}.  We leave a more detailed analysis of these states and their phenomenology for future work.

\section{Discussion}
\label{sec:discussion}

In this paper, we have shown how axion self-interactions can lead to nongravitational DM structure growth resulting in compact halos, and we have proposed several observational signatures of these halos. This growth is driven by parametric resonance for modes of order the axion Compton wavelength at the time when the axion field starts oscillating. The effect on DM density perturbations is bigger when the axion field starts with ``large misalignment,'' that is at a flat portion of its potential. This enhancement of structure formation thus presents a qualitative and quantitative production mechanism for compact DM halos, which in extreme cases can be scalar field configurations such as solitons and oscillons.

The possibility of the existence of such objects in our universe was considered before but without a concrete cosmological production mechanism (with the exception of the aforementioned QCD minicluster literature). Here we outline a framework that can make quantitative predictions for the production of compact axion structures in terms of fundamental parameters of the theory. The totality of all our observational predictions is summarized in Fig.~\ref{fig:summary}, covering an enormous range of axion masses from $10^{-22}\,\mathrm{eV}$ to $10\,\mathrm{eV}$.

For axion masses larger than $10^{-5}\,\mathrm{eV}$---and for the QCD axion with decay constants near $f_a \sim 10^{10}\,\mathrm{GeV}$---a large fraction of the axion DM is in compact dense halos that episodically visit Earth, substantially affecting contemporary and future experiments that target this parameter space. Axion DM experiments operating in this mass range should reconsider their search strategies and their method of data analysis to account for these effects. In fact, if the vast majority of the axions is in dense halos, then regions of the axion parameter space that are now considered excluded because of negative direct axion DM searches could actually be allowed, so a re-interpretation of the present axion exclusion plots may become necessary. A related theoretical challenge is to compute precisely the fraction of DM axions that is in compact halos, which will likely require numerical simulation.

Axions lighter than $10^{-5}\,\mathrm{eV}$ can be probed through various types of gravitational lensing searches as well as measurements of our local DM distribution. Axions lighter than $10^{-18}\,\mathrm{eV}$ can lead to substantially earlier star formation, possibly accelerating the formation of early supermassive black hole seeds, or providing new signatures accessible with better measurements of re-ionization. Understanding such signatures will also probably require numerical simulation, as will the computation of the compact halo spectrum and the relevance of the gravothermal catastrophe for the late-time properties of these halos.  Although we have not studied their signatures in this paper, axions heavier than $10~\mathrm{eV}$ can also form gravitationally bound structures that survive to the present day, and this could potentially spur new ideas for direct detection experiments in this range.  We leave this possibility for future work.

In the extreme case where nonlinear DM structure forms well before matter-radiation equality, we are presented with the exciting possibility of oscillon DM. Of course this would require that the oscillon is cosmologically long lived, which is not the case for the cosine potential. Nevertheless, numerical simulations reveal that oscillons of other well-motivated, flatter potentials can live for at least $10^8$ cycles~\cite{Olle:2019kbo}, corresponding to lifetimes of tens of millions of years for fuzzy DM. This suggests the idea of oscillon DM may be realizable for some potentials.

Axions are extremely well-motivated DM candidates, and are a main focus of research beyond the Standard Model. With this work, we hope to bring into focus a largely overlooked property of axions that changes our notions of DM structure and its signatures. The large-misalignment mechanism for axion DM production points to previously unexplored possibilities for the properties of axion DM and its experimental signatures over tens of orders of magnitude in parameter space.

\begin{acknowledgments}
We would like to thank Eva Silverstein, Mustafa Amin, Gray Rybka, Kendrick Smith, Will East, Eric Braaten, Alberto Sesana, and Philip Mocz for useful discussions. AA is grateful to Neal Dalal and Gilbert Holder for illuminating discussions on the gravothermal catastrophe. 
LL is grateful to Steve Liebling and Carlos Palenzuela, who are also main developers of the computational infrastructure employed for the 3D simulations described here. 
KVT is indebted to Masha Baryakhtar, Nikita Blinov, Nick Gnedin, Victor Gorbenko, Siddarth Mishra-Sharma, Neal Weiner, and Edward Witten for conversations and several insights. We are grateful to Neal Dalal, Junwu Huang, Davide Racco, Masha Baryakhtar, and Gray Rybka for useful feedback on our manuscript.
AA acknowledges the support of NSERC and the Stavros Niarchos Foundation. 
SD is grateful for support from the National Science Foundation under Grant No.~PHYS-1720397, the Gordon and Betty Moore Foundation Grant GBMF7946. SD, JT and MG would like to thank Perimeter Institute for their hospitality during the completion of this work. Research at Perimeter Institute is supported by the Government of Canada through Industry Canada and by the Province of Ontario through the Ministry of Economic Development \& Innovation.
LL acknowledges the support of NSERC and CIFAR.
KVT was supported by a Schmidt Fellowship funded by the generosity of Eric and Wendy Schmidt, by a grant-in-aid (No.~DE-SC0009988) from the U.S.~Department of Energy, and funding by the Gordon and Betty Moore Foundation through Grant GBMF7392. AA and KVT thank KITP for multiple long-term visits during the completion of this work, supported in part by the National Science Foundation under Grant No.~NSF PHY-1748958. 
\end{acknowledgments}

\appendix

\section{Bound states} \label{sec:boundstates}

In order to understand the collapse dynamics and resulting objects, we must first recall the spectrum of self-bound axion field configurations.  These are well-known in the literature \cite{Seidel:1991zh, Braaten:2015eeu, Braaten:2016dlp, Chavanis:2011zm, Chavanis:2011zi, Eby:2017azn, Eby:2017teq, Eby:2018ufi, Visinelli:2017ooc, Schiappacasse:2017ham, Mukaida:2016hwd, Salmi:2012ta} and can be split into two categories: \textit{solitons} and \textit{oscillons}.  The former are diffuse objects (with size $\gg 1/m$) held together by their self-gravity and stable over cosmological times.  The latter, on the other hand, are far more compact (with size $\mathcal{O}(1/m)$) and only metastable.

Since they are not our main focus in this paper, we do not go into much detail about these solutions except to recall a few important results about them.  First, solitons: field configurations bound by self-gravity and stabilized by kinetic pressure.  For $f \ll M_{\mathrm{Pl}}$, we can treat these configurations with a nonrelativistic approximation to the scalar Klein-Gordon equation.  This yields a Schr\"odinger equation that can be solved numerically \cite{Chavanis:2011zm} (and approximated analytically \cite{Chavanis:2011zi}) to yield the following:
\begin{enumerate}
\item Solitons are long-lived, with lifetimes far longer than the age of the Universe~\cite{Eby:2017azn, Eby:2018ufi}.  Fundamentally, this is because they are diffuse objects, with sizes much larger than $1/m$ and accordingly small velocities $v_\mathrm{sol}$.  Since self-interaction-induced radiation (i.e.~outgoing axions) is a relativistic effect, it is exponentially suppressed by a form factor $\sim e^{-1/v_\mathrm{sol}^2}$.

\item Solitons have a well-defined mass-radius relation given by \cite{Chavanis:2011zm}:
\begin{equation} \label{eq:solitonMassRadius}
R_{99} \sim \frac{9.95}{G_N m^2 M}
\end{equation}
where $R_{99}$ is the radius containing 99\% of the mass of the soliton.

\item Solitons have a fixed density profile which can be numerically obtained by solving the Schr\"odinger-Poisson equation.  Here we give an approximation to this profile.  Defining the scale radius as in Sec.~\ref{sec:gravcollapse} by $r_s^{\mathrm{sol}} \equiv \{r | \partial \ln \rho(r) / \partial \ln r = -2 \}$ and the scale density $\rho_s^{\mathrm{sol}} \equiv \rho(r_s^{\mathrm{sol}})$, the soliton's density profile is well-approximated by:
\begin{align}
\rho(r)  &\simeq \rho_0 \exp \left( - \frac{r^2}{2 R_0^2} \right) \qquad \mathrm{or} \nonumber \\
\rho(r)  &\approx \frac{\rho_0}{\left[ 1 + \frac{r^2}{2 n R_0^2} \right]^n }
\end{align}
where $\rho_0 \approx 2.945 \rho_s^{\mathrm{sol}}$ and $R_0 \approx 0.6530 r_s^{\mathrm{sol}}$.  The first of these approximations is accurate at small radii, while the second with $n=8$ is accurate to 10\% for $r \lesssim 3.2 r_s^{\mathrm{sol}}$. At asymptotically large radius, $\ln \rho(r) \propto -r$, as for e.g.~hydrogenic wavefunctions.

\item There is a maximum mass for solitons in potentials with attractive self-interactions.  At larger masses, configurations are unstable to a violent collapse and subsequent explosion due to the attractive self-interactions of the cosine potential (which can be seen in, for example, \cite{Chavanis:2016dab,Chavanis:2011zm,Chavanis:2011zi} and the simulations of Ref.~\cite{Levkov:2016rkk}).  For the cosine potential of Eq.~\ref{eq:cosinepot}, the critical mass $M_\mathrm{crit}$ can be estimated analytically to be:
\begin{equation} \label{eq:critsolmass}
M_{\mathrm{crit}} \simeq \sqrt{24 \pi^3} \frac{f M_{\mathrm{Pl}}}{m}.
\end{equation}

\end{enumerate}

We also briefly review oscillons, dense relativistic structures bound together and stabilized solely by self-interactions.\footnote{There has been disagreement in the literature over what to call these objects.  We use the word \textit{soliton} to refer to those objects which are bound by gravity and stabilized by kinetic pressure, while we use the word \textit{oscillon} to refer to those objects which are both bound and stabilized by self-interactions and kinetic pressure.  The former have also been referred to as ``dilute axion stars'' (e.g. in Ref.~\cite{Visinelli:2017ooc}).  The latter, meanwhile, have been referred to as ``dense axion stars'' (e.g.~in Refs.~\cite{Copeland:1995fq, Honda:2001xg, Visinelli:2017ooc, Braaten:2015eeu, Gleiser:2019rvw}) and, in a particularly confusing turn of events, ``solitons'' (e.g.~in Ref.~\cite{Bogolyubsky:1976yu}).}  They have also been studied in the literature (see e.g.~Refs.~\cite{Bogolyubsky:1976yu, Farhi:2005rz, Visinelli:2017ooc, Braaten:2015eeu,PhysRevD.98.023009}), although they are not nearly as well-understood as solitons.  For our purposes, however, we only need a few empirical observations about them, all of which we checked for a wide variety of initial conditions via numerical simulations described in App.~\ref{sec:numerics}:
\begin{enumerate}

\item The internal density of the oscillon is $\mathcal{O}(m^2 f^2)$, the natural scale associated with the potential of Eq.~\ref{eq:cosinepot}.

\item For small oscillons with sizes of $\mathcal{O}(\mathrm{few}/m)$, we can use the above density to obtain a rough estimate of the oscillon mass: $\mathcal{O}(10^{2-3} f^2/m)$, in agreement with our simulations.  Initial field configurations with substantially more mass tend to radiate it away in a transient burst and initial field configurations with substantially less tend to immediately disperse.

\item Because of their approximately constant internal density, oscillons have a mass-radius relation given by $M \propto R^3$.

\item The per-particle binding energies of the oscillons are not too large, of order $\mathcal{O}(0.1 m)$.  This can be inferred from the spectrum of emitted radiation at large distances (see App.~\ref{sec:numerics}).

\item Perhaps most importantly, for the cosine potential of Eq.~\ref{eq:cosinepot}, oscillons are only metastable, with relatively short lifetimes $\tau_{\mathrm{osc}} \lesssim \mathcal{O}(10^3/m)$.  They decay by emitting axion radiation until they reach a small enough mass such that self-interactions are no longer able to bind them.  At that point, they begin dispersing outward due to the repulsive kinetic pressure (and the expansion of the Universe).

This is observed in numerical simulations (see App.~\ref{sec:numerics}), but it is also known that different axion potentials can give support far longer-lived oscillons.  Potentials with lifetimes $\tau_{\mathrm{osc}} > \mathcal{O}(10^8/m)$ are known \cite{Olle:2019kbo}, and there is no clear upper bound.  Such long-lived objects may be cosmologically relevant, but at the moment we defer these questions for later work.  Since the longest-lived oscillons of the cosine potential have $\tau_{\mathrm{osc}} \lesssim \mathcal{O}(10^3/m)$, they will decay before matter-radiation equality if they are formed in the early Universe.

\end{enumerate}

\section{Numerical results} \label{sec:numerics}

To understand the dynamics of the axion field and extract its generic behavior under the conditions of interest, we employed several fully nonlinear, relativistic numerical simulations. These allow us in particular to develop and sharpen our analytic estimates for the time and length scales involved in a self-interaction-induced 
collapse (a highly nonlinear process) as well as to explore potential observable opportunities in gravitational waves. 
In Appendix~\ref{sec:numericspherical}, we discuss a set of spherically-symmetric studies, which we used primarily to probe the stability and lifetimes of oscillons in our various potentials. These studies are complemented with corresponding
analysis in an expanding Universe of which implementation details and results are
presented in~\ref{sec:numericexpand}.
We also employ fully three-dimensional simulations to confirm our estimates for the gravitational power radiated during a self-interaction-induced collapse.  These are discussed in Appendix~\ref{sec:numericgrav}.

\subsection{Spherically-symmetric simulations} \label{sec:numericspherical}

\begin{center}
\emph{Implementation}
\end{center}
For simplicity, we adopt Schwarzschild coordinates where the metric can be written as,
\begin{equation}
ds^2 = - \alpha^2 dt^2 + a^2 dr^2 + r^2 d\Omega^2\, \label{G2metric}.
\end{equation}
Thus the only relevant metric functions are the lapse function $\alpha(t, r)$
and $a(t,r)$. These coordinates become singular when a horizon forms but we study
weak regimes so this issue does not arise. In our implementation, we employ ``standard'' first order variables
as used in e.g.~Ref.~\citep{Choptuik:1992jv}, 
\begin{equation}
\Phi \equiv \phi'\, , \qquad \Pi \equiv \frac{a}{\alpha} \dot \phi \label{G2firstorderdef}\, ,
\end{equation} 
using the notation $\dot f = \partial_t f$ and $f' = \partial_r f$; 
rescaling both $(r,t)$ by $m^{-1}$ and $\phi$ by $f^{-1}$; and, for convenience, we also
introduce ${\cal R} \equiv \frac{f}{\sqrt{8 \pi}M_{P}}$. From the $rr$ and $rt$ components of Einstein's equations, we obtain
\begin{eqnarray}
\alpha' &=& \frac{\alpha}{2} \left[ r 8 \pi {\cal R}^2 \left( \frac{(\Phi^2+\Pi^2)}{2} - V a^2 \right) +\frac{a^2-1}{r} \right]\label{G2alpha}\, ,  \\
\dot a &=& 4 \pi \, \Phi \,\Pi \,\alpha \,a \, {\cal R}^2 \label{G2a}\, .
\end{eqnarray}
The first-order variables of the axion field then obey
\begin{eqnarray}
\dot \Pi &=& \left(\frac{\Phi \alpha}{a}\right)' + \frac{2 g \alpha}{r a} - \alpha a V'  \label{G2Pi}\, , \\
\dot \Phi &=& \left(\frac{\alpha}{a}\Pi\right)' \label{G2Phi}\, ,   \\
\dot \phi &=& \frac{\alpha}{a} \Pi\, .\label{G2phi}
\end{eqnarray}

To efficiently cover the large range of scales relevant in the problem, we employ a nonuniform radial grid defined by 
$r = \upsilon \tan(x)$ with $x\in[0,2\pi)$; $x$ is then uniformly discretized with $dx = \pi/(2 (N_x-1))$. Here,
$\upsilon=20$ is included for convenience and $N_x$ the number of points in our discretization.
The radial equation \ref{G2alpha} is solved at each given time while the evolution equations (\ref{G2a}-\ref{G2phi}) are employed to
obtain the scalar field behavior and the metric field $a$.  The radial integration is done through a Runge-Kutta 4th-order algorithm
integrating inwards with the asymptotic boundary condition $\alpha=1$; integration in time is performed with a  Runge-Kutta 3rd-order in time using the method of lines.
Spatial derivatives are computed with second (third, or fourth) order finite-difference operators satisfying 
summation by parts~\cite{STRAND199447,Calabrese:2003vx}. 
Regularity at the origin is addressed by using l'H\^opital's rule at $r=0$ to
regularize the equation. We employ maximally dissipative boundary conditions at the outer radial boundary. A small amount of 
artificial dissipation is added for convenience (for stabilility and convergence as well as for ensuring spurious high frequency behavior does not affect low frequency physics). For further details see \cite{Calabrese:2003yd,Calabrese:2003vx,Guzman:2007ua}.
Finally, in our simulations where we typically employ ${\cal R}=10^{-3},10^{-2}$, the timestep spacing is chosen as $dt = 10^{-1} dx$ to satisfy the Courant-Friedrichs-Levy (CFL) condition and accurately capture the rapid time-scale variations of the field dynamics.

\begin{figure}[tbp]
\includegraphics[width=0.46\textwidth]{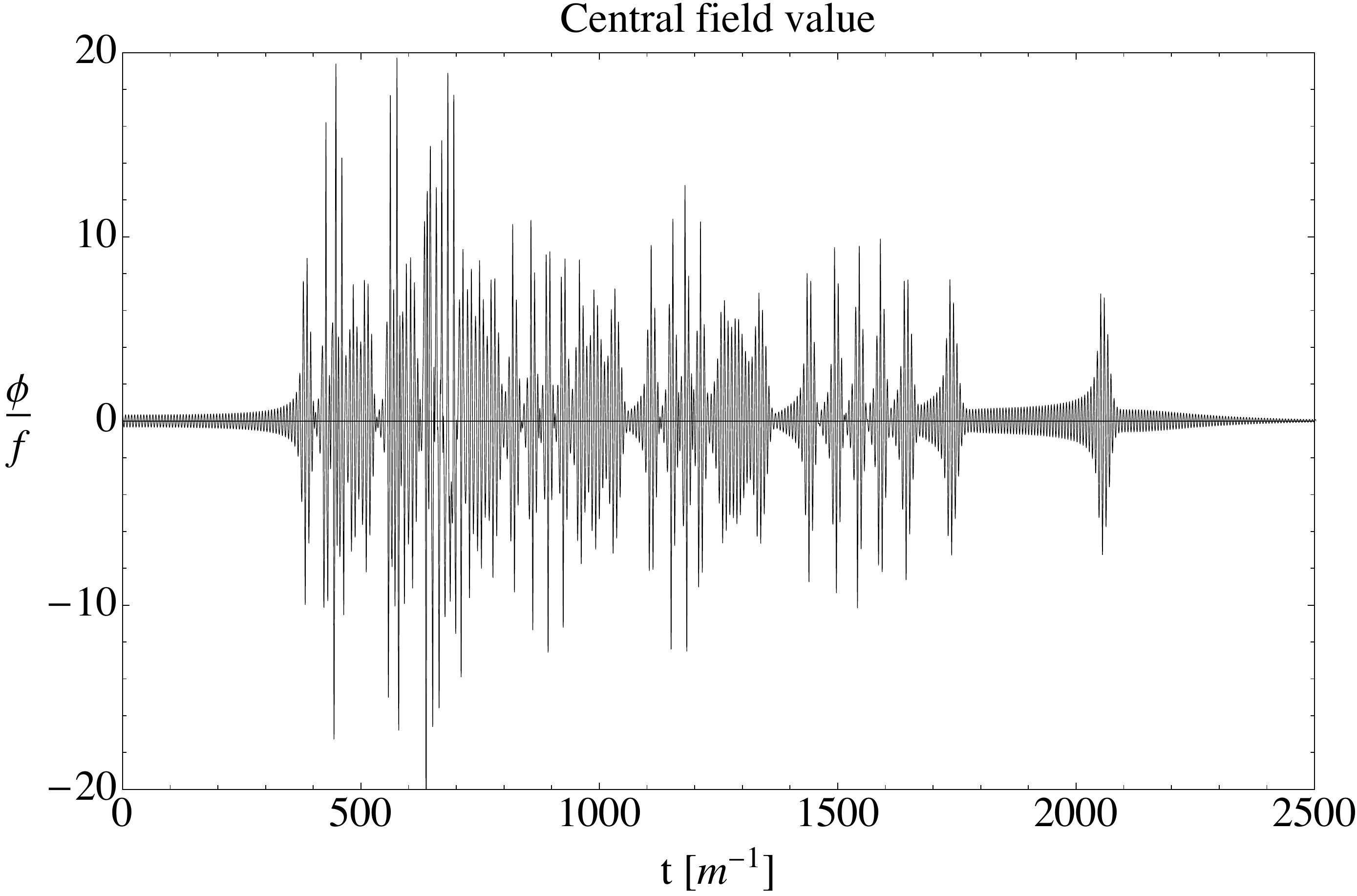}
\caption{Central value of the axion field for ${\cal R}=5 \times 10^{-3}$ using an initial condition given by Eq.~\ref{eq:initcondssim1D} with $M_0 = 2 \times 10^4$, $\sigma = 40$.  This simulation used a spherically-symmetric code which required far fewer computational resources than the 3D code used to generate Fig.~\ref{fig:fig_nonlinear3D_fieldcenter}, but the results are in agreement both qualitatively and quantitatively in terms of rough timescales and field excursions.}
\label{fig:fig_nonlinear1D_fieldcenter}
\end{figure}

\begin{figure}[tbp]
\includegraphics[width=0.46\textwidth]{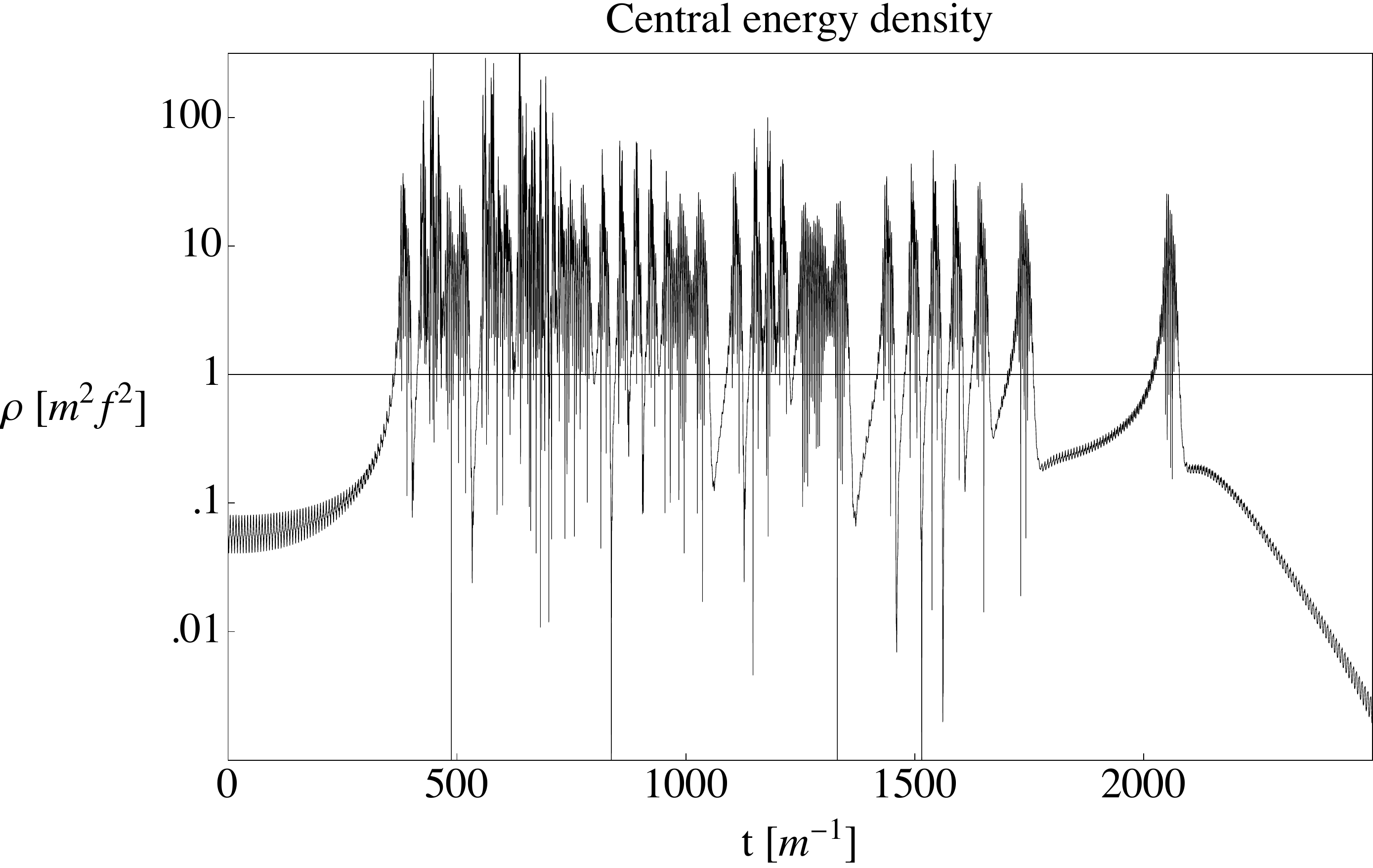}
\caption{Central value of the axion energy density for ${\cal R}=5 \times 10^{-3}$ using an initial condition given by Eq.~\ref{eq:initcondssim1D} with $M_0 = 2 \times 10^4$, $\sigma = 40$.  Note the spikes in density visible during the collapse and the eventual fuzzing out after enough energy has been lost.  While the configuration is far from perfectly periodic or regular, it is quite long-lived compared to its natural timescale of $m^{-1}$.}
\label{fig:fig_nonlinear1D_rhocenter}
\end{figure}

\begin{center}
\emph{Results and observations}
\end{center}
We first ran a set of spherically-symmetric simulations with initial conditions corresponding to subcritical and supercritical solitons.  The subcritical solitons remained stable for as long as we simulated ($>\mathcal{O}(10^6/m)$), while the supercritical solitons collapsed under the influence of self-interactions to an {\em oscillon of radius $R_o \approx 3/m$,} before violently radiating away enough of their energy to become subcritical and then fuzzing out to a subcritical soliton.  A typical central value of scalar field and density profile for a collapsing supercritical soliton is shown in 
Figs.~\ref{fig:fig_nonlinear1D_fieldcenter} and \ref{fig:fig_nonlinear1D_rhocenter} respectively. These have been obtained with initial
data defined as:
\begin{eqnarray}
\phi(t=0) &=& \sqrt{\pi/10} \, \sqrt{M_0/\sigma^3} e^{-r^2/(2\sigma^2)}, \nonumber \\ 
\partial_t \phi(t=0) &=& 0 \, . \label{eq:initcondssim1D}
\end{eqnarray}
In the plots here, we adopt $M_0 = 2 \times 10^4$, $\sigma = 40$, but we also simulated a variety of other masses and initial sizes and obtained qualitatively similar results in all cases. Our results here should be compared with those of Ref.~\cite{Levkov:2016rkk}, with which they are broadly consistent.

To establish that no ``stable oscillons'' (i.e.~states bound by self-interactions and long-lived on cosmological scales) could form, we also simulated several high-energy-density initial conditions, including initial states corresponding to solutions of the Schr\"odinger equation under a nonrelativistic assumption.  For the cosine potential (Eq.~\ref{eq:cosinepot}), we found metastable states for a wide variety of initial conditions, but no states that lived longer than $\mathcal{O}(10^3/m) = \mathcal{O}(1000) \, \mathrm{yr} \, \frac{10^{-22} \, \mathrm{eV}}{m}$.  They are thus not cosmologically long lived, so we do not expect any of them to be present in the late-time Universe. We note in passing that in finely-tuned configurations significantly longer lifetimes are in principle achievable (e.g.~\cite{Copeland:1995fq,Honda:2001xg}) though this possibility would not be generic.  For other potentials, such as those of Eqs.~\ref{eq:ratiopot} and \ref{eq:linwingspot}, we also simulated such initial conditions, and for these we were able to find metastable states with lifetimes at least $\mathcal{O}(10^5/m)$, at which point the simulations became computationally costly.  It is unknown what leads to such longevity in these potentials, and whether there is an upper bound on the lifetime of such oscillons. We reserve a careful study of this for future work, limiting ourselves in this paper only to outlining some of the observable consequences of cosmologically long-lived oscillons should they exist.

\begin{figure}[tbp]
\includegraphics[width=0.48\textwidth]{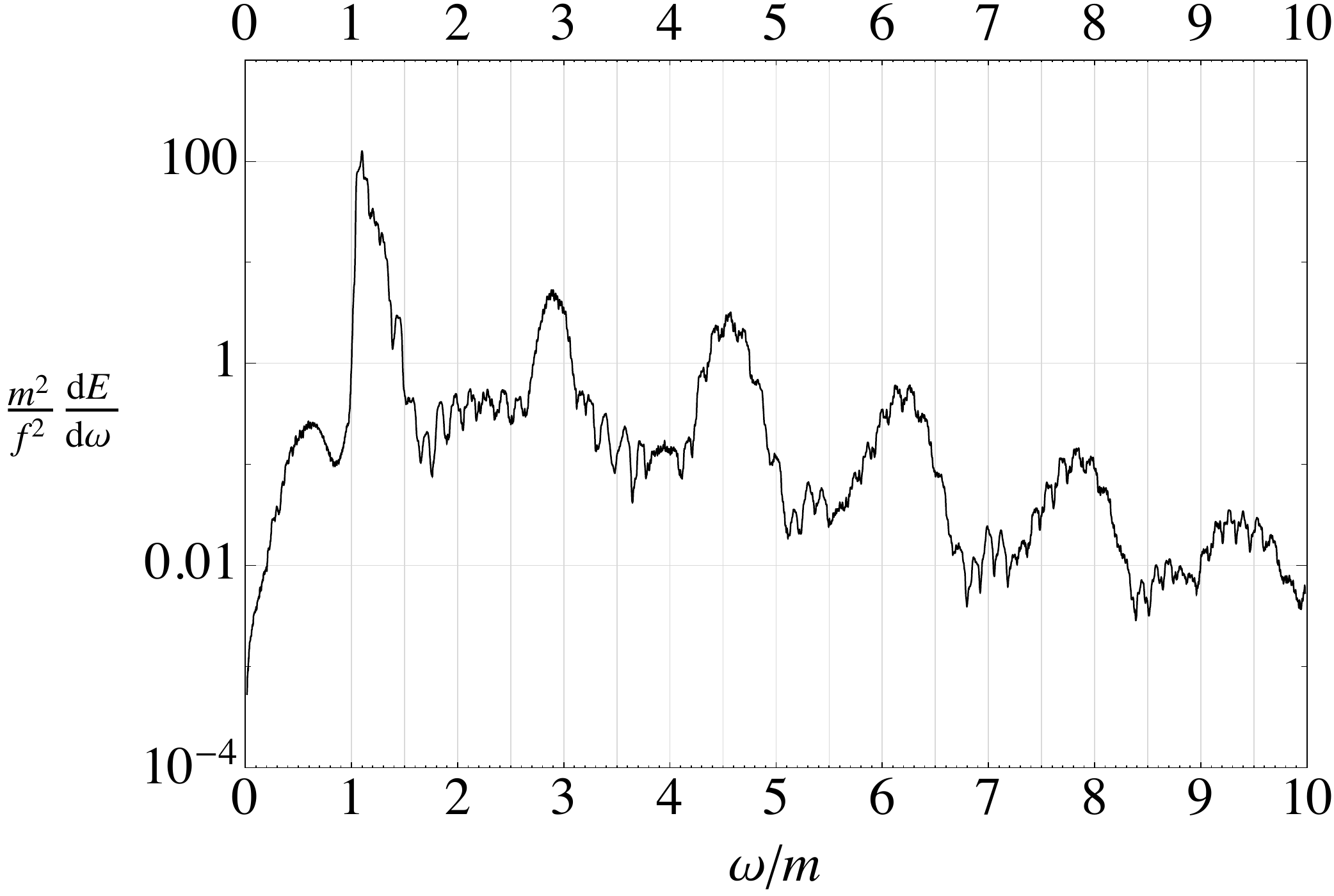}
\caption{Spectrum of outgoing radiation for ${\cal R}=5 \times 10^{-3}$ using an initial condition given by Eq.~\ref{eq:initcondssim1D} with $M_0 = 2 \times 10^4$, $\sigma = 40$.  This simulation used a spherically-symmetric code.  Note the peaks at peaks at $\approx 3 \omega_0, 5 \omega_0, 7 \omega_0,...$ where $\omega_0 \approx 0.9 m$.  These indicate that $3 \to 1$, $5 \to 1$, $7 \to 1$ etc.~processes are dominant contributors to scalar wave emission from oscillon-like field configurations.  The large nonrelativistic peak just above $m$ is due partly to transient radiation still present from our initial state and partly to the fact that the oscillon has not yet settled to its ground state (and because of its short lifetime, does not have time to before dispersing).}\label{fig:simplot_spectrum}
\end{figure}

By measuring the scalar field at large distances from the center, we were also able to extract the spectrum of outgoing scalar radiation.  We performed this analysis both for collapsing supercritical solitons and for the longest-lived metastable oscillon states we could produce, and in all cases the results showed clear peaks at energies $\omega$ approximately $3\omega_0$, $5\omega_0$, $7\omega_0$, \dots, where $\omega_0$ is the energy of the soliton or oscillon state and is slightly less than $m$ due to the state's binding energy.  A representative spectrum is shown in Fig.~\ref{fig:simplot_spectrum}.  This is consistent with self-interaction-induced $3 \to 1$, $5 \to 1$, $7 \to 1$, \dots processes being the dominant contributors to scalar emission from oscillons and collapsing solitons, which is in turn consistent with the fact that all metastable oscillon states we observed were small in size (with radius of order $1/m$).

\subsection{Expanding-universe simulations} \label{sec:numericexpand}

\begin{center}
\emph{Implementation}
\end{center}

In Sec.~\ref{sec:quarticcollapse}, we simulated the collapse of fractionally small, spherically symmetric overdensities in an expanding Universe, according to Eq.~\ref{eq:wavepacketpde} with the initial conditions of Eq.~\ref{eq:wavepacketic}. We used Mathematica 11.3's \texttt{NDSolve} routine \cite{Mathematica} with the adaptive time resolution (in $t_m$ space) of \texttt{MethodOfLines}. Anticipating the need for higher spatial resolution near the origin due to the collapse of the wavepackets, we transformed the partial differential equation on a discretized spatial grid uniform in $r_m^{1/2}$ using \texttt{SpatialDiscretization} and \texttt{TensorProductGrid}. The spatial resolution was allowed to float dynamically up to $\texttt{MaxPoints} = 2 \times 10^6$: for the simulation in Fig.~\ref{fig:quarticcollapse}, we started with minimum number of $\texttt{MinPoints} = 600$ initial spatial lattice points evenly distributed in $r_m^{1/2}$ space between $r_{m,\mathrm{min}} = 2.5 \times 10^{-3}$ and $r_{m,\mathrm{max}} = 25$.  We employed Neumann boundary conditions at spatial boundaries, and checked that results were independent of the box size, i.e.~$r_{m,\mathrm{min}}$ and $r_{m,\mathrm{max}}$. (Because of the presence of the zero mode at the outer boundary, dissipative boundary conditions would lead to unwanted artefacts.) The time resolution was also dynamically variable but was never allowed to exceed a time step in $t_m$ space larger than $\texttt{MaxStepSize} = 2 \times 10^{-3}$. Numerical convergence and robustness of the obtained results was verified by varying the spatio-temporal resolution as well as slightly changing the initial conditions, and inspecting if the qualitative features of the numerical solution were the same. 

\begin{center}
\emph{Results and observations}
\end{center}
The results of the simulation in Fig.~\ref{fig:quarticcollapse} are described in Sec.~\ref{sec:quarticcollapse}, and serve as a bridge between the linear theory, the collapse into an oscillon-like configuration, and the subsequent evaporation. We have performed similar simulations for larger-radius wavepackets, which we found to collapse into larger-mass oscillon states with somewhat longer lifetimes, and more complicated behavior in the nonlinear regime. Another simulation with exactly the same parameters as in Fig.~\ref{fig:quarticcollapse} save for the opposite sign of $\delta_0 = - 10^{-2}$, i.e.~a linear \emph{underdensity}, reveals that underdensities also grow via parametric resonance but do not produce implosions, instead the growth of fluctuations turns off smoothly when nonlinearity is reached.

We have also run simulations for a handful of multi-scale configurations, e.g.~two superimposed wavepackets of different radii. In those cases, we found that the collapse of the small wavepacket did not prevent the collapse of the larger wavepacket. In order to study the interactions between oscillons and to understand the mode mixing over a large range of scales, simulations with a larger dynamic range in both time and space would be helpful.

\subsection{Three-dimensional simulations} \label{sec:numericgrav}

We also ran a few fully three-dimensional simulations incorporating full general relativity in order to study gravitational wave radiation from an asymmetric collapsing cloud.  As discussed in Sec.~\ref{sec:gravcollapse}, an axion cloud collapsing under the influence of self-interactions in the early Universe will in general be asymmetric.  This asymmetry will lead to gravitational wave radiation, but in order to estimate the actual power radiated, we must know how long it takes for the collapsing structure to radiate away its asymmetry.
Our fully consistent 3D simulations allow us to follow the
behavior of the scalar field and compute, in particular, the gravitational radiation emitted by the system and contrast it with our analytical estimates discussed in Sec.~\ref{sec:GW}.

\begin{center}
\emph{Implementation}
\end{center}
We employ the \textsc{had} \cite{had} computational infrastructure to efficiently study our system of interest, described by a scalar field minimally coupled to the Einstein
equations in 3D (subject to the cosine potential for concreteness). 
This infrastructure provides distributed,  adaptive mesh refinement Berger-Oliger style 
AMR \cite{had,lieb} with full sub-cycling in time, together with an improved treatment of artificial boundaries \cite{lehner}.
We adopt the CCZ4 formulation of Einstein equations (for details see Ref.~\cite{PhysRevD.95.124005}). Discretization is achieved
through finite difference schemes based on the Method of Lines on a regular Cartesian grid. A fourth-order accurate spatial discretization satisfying the summation by parts rule, together with a third order accurate (Runge-Kutta) time integrator, are 
used to achieve stability of the numerical implementation~\cite{Calabrese:2003vx,Calabrese:2003yd,ander}.

Our simulations are performed in a domain $[-1600/m,1600/m]^3$ with a coarse resolution of $\Delta x_{1}=40/m$ and allow up to 7 levels of refinement which automatically adapt through a self-shadow hierarchy to ensure the error in the solution
is kept below $4 \times 10^{-4}$ (thus, the minimum resolution is $0.3125/m$). As observed in the spherically symmetric studies,
the system goes through a rather violent temporal oscillation --even when relevant spatial wavelengths are relatively long--,
due to the source dependence on ${\cal R}$. We thus adopt a small Courant parameter of $\lambda_c \approx 10^{-2}$ such that $\Delta t_{l} = \lambda_c \, \Delta x_{l}$ on each refinement level $l$ to guarantee that the CFL condition is satisfied
and relevant physical behavior is accurately captured. Previous related work with this infrastructure 
(e.g.~\cite{Barausse:2012da,Hirschmann:2017psw,Sagunski:2017nzb,Palenzuela:2017kcg}) have thoroughly tested
the implementation. Here we have further verified its suitability for our current purposes through convergent studies.
Armed with this implementation, we study asymmetric initial configurations---which are also weakly gravitating---defined 
in a similar way as in Eq.~\ref{eq:initcondssim1D},
\begin{eqnarray}
\phi(t=0) &=& \sqrt{\pi/10} \, \sqrt{M_0/\sigma^3} e^{-\hat r^2/(2\sigma^2)} \nonumber \\ 
\partial_t \phi(t=0) &=& 0 \, ; \label{eq:initcondssim}
\end{eqnarray}
with $\hat r^2 = (o_1 x)^2 + (o_2 y)^2 + z^2$. The parameters $\{o_1,o_2\}$ are chosen to define nonspherical initial
configurations and explore the radiative properties of the system in the gravitational and scalar sectors.

\begin{center}
\emph{Results and observations}
\end{center}
We have run several cases described by $M_0=\{ 5 \times 10^3, 2 \times 10^4 \}$, $o_1=o_2=\{1, 1.25, 1.5\}$, $\sigma=\{20,40\}$ and $o_1=1, o_2=\{ 1.25, 1.5\}$, for 
${\cal R}=\{ 10^{-2}, 5 \times 10^{-3}, 10^{-3}\}$ to scan a range of relevant cases that could be studied with reasonable computational resources---typically a month of running employing 40 processors. As we show below, the overall
behavior follows closely that observed in our extensive 1D studies and the combined information provides a clear
picture of the axion field's dynamics in the nonlinear regime.

\begin{figure}[tbp]
\includegraphics[width=0.46\textwidth]{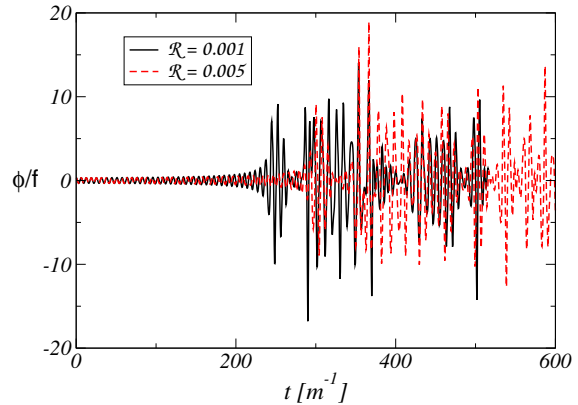}
\caption{Central value of the axion field for ${\cal R}=10^{-3}, 5\times 10^{-3}$.
Note this plot shows the same qualitative features (and rough quantitive timescales and field ranges) 
as those of Fig.~\ref{fig:fig_nonlinear1D_fieldcenter}.
}\label{fig:fig_nonlinear3D_fieldcenter}
\end{figure}

All cases progress in a similar manner. Initially, much like what is seen in the spherically symmetric case, 
a transient stage lasting a few $\approx 100/m$ 
shows the field oscillating with frequency $2 \pi m$ and slowly radiating---mainly through the scalar channel. 
Then, through a rather sudden change, the scalar field extent
of initial size $\approx \sigma$ collapses to a size of $\approx 3/m$, which is followed by strong oscillations interspersed with phases describing a modest expansion and recollapse. Figure \ref{fig:fig_nonlinear3D_fieldcenter} illustrates such behavior by showing
the central value of the scalar field for  $\{M_0=2 \times 10^4,\sigma=40,o_1=1, o_2=1.5\}$. It is during the collapsing 
stages that gravitational radiation is mainly produced at bursts due to sudden changes in the source.

\begin{figure}[tbp]
\includegraphics[width=0.46\textwidth]{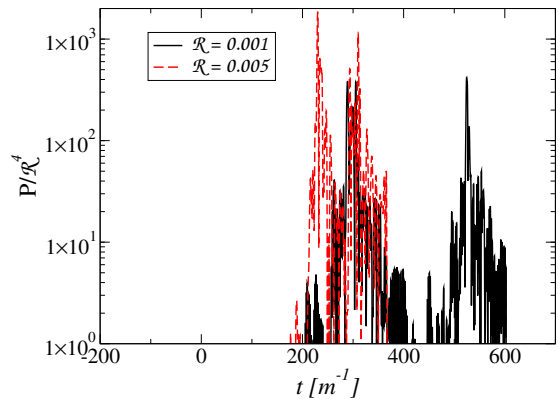}
\caption{Emitted power of gravitational waves vs time for two representative
cases (with ${\cal R}=10^{-3}, 5 \times 10^{-3}$ and $o_1=o_2=1.25$) normalized with the 
expected ${\cal R}^4$ dependency.}\label{fig:fig_nonlinear3D_powerrad}
\end{figure}

\begin{figure}[tbp]
\includegraphics[width=0.46\textwidth]{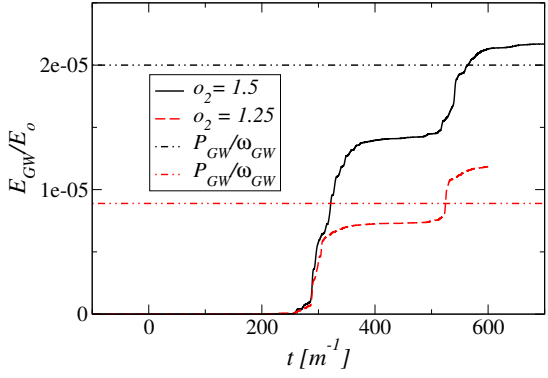}
\caption{Total radiated energy, relative to the initial mass of the field configuration, 
vs time for $M_0=2 \times 10^4$, $\sigma=40$ and $o_1=o_2=\{1.25, 1.5\}$ together with their corresponding estimates using the quadrupole formula and the leading gravitational wave frequency.}\label{fig:fig_nonlinear3D_enerrad}
\end{figure}

\begin{figure}[tbp]
\includegraphics[width=0.22\textwidth]{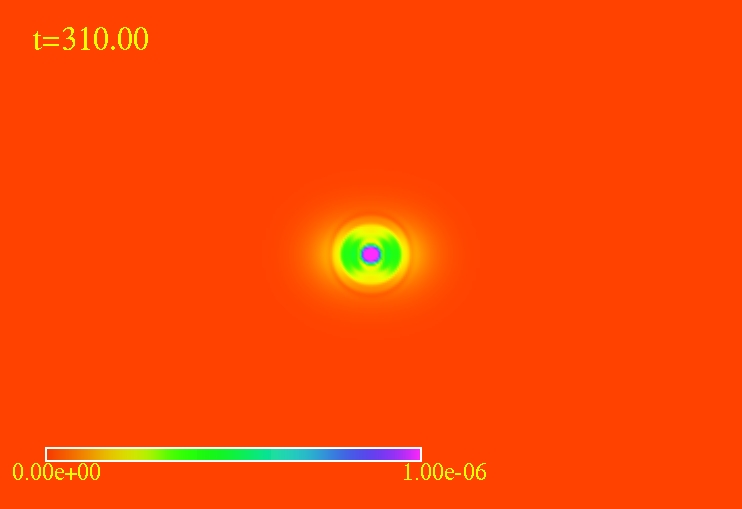}
\includegraphics[width=0.22\textwidth]{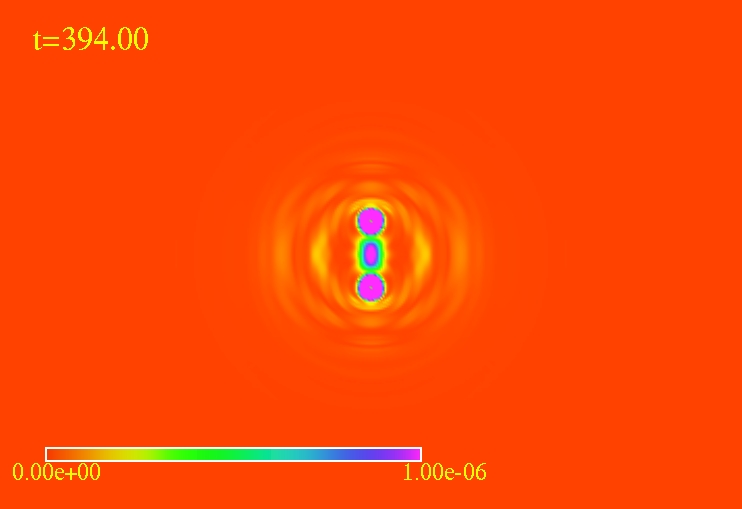}\\
\includegraphics[width=0.22\textwidth]{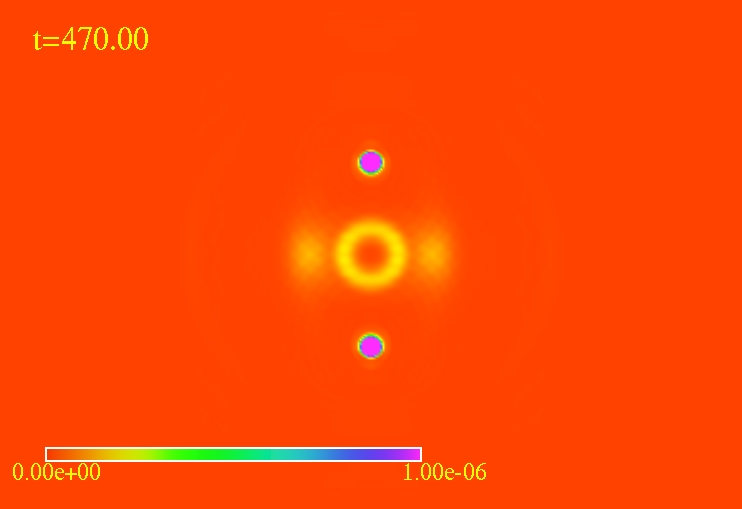}
\includegraphics[width=0.22\textwidth]{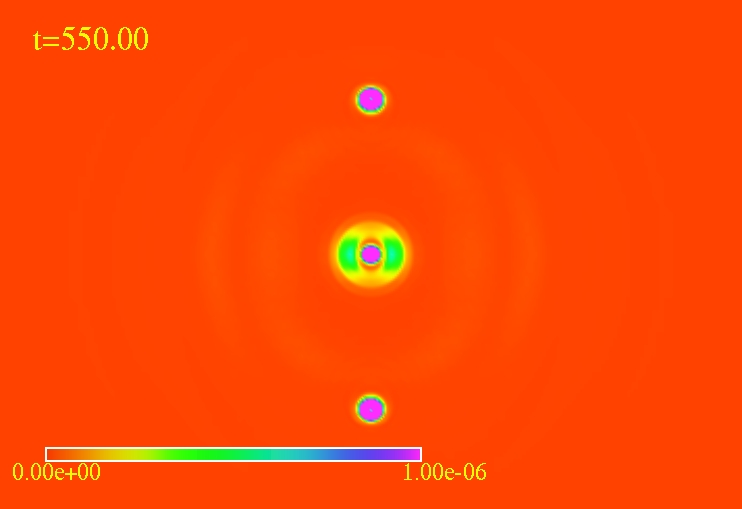}
\caption{Representative snapshots at the equatorial plane of the
scalar field density $\rho=T_{00}$ at four different times $t_m=310,394,470,500$. As the scalar field in the
central region oscillates, two scalar field ``blobs'' are expelled from the central region.}\label{fig:rhosnaps}
\end{figure}

The power emitted in gravitational waves agrees with Eq.~\ref{eq:OmegaGW} presented in Sec.~\ref{sec:GW}. This is illustrated in 
Fig.~\ref{fig:fig_nonlinear3D_powerrad} which shows the power radiated in gravitational waves for asymmetric initial configurations ($o_1=1, o_2=1.5$) normalized by ${\cal R}^{-4}$ ($\propto f^{-4}$). Also, Fig.~\ref{fig:fig_nonlinear3D_enerrad} depicts the cumulative energy radiated (until time $t$) versus time
for two asymmetric configurations ($o_1=1, o_2=\{1.25,1.5\}$) normalized with respect to the initial mass of the
axion configuration (described by $M_0=2 \times 10^4, \sigma=40$).
Importantly, we note that as time progresses the matter distribution approaches a spherical shape mainly due to: (i) 
significant scalar field radiation, and (ii) ``gravitational cooling'', where scalar field ``blobs''
with masses $\approx 1\%$ of the initial mass are shed and propagate 
away\footnote{We note in passing that analogue behavior has also been observed
in other settings involving scalar field nonlinear interactions, e.g. scalar field 
collapse~\cite{Seidel:1991zh,Sanchis-Gual:2019ljs} and boson star 
collisions~\cite{Palenzuela:2006wp,Palenzuela:2007dm,Palenzuela:2017kcg}.} from the oscillon  
at $v \approx 0.2 \, c $. 
This latter behavior is illustrated in Fig.~\ref{fig:rhosnaps}, corresponding to the case
$M_0=2 \times 10^4$, $\sigma=40$ and $o_1=o_2=1.5$. Both of those
processes together with gravitational wave emission weaken gravitational radiation as time progresses, as can
be appreciated in Figs.~\ref{fig:fig_nonlinear3D_powerrad} and~\ref{fig:fig_nonlinear3D_enerrad}.

\section{Isocurvature constraints} \label{sec:isocurvature}

\begin{figure}[ht]
\includegraphics[width=0.47\textwidth]{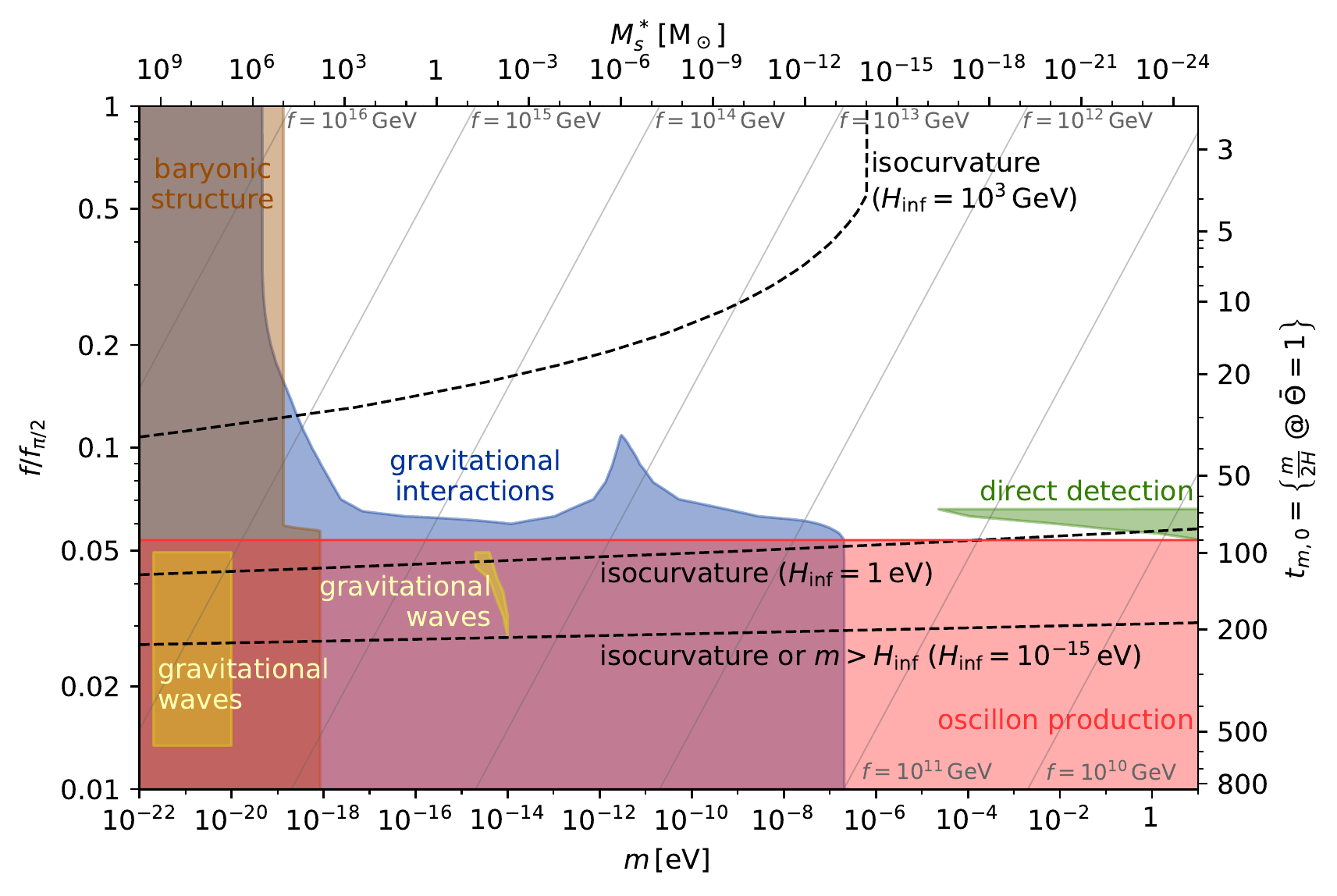}
\caption{Same as Fig.~\ref{fig:summary}, with isocurvature constraints (lower bounds on $f$) indicated by dashed black lines for three different inflationary scales $H_\mathrm{inf}$.}
\label{fig:iso}
\end{figure}

Isocurvature fluctuations may also place constraints on large-misalignment axions in some models. Provided the field is light during inflation ($m \ll H_{\mathrm{inf}}$), we compute constraints on the axion parameter space as a function of $H_{\mathrm{inf}}$, shown in Fig.~\ref{fig:iso}.  However, as we discuss briefly in Sec.~\ref{sec:flat}, the axion can be much heavier during inflation ($m \gg H_{\mathrm{inf}}$) if it has a temperature-dependent potential, and in this case we will see that isocurvature fluctuations are substantially suppressed and thus provide no constraint on the axion parameter space.  The dashed lines shown in Fig.~\ref{fig:iso} assume the former, but should not be interpreted as absolute constraints given the above discussion.

Any scalar field $\phi$ with $m \ll H_{\mathrm{inf}}$ present during inflation will pick up fluctuations on all scales of order
\begin{align} \label{eq:inflationfluctuations}
\delta \phi \sim \frac{H_{\mathrm{inf}}}{2 \pi}
\end{align}
where $H_{\mathrm{inf}}$ is the Hubble scale during inflation~\cite{Bunch:1978yq}.  In our case, where $\phi$ is the axion field, this translates into fluctuations in the misalignment angle of order $\delta \Theta \sim H_{\mathrm{inf}} / (2 \pi f)$.  The \textit{Planck} collaboration constrains such fluctuations to be small \cite{Akrami:2018odb}, and requiring this will constrain $f$ to be larger than some minimal value that depends on $H_{\mathrm{inf}}$.

We are primarily interested in the case where the field starts near the top of the potential since this is where all our signatures come from.  For the pure cosine potential of Eq.~\ref{eq:cosinepot}, we have from Section \ref{sec:linGR} that when the field starts with an initial misalignment angle $\Theta_0$ near the top, the late-time density $\rho$ is proportional to:
\begin{align}
\rho &\propto 0.2 \left[ t_m^{\text{osc}} + 4 \ln t_m^{\text{osc}} \right]^2; \\
t_m^{\text{osc}} &\equiv \ln \left[ \frac{1}{\pi - | \Theta_0 |} \frac{2^{1/4} \pi^{1/2}}{ \Gamma \left( \frac{5}{4} \right) }\right] .
\end{align}
Fluctuations of order $\delta \Theta_0$ in the initial misalignment angle $\Theta_0$ translate into late-time density fluctuations $\delta_{\mathrm{iso}}$ of order:
\begin{align}\label{eq:isocurvature}
\delta_{\mathrm{iso}} \equiv \frac{\delta \rho}{\rho} = \frac{\mathrm{d} \rho}{\mathrm{d} \Theta_0} \frac{\delta \Theta_0}{\rho} \simeq C \frac{\delta \Theta_0}{(\pi - |\Theta_0|) \ln \pi / (\pi - |\Theta_0|)}
\end{align}
where $C$ is a constant that varies between roughly 1.5 and 2.5 with weak dependence on $\Theta_0$.

\textit{Planck} requires that isocurvature fluctuations in the power spectrum be subdominant to the measured adiabatic fluctuations by a factor of roughly $10^{-2}$ \cite{Akrami:2018odb}.  Since the adiabatic fluctuations in the power spectrum are $\mathcal{O}(10^{-9})$ this means $\delta_{\mathrm{iso}} \lesssim \sqrt{10^{-2} 10^{-9}} \sim 10^{-5.5}$.  For fixed $H_{\mathrm{inf}}$, this translates into a constraint on the smallest allowable $f$ or, equivalently, a constraint on the maximum allowable tuning for $\Theta_0$.  

In the regime where $|\Theta_0|\ll 1$, the energy density in axions is $\rho \propto m^2 f^2 \Theta_0^2$ and Eq.~\ref{eq:isocurvature} reduces to:
\begin{align}
\delta_{\mathrm{iso}}=\frac{H_{\mathrm{inf}}}{\pi f |\Theta_0|}
\end{align}
The product $f |\Theta_0|$ for a given axion mass $m$ is fixed by the requirement that the axion carries all the DM density today. Taking into account the analysis in Sec.~\ref{sec:linGR}, a bound on $\delta_{\mathrm{iso}}$ is thus equivalent to a constraint on $f_{\pi/2}$. In turn, this translates to an upper bound on the axion mass $m$ which is now a function of  $H_{\mathrm{inf}}$, scaling roughly roughly as $H_{\mathrm{inf}}^{-4}$.  The above discussion of the two extremes, $\pi-|\Theta_0| \ll1$ and $|\Theta_0| \ll 1$, explains the asymptotic behavior of the exact bounds shown in Fig.~\ref{fig:iso} which have been derived for an arbitrary $\Theta_0$.

With an understanding of the above, we turn to the case where the temperature-dependence of the axion potential causes it to be heavy during inflation ($m \gg H_{\mathrm{inf}}$).  In this case it still picks up fluctuations, but they are substantially suppressed when averaging over the scales measured in the CMB~\cite{Vilenkin:1982wt}:
\begin{equation} \label{eq:inflationfluctuationsheavy}
\delta \phi \propto \frac{H_{\mathrm{inf}}}{m \ell_\mathrm{CMB}^{3/2}}
\end{equation}
where $\ell_\mathrm{CMB} \sim 10\,\mathrm{Mpc} \sim 1/(10^{-30}\,\mathrm{eV})$ is the smallest length scale that can be probed with the CMB.  In this case Eq.~\ref{eq:isocurvature} picks up a similar suppression, and so for $m \gtrsim H_{\mathrm{inf}}$, isocurvature fluctuations are suppressed far below any level of detectability and thus provide no constraint on the axion parameter space.

\section{Low-frequency gravitational wave detection}\label{sec:gwprospects}

For gravitational waves of frequency below $10^{-7}\,\mathrm{Hz}$, there are three known detection methods: astrometry, pulsar timing arrays, and the CMB. Here, we briefly review each method and discuss their sensitivity as presented in Fig.~\ref{fig:GWest}.

\begin{center}\emph{Astrometry} \end{center}

Stochastic gravitational waves cause an apparent distortion of the position of background sources on the celestial sphere~\cite{braginsky1990propagation}. At low frequencies, where the GW frequencies are smaller than the inverse integration time of the observations, the time derivative of this distortion will manifest itself as a stochastic proper motion of e.g.~extragalactic sources, which should otherwise appear nearly stationary by account of their large line-of-sight distance. The GW abundance is related to this stochastic proper motion as~\cite{gwinn1997quasar,book2011astrometric,mignard2012analysis}:
\begin{align}
\Omega_\mathrm{GW} = \frac{\langle \mu^2 \rangle}{H_0^2} = \frac{6}{5}\frac{1}{4\pi H_0^2} \sum_{m = -2}^2 \sum_{i=1}^2 \left \langle \big|s_{\ell=2,m}^{(i)}\big|^2 \right\rangle.
\end{align}
In the second equation, we used the fact that $5/6$ of the expected signal is contained in the quadrupole ($\ell = 2$) modes, if one decomposes the proper motion field as $\vect{\mu} = \sum_{\ell, m}  s_{\ell m}^{(1)} \vect{\Psi}_{\ell m} + s_{\ell m}^{(2)} \vect{\Phi}_{\ell m}$, where $\vect{\Psi} = \nabla Y_{\ell m} / \sqrt{\ell (\ell + 1)}$ and $\vect{\Phi} = \hat{\vect{r}} \times \vect{\Psi}$ are the (orthonormal) spheroidal and toroidal vector spherical harmonics, respectively. 

The variance at which any low-$\ell$ mode coefficient can be measured with $N$ uniformly distributed sources measured with proper motion standard deviation $\sigma_{\mu}$, is $\sigma^2(s_{\ell m}^{(i)}) \simeq 4\pi \sigma_\mu^2 / N$. Therefore, the expected precision $\delta \Omega_\mathrm{GW}$ to which one could measure the stochastic background is:
\begin{align}
    \delta \Omega_\mathrm{GW} \simeq \frac{12}{H_0^2} \frac{\sigma_\mu^2}{N} \approx 6\times 10^{-8} \left(\frac{\sigma_\mu}{\mathrm{\mu as\,y^{-1}}}\right)^2 \left(\frac{10^6}{N}\right). \label{eq:OmegaGWastrometry}
\end{align}
If low-$\ell$ systematics can be held under control, which is a challenge~\cite{lindegren2018gaia,mignard2018gaia}, then \textit{Gaia} is projected to reach a limit of $\Omega_\mathrm{GW} < 0.006$ after its nominal 5-year mission time with its current catalogue of 556,869 quasars~\cite{Darling:2018hmc}. With a likely quadrupling of the catalogue size and a mission extension to 10 years, further improvements by a factor of 1/32 in $\Omega_\mathrm{GW}$ can be
expected. (Note that a statistics-limited $\sigma_\mu^2 \propto \tau_\mathrm{int}^{-3}$ scales as the inverse cube of the integration time $\tau_\mathrm{int}$.) Astrometry with radio interferometers is also a promising avenue, as evidenced by the constraint $\Omega_\mathrm{GW} < 0.0064$ at 95\% confidence level (CL) with 711 radio sources observed by the Very Large Baseline Array (VLBA)~\cite{Darling:2018hmc}.
Future astrometric missions---either space-based, optical satellites~\cite{hobbs2016gaianir,malbet2012high,boehm2017theia} or ground-based, radio interferometers such as SKA~\cite{fomalont2004microarcsecond}---can potentially attain sensitivities of $\delta \Omega_\mathrm{GW} \sim 10^{-8}$ with large and precise catalogues over long integration times.
Proper accelerations of quasars (SKA) or galactic stars (\textit{Gaia}, \textit{Theia}) $\vect{\alpha} \equiv \dot{\vect{\mu}}$ can also be used to search for stochastic gravitational waves at low frequencies $f \lesssim 1/\tau_\mathrm{int}$.\footnote{This fact has, to our knowledge, not yet been appreciated in the literature.} Their sensitivity in terms of $\delta \Omega_\mathrm{GW}$ is parametrically worse by a factor of $\sim 1/(f \tau_\mathrm{int})^2$, but they offer the possibility of much larger and more precise catalogues, as Galactic stars have tiny intrinsic proper accelerations (but generally large proper motions).

\begin{center} \emph{Pulsar timing arrays (PTA)} \end{center}

Stochastic gravitational waves produce random changes in the times-of-arrival of pulses from individual pulsars. The effects can be inferred from cross-correlation of timing residuals of two pulsars~\cite{Siemens_2013}. The sensitivity improves with increasing pulsar stability $\sigma$, observation time $t_\mathrm{int}$ and decreasing cadence
(i.e.~the time $\Delta t$ between two observations of the same pulsar). Using the prescription of Refs.~\cite{maggiore1,maggiore2} and \cite{Moore_2014,Moore_2015}, and assuming that our signal is peaked around frequency $f_\text{GW}$ with a spread of $\Delta f\sim f_\text{GW}$, the sensitivity of a pulsar network consisting of $N_p$ pulsars is given by:
\begin{align}
\begin{split}
    H_0^2\Omega_{\text{GW}}\approx 6\times 10^5\frac{\Delta t\,\sigma^2}{N_p \sqrt{t_\mathrm{int}}}f_{\text{GW}}^{9/2}.
    \end{split}
    \label{eq:OmegaTh}
\end{align}

The above equation applies when $t_\mathrm{int}^{-1}\lesssim f_\text{GW}\lesssim \Delta t^{-1}$; outside this frequency range there is essentially no sensitivity to GW radiation. In Eq.~\ref{eq:OmegaTh}, we have assumed a detection SNR threshold $\varrho_{\text{th}}=3$~\cite{Moore_2015}. In Fig.~\ref{fig:GWest}, we present our estimates for current and future pulsar timing experiments. In particular, we indicate sensitivities corresponding to EPTA~\cite{Kramer:2013kea},
IPTA~\cite{Perera:2019sca}, and SKA~\cite{Carilli:2004nx} assuming 5, 20 and 100 pulsars followed for 10, 15
and 30 years respectively. The apparent steady improvement
in sensitivity of PTA efforts indicate tantalizing prospects for
detection/constraints in the $10^{-15}\text{--}10^{-14}\,\mathrm{eV}$ range. 

\begin{center} \emph{CMB, BBN, and large-scale structure} \end{center}

GWs produced deep in the radiation dominated era contribute to the total radiation that drives the expansion of the Universe and can have an imprint on the CMB as well as on Big Bang Nucleosynthesis (BBN). Their energy contribution is indistinguishable from that of relativistic neutrinos and can thus be parametrized as a relativistic degree of freedom $N_{\text{GW}}$, contributing to $N_{\text{eff}}$. The \textit{Planck} \cite{Aghanim:2018eyx} limit  on $N_{\text{eff}}$ can then be translated into a bound on $\Omega_{\text{GW}}$. In Fig.~\ref{fig:GWest}, we plot the limits calculated by Ref.~\cite{PAGANO2016823}, where the 2015 \textit{Planck} polarization data in the SimLow likelihood together with the Planck Lensing likelihood and BAO observations at $95\%$ C.L.~was used. Future satellite missions such as EUCLID~\cite{euclid} will improve the bound by more than one order of magnitude. The corresponding dashed lines on Fig.~\ref{fig:GWest} come from simulations of mock data (see Ref.~\cite{PAGANO2016823} for further details). The BBN bound is relevant only for structures that collapse at $z\gtrsim 4\times 10^{8}$ and is of the order of the CMB bound.

\bibliography{compact_axion_structures}
\end{document}